\documentclass[superscriptaddress,aps,pra,twocolumn, showpacs,nofootinbib,longbibliography,notitlepage]{revtex4-2}

\UseRawInputEncoding
\usepackage{pgfplots}
\usepackage{etex}
\usepackage{physics}
\usepackage{amsmath,amssymb,amsthm,amstext,mathrsfs}
\usepackage{times,txfonts}
\usepackage{braket}
\usepackage{subcaption}
\usepackage{color}
\usepackage{natbib}
\usepackage{blkarray}
\usepackage{mathtools}
\usepackage{latexsym}
\usepackage{tabularx, booktabs}
\usepackage{graphics,epstopdf}
\usepackage{float}
\usepackage{graphicx}
\usepackage{amsfonts}
\usepackage{subcaption}
\usepackage{enumerate}
\usepackage{color,soul}
\usepackage{comment}
\usepackage[inkscapelatex=true]{svg}
\usepackage{hyperref}
\hypersetup{
	colorlinks=true,
	linkcolor=red,
	filecolor=blue,      
	urlcolor=blue,
	citecolor=purple,
}

\newcommand{\be}{\begin{equation}}
	\newcommand{\ee}{\end{equation}}
\newcommand{\ba}{\begin{eqnarray}}
	\newcommand{\ea}{\end{eqnarray}}

\newtheorem{defi}{Definition}

\newtheorem{thm}{Theorem}

\newtheorem{Claim}{Claim}

\DeclareMathOperator*{\Motimes}{\text{\raisebox{0.25ex}{\scalebox{0.65}{$\bigotimes$}}}}

\begin{document}

\title{Robust self-testing of the $m-$partite maximally entangled state and observables } 

\author{Ritesh K. Singh}
\email{riteshcsm1@gmail.com}
\affiliation{Department of Physics, Indian Institute of Technology Hyderabad, Telengana-502284, India}

\author{Souradeep Sasmal}
\email{souradeep.007@gmail.com}
\affiliation{Institute of Fundamental and Frontier Sciences, University of Electronic Science and Technology of China, Chengdu 611731, China}

\author{A. K. Pan}
\email{akp@phy.iith.ac.in}
\affiliation{Department of Physics, Indian Institute of Technology Hyderabad, Telengana-502284, India}


\begin{abstract}
As quantum technologies continue to advance rapidly, the device-independent testing of the functioning of a quantum device has become increasingly important. Self-testing, a correlation based protocol, enables such  certification of a promised quantum state as well as measurements performed on it without requiring knowledge of the device's internal workings. This approach typically relies on achieving the optimal quantum violation of a suitable Bell inequality. Self-testing has been extensively investigated in the context of bipartite Bell experiments. However, its extension to multipartite scenarios remains largely unexplored, owing to the intricate nature of multipartite quantum correlations. In this work, we propose a simple and efficient self-testing protocol that certifies the state and observables based on the optimal quantum violation of the Svetlichny inequality involving an arbitrary number of parties, each with two inputs. Our method leverages an elegant sum-of-squares approach to derive the optimal quantum value of the Svetlichny functional, devoid of assuming the dimension of the quantum system. This enables the self-testing of the $m-$partite maximally entangled state and local anti-commuting observables for each party. Moreover, we develop a swap circuit isometry to assess the proximity of reference states and measurements to their ideal counterparts in the presence of noise and imperfections in real experiments, thereby demonstrating the robustness of our self-testing protocol. Finally, we illustrate how our self-testing protocol facilitates the generation of certified genuine randomness from correlations that enable the optimal violation of the Svetlichny inequality.
\end{abstract}


\maketitle

\section{Introduction}

The rapid advancements in quantum technologies underscore their potential to outperform classical information processing tasks. Subsequently, these developments raise a significant practical concern - the reliability of the quantum devices in use. This issue becomes pertinent to wide areas of information processing tasks ranging from cryptography \cite{Ekert1991, Pironio2009, Wooltorton2024} to randomness expansion \cite{Pironio2010, Colbeck2012, Liu2021, Wooltorton2022}.  It also becomes relevant in the context of delegated quantum computation \cite{Fitzsimons2017, Urmila2018, Gheorghiu2019}, where a party lacking quantum resources (the verifier) delegates a computational task to multiple parties claiming to possess quantum computers (the provers).  It has been demonstrated that when multiple non-communicating provers share entanglement, constructing a certification protocol guarantees the blindness of the problem \cite{Fitzsimons2017, Urmila2018, Gheorghiu2019}. Such applications necessitate the development of certified quantum devices.

One approach to certification is through tomographic \cite{Banaszek2013, Donnel2016, Goh2019} or interferography \cite{Sahoo2020} protocols. However, tomographic methods face two major limitations. First, quantum devices engaged in complex tasks, such as quantum computation, often involve multipartite quantum states, where the complexity of tomographic methods escalates exponentially with the number of particles involved \cite{Pal2014}. Second, tomography requires complete trust in the measurement settings, which may themselves be susceptible to errors. Therefore, a more practical and reliable certification technique is needed - one that tests the properties of quantum devices solely based on input-output statistics, allowing users to verify the device without needing to understand its internal workings.

To address this, self-testing protocol, first introduced by Mayers and Yao \cite{Mayers2004}, offers a device-independent certification method, enabling the certification of quantum states and measurements solely from input-output statistics, independent of the device's internal workings or the quantum system's dimension. Such a protocol is fundamentally linked to the Bell inequality \cite{Bell1964}, whose quantum violation establishes the existence of quantum correlations that cannot be reproduced by local realist models - a phenomenon known as Bell nonlocality \cite{Brunner2014rev}.  Beyond its foundational implications, Bell nonlocality becomes a powerhouse for a myriad of implications in quantum information theory, underpinning various quantum technologies. A key consequence of Bell nonlocality is the ability to certify quantum correlations device-independently.

As a result, interest in designing self-testing methods and studying their robustness has grown significantly. For a recent review, we refer to Ref. \cite{Supic2020}. Following initial breakthroughs in self-testing maximally entangled two-qubit state \cite{Mayers2004} using the Clauser-Horne-Shimony-Holt (CHSH) inequality \cite{Clauser1969}, robust versions of this protocol were also developed \cite{McKague2012,Wu2016,Sarkar2023a}. Recent advances have further expanded the scope of self-testing, demonstrating that all pure entangled bipartite states can be self-tested \cite{Coladangelo_ncomms_2017}. Additionally, self-testing has been extended to certify unsharpness parameters \cite{Roy_Pan2023, Rajdeep2024}, quantum channels \cite{pawel2018,Sekatski2023,Wagner2020} and quantum memory \cite{Sekatski2023}. In networked scenarios involving multiple sources, self-testing has been used to verify Pauli observables \cite{Bowles2018a, Bowles2018}, all entangled states \cite{Šupić2023}, commuting observables \cite{Sneha2023} and even falsification of the real quantum world \cite{Renou2021}. Experimental investigations have also made significant progress in self-testing single quantum systems \cite{Hu2023}, bipartite states \cite{Zhang2019}, multipartite entangled states \cite{Zhang2018, Dian2021}, and states generated by quantum networks \cite{Agresti2021}. Based on robust self-testing of graph states, a protocol was introduced in \cite{McKague2016a} that certifies that the provers are solving bounded-error quantum polynomial time problems. In the prepare-measure scenario, the semi-device-independent certification of the state \cite{Tavakoli2018}, measurement \cite{Farkas2019, Mohan2019}, and unsharpness parameters have been studied \cite{Uola2020, supic2016, Cavalcanti_2017, Šupić_2020, Sumit2021, Pan2021}. However, the present work considers self-testing in a device-independent scenario, without assuming the dimension of the system.

In comparison, self-testing in multipartite scenarios has received less attention. Although the multipartite generalization of the bipartite Bell experiment seems an extension of the bipartite scenario, quantum correlations for the former display a much richer and intricate structure than that of the latter \cite{Svetlichny1987,Roy1989,Roy1991,Mermin1990,Uffink2002,Seevinck2002,Collins2002}. Note that, the quantum violation of a multipartite Bell-type inequality may not warrant the distribution of nonlocal quantum correlation for each observer, as it may be possible that some of the observers share a general no-signalling nonlocal correlation, and the rest share local correlations. In such a case, the nonlocality is not genuinely distributed to all parties. Svetlichny first introduced the notion of genuine nonlocality in the multipartite scenario \cite{Svetlichny1987} by characterizing the correlation as fully local, bilocal, and genuine nonlocal. Later, various other inequalities have been proposed in multipartite scenarios \cite{Gallego2012,Sagnik2020,Bancal2013}. Genuine multipartite nonlocality is the strongest form of nonlocality argument, in which nonlocal correlation is distributed for each involved party, thus forbidding the existence of a local-nonlocal model that can simulate the optimal quantum violation of a multipartite inequality. 

Beyond the bipartite scenario, the robust self-testing of the tripartite $W$ state was proposed using the swap circuit method \cite{wu2014}. Self-testing of all multipartite states admitting a Schmidt decomposition was given in \cite{Supic2018}. The Greenberger-Horne-Zeilinger (GHZ) state was self-tested \cite{Sarkar2022} based on the optimal quantum violation of an inequality proposed in \cite{Augusiak2019}, and the existence of local isometry is argued, however the specific swap circuit is not provided. In a recent work \cite{Panwar2023}, based on the classes of multipartite Bell inequalities \cite{Mermin1990,Uffink2002,Seevinck2002,Werenr2001,Weinfurter2001,Zukowski2002},  self-testing of the GHZ state is demonstrated. However, in \cite{Sarkar2022, Panwar2023}, the robustness analysis of their self-testing argument was not studied - a crucial ingredient for implementing the protocol in real experimental scenarios. 

In this work, we develop a device-independent self-testing protocol for the $m-$partite maximally entangled state and observables, based on the optimal quantum violation of the Svetlichny inequality. By employing the sum-of-squares (SOS) approach, we derive the optimal quantum value of the Svetlichny functional without assuming the dimension of the quantum system. This optimal value serves as the basis for self-testing $m-$partite maximally entangled states and local anti-commuting observables.

To support our claim, we introduce a swap circuit method that provides an explicit construction of the local isometry required for self-testing the state and observables. The swap-circuit approach not only establishes the existence of a local isometry but also offers a specific arrangement of local unitary gates necessary to physically implement the self-testing protocol. Additionally, we present a detailed methodology for evaluating the robustness of the self-testing statements in the presence of noise and imperfections, ensuring the protocol’s practicality in real-world conditions. Furthermore, we explore the generation of certified randomness based on the quantum violation of the Svetlichny inequality.

The paper is organized as follows. In Sec. II, we provide a preliminary discussion of the multipartite Bell scenario, including different notions of locality, and a re-expressed version of the Svetlichny functional with convenient notational indices. In Sec. III, we derive the optimal quantum value of the Svetlichny functional without assuming the dimensions of the quantum system, as well as establish the conditions on state and observables necessary to achieve this value. In Sec. IV, we provide an explicit construction of the swap circuit isometry and demonstrates its role in self-testing $m-$partite GHZ state and observables. Sec. V examines the robustness of the swap-circuit-based self-testing protocol when the physical state and measurements deviate from the ideal ones in the presence of noise and imperfection, with a specific case considering errors arising from the imperfect implementation of observables by one party. In Sec. VI,  we derive the global randomness that can be generated when the behavior produces maximal quantum violation of Svetlichny inequality. Finally, we summarize our findings and provide a few future directions.


\section{Preliminaries}
 To begin with, we encapsulate the notion of the genuine multipartite nonlocality and corresponding multipartite Bell inequality to detect it. Then, we define the concept of self-testing and how it is connected with the optimal quantum violation of a Bell inequality. 


\subsection{Multipartite Bell inequality}

A multipartite Bell experiment features $k\in[m]$ number of spatially separated parties who share a joint $m-$partite quantum state $\rho$ originating from a common source. Each party randomly performs one of two possible local measurements, denoted by $A^k_{x_k} $ with $x_k \in\{0,1\}$ and produces an output $a_k \in \{+1,-1\}$. Here, the superscript `$k$' indicates the party number, and the subscript `$x_k$' represents the choice of measurement setting chosen by the $k^{th}$ party. The statistics obtained in such an experimental scenario are characterised by a vector, commonly referred to as a behaviour $\mathbb{P} \in \mathbb{R}^{2^{2k}}$, defined by
\begin{equation}
   \mathbb{P}\equiv \Big\{p(\textbf{a}|\textbf{x}) : \textbf{a}= (a_1,a_2,...,a_m); \textbf{x} = (x_1,x_2,...,x_m)\Big\}
\end{equation}
where $p(\textbf{a}|\textbf{x})$ denotes the joint probability of outcomes. In quantum theory, the joint  probability is given by
\begin{equation}
    p_{\mathcal{Q}}(\textbf{a}|\textbf{x},\rho) = \Tr[\rho \ \qty(\Motimes_{k=1}^{m} M^k_{a_k|x_k})]
\end{equation}
where $\qty(\Motimes_{k=1}^{m} M^k_{a_k|x_k}) = M^1_{a_1|x_1} \otimes M^2_{a_2|x_2} \otimes \cdots \otimes M^m_{a_m|x_m}$; $M^k_{a_k|x_k}$ are the measurement operators corresponding to the outcomes $a_k$ when observables $A^k_{x_k} \equiv\{M^k_{a_k|x_k}|M^k_{a_k|x_k}\geq 0, \sum_{a_k} M^k_{a_k|x_k} = \openone\}$ are measured by $k^{th}$ party.

Now, in a realist description of quantum theory, the source produces a hidden variable $\lambda\in\Lambda$, where $\Lambda$ denotes the state space of $\lambda$. The variable $\lambda$ determines the outcome of a measurement of each party, which is independent of the choice of measurement settings of the other parties, a notion widely termed as locality condition \cite{Bell1964, Brunner2014rev}. In such a local realist model, the multipartite joint probability distribution is factorisable in the following way \cite{Mermin1990}
\begin{equation}
p(\textbf{a}|\textbf{x},\lambda) = \prod\limits_{k=1}^m p(a_k|x_k,\lambda) \ \ \forall \lambda
\end{equation}
In order to reproduce quantum theory by this local model, the following condition needs to be satisfied
\begin{equation}\label{jp}
p_{\mathcal{Q}}(\textbf{a}|\textbf{x},\rho)=\int \mu(\lambda) \prod\limits_{k=1}^m p(a_k|x_k,\lambda) \ d\lambda
\end{equation} 
where $\mu(\lambda)$ denotes the probability distribution of the ontic variable $\lambda$ in the ontic state space $\Lambda$ with $\mu(\lambda)\geq0$ and $\int_{\lambda \in \Lambda} \mu(\lambda)d\lambda=1$.
Any behaviour $\mathbb{P}$ that is reproduced by Eq.~(\ref{jp}) is called \textit{fully local}. If the distribution cannot be expressed as Eq.~(\ref{jp}), it is called $m-$partite \textit{standard nonlocal} distribution. 

Unlike the bipartite scenario, in a multipartite Bell experiment ($m>2$), it is crucial to recognise that even if a behaviour lacks the local realist model described by Eq.~(\ref{jp}), it may still admit a local realist description if all but one of the parties collaborate. For instance, in the case of tripartite correlations involving Alice, Bob and Charlie, the correlation might lack the factorisability structure described by Eq.~(\ref{jp}). However, if two parties, such as Bob and Charlie, collaborate within the same laboratory (this means that if signalling is allowed between Bob and Charlie), the behaviour $\mathbb{P}$ might be describable by a local realist description across the bi-partition of Alice and the Bob-Charlie pair. This type of behaviour is referred to as bilocal \cite{Svetlichny1987}. Generally, in the $m-$partite case, a behaviour is defined as $m$-local if the $m-$partite joint probability distribution can be described by a local realist model when all but one of the parties collaborate, as follows
\begin{equation}\label{gn1}
p(\textbf{a}|\textbf{x},\lambda) =  p(a_{k}|x_{k},\lambda) \ p(\textbf{a}_{\neq k}|\textbf{x}_{\neq k},\lambda)  \ \ \forall k \in \{1,2,...,m\}
\end{equation}
where $\textbf{a}_{\neq k}=(a_1,a_2,...,a_{n\neq k},...,a_m)$  and $\textbf{x}_{\neq k}=(x_1,x_2,...,x_{n\neq k},...,x_m)$.
In order to reproduce quantum theory by such a $m$-local model, the following condition needs to be satisfied
\begin{equation}\label{gn}
   p_{\mathcal{Q}}(\textbf{a}|\textbf{x},\rho)= \sum_{k=1}^m \int_{\lambda} \mu_k(\lambda) \qty[p(a_{k}|x_{k},\lambda) \ p(\textbf{a}_{\neq k}|\textbf{x}_{\neq k},\lambda)] d\lambda
\end{equation}
where $0 \leq \int \mu_k(\lambda) d\lambda \leq 1$ and $\sum_{k=1}^m \int_{\lambda} \mu_k(\lambda)d\lambda=1$. Any behaviour that cannot be reproduced by a model satisfying Eq.~(\ref{gn}), is called $m-$partite \textit{genuine nonlocal} distribution \cite{Svetlichny1987, Collins2002}. While any genuine nonlocal behaviour must be standard nonlocal, there exist standard nonlocal behaviours having a local realist model of the form of Eq.~(\ref{gn}).

For $m=3$, a genuine tripartite nonlocality is detected by using the Svetlichny functional \cite{Svetlichny1987, Seevinck2002}, rewritten in the following form
\begin{eqnarray}\label{tpsv}
    \mathscr{S}_3&=& \qty(A^1_0 - A^1_1)\otimes A^2_0 \otimes A^3_0 - \qty(A^1_0 + A^1_1)\otimes A^2_0 \otimes A^3_1 \nonumber \\
    &-& \qty(A^1_0 + A^1_1)\otimes A^2_1 \otimes A^3_0 - \qty(A^1_0 - A^1_1)\otimes A^2_1 \otimes A^3_1
\end{eqnarray}
The local realist value of $\mathscr{S}_3$ is $\langle \mathscr{S}_3 \rangle_{\lambda}=4$ \cite{Svetlichny1987} and the optimal quantum value corresponding to the tripartite qubit states and observables is found \cite{Svetlichny1987} to be $\langle \mathscr{S}_3 \rangle_{\mathcal{Q}}=\Tr[\rho \mathscr{S}_3] = 4\sqrt{2}$. Such optimal quantum value occurs when the shared state is GHZ state, $\ket{GHZ}=\frac{1}{\sqrt{2}}(\ket{000}+\ket{111})$ and each party implements local anti-commuting measurements.      

We express the $m-$partite Svetlichny functional \cite{Seevinck2002} in the following way.
\begin{equation}\label{svm}
    \mathscr{S}_m = \sum_{\mu=1}^{2^m} v^n_{\mu} \  \Motimes_{k=1}^{m} A^{k}_{i^{k}_{\mu}}
\end{equation}
In order to fix the value of the coefficient $v^n_{\mu}$, let us define a matrix $\mathscr{M}$ of $(2^{m}\times m)$ dimension whose each row represents one of possible $2^m$ number of $m$-bit strings sampled from $\{0,1\}^m$. We denote this matrix as $\{i_{\mu}^{k}\}$, where $\mu \in [2^m]$ represents index of rows and $k \in [m]$ represents index of columns. Here $i_\mu \in \{0,1\}^m$, \textit{i.e.}, $i_\mu^k \in \{0,1\}$. Note that, each row contains some $n$ number of $1$'s and hence $(m-n)$ number of $0$'s. 
The entries of a particular row $\mu$ are decided as follows. For each increment in the index $\mu$, we add binary number $1$ to the binary number $i_\mu$ so that $i_{\mu+1} = i_\mu +_b 1$, where $+_b$ is addition operation on binary numbers. Starting with, $\mu = 1$, (\textit{i.e.}, $1$st row in the defined matrix $\mathscr{M}$), $i^k_1 = 0 \ \forall k\in[m]$. For $\mu=2$, $i^k_2 = 0 \ \forall k\in [m-1]$ and $i^m_2 = 1$. This procedure of adding binary number $1$ to the previous entry is repeated till $\mu = 2^{m}$. Finally, coefficient $v^n_{\mu}$ is defined as $v^n_{\mu} = (-1)^{n(n+1)/2} \in \{-1,1\}$ which depends on the number of $1$'s contained in the $\mu^{th}$ row. 

The local realist bound of $\mathscr{S}_m$ has been obtained \cite{Seevinck2002} to be $\langle \mathscr{S}_m \rangle_{\lambda}=2^{m-1}$ and the optimal quantum bound was shown to be $\langle \mathscr{S}_m \rangle^{opt}_{Q}=2^{m-1}\sqrt{2}$. Crucially, $\langle \mathscr{S}_m \rangle^{opt}_{Q}$ is commonly derived \cite{Seevinck2002} by assuming the local system to be qubit. We derive the optimal quantum bound of the Svetlichny functional without assuming the dimension of the system which enables self-testing of state and observables.


\subsection{Self-testing of state and observables}

Let us now define the notion of self-testing for the purpose of our main findings.

{\defi Any observed behaviour $\mathbb{P}_{*} \equiv \{p_{*}(\textbf{a}|\textbf{x},\rho)\}$ self-tests a quantum strategy $\qty{\rho; \ \qty(\Motimes_{k=1}^{m} M^k_{a_k|x_k}) }$, up to local isometries, \textit{iff}, that is the only strategy attaining $\mathbb{P}_{*}$, such a behaviour is extremal and fixes the optimal quantum bound of a Bell functional, i.e., $\mathscr{F}.\mathbb{P}_{*} = \mathscr{F}^{opt}_{\mathcal{Q}}$, where $\mathscr{F}^{opt}_{\mathcal{Q}}$ denotes the optimal quantum bound of a Bell functional $\mathscr{F}$. }\\

A behaviour $\mathbb{P}_{*}$ that maximally violates a Bell inequality, is the extremal behaviour of the set of quantum correlations \cite{Supic2020}. This in turn implies that the behaviour $\mathbb{P}_{*}$ is generated by a unique quantum strategy \cite{Supic2020}. 

In the following section, we derive the optimal quantum bound of $\mathscr{S}_m$ in Eq. (\ref{svm}) and obtain the self-testing statements of $m-$partite maximally entangled state and corresponding observables without reference to the dimension of the quantum systems involved.


\section{Result: Derivation of the optimal Quantum value for $m-$partite Svetlichny functional}

Before we proceed to evaluate the optimal quantum value $(\mathscr{S}_m)^{opt}_Q$, we re-express the Svetlichny functional $\mathscr{S}_m$ in Eq.~(\ref{svm}), by factoring out the common terms of the type $(A_0^1 \pm A_1^1)$. This adjustment results in a reduction of the number of terms appearing from $2^m$ to $2^{m-1}$ (for instance, see Eq.~\ref{tpsv}), 
\begin{equation}\label{svm_mod}
    \mathscr{S}_m = \sum_{\mu = 1  }^{2^{m-1}} v^n_{\mu} \ \qty[A_0^1 + (-1)^{n+1} A_1^1] \Motimes_{k=2}^{m} A^{k}_{i^{k}_{\mu}}
\end{equation}
We then derive the optimal quantum value $(\mathscr{S}_m)^{opt}_Q$, by employing the SOS technique introduced in \cite{Roy_Pan2023}. This derivation is devoid of the assuming the dimension of the quantum system. 
\subsection{SOS technique}
For our purpose, we introduce an operator $\Gamma_{m}$ which can be written as
\begin{equation}\label{mparty_Svetlichny}
    \Gamma_{m} = \frac{1}{2} \sum_{\mu = 1}^{2^{m-1}} \omega_\mu \ L^{\dagger}_\mu L_\mu
\end{equation}
where we define
\begin{equation}\label{m_party_li}
        L_{\mu} = \frac{1}{\omega_\mu}\qty[A^1_0 + \qty(-1)^{n+1} A^1_1]\Motimes_{k=2}^{m}\openone - v^n_{\mu} \ \openone \Motimes_{k=2}^{m} A^k_{i^k_{\mu}} 
\end{equation}
We define the normalisation factor $\omega_{\mu}$ as $\omega_{\mu}=\norm{A^1_0 + \qty(-1)^{n+1} A^1_1}_{\rho}$.
Since $n$ can assume any value from $0$ to $m$, then $\qty(-1)^{n+1} \in \{+1,-1\}$. Hence there exists only two possible values of $\omega_\mu$'s contingent upon whether $n$ is even or odd. As already mentioned, the value of $n$ is determined by the choice of row ($\mu$) from the matrix $\mathscr{M}$ defined in the paragraph succeeding Eq.~(\ref{svm}).

By construction, $L_\mu$'s are hermitian which makes $\Gamma_m \succeq 0$, \textit{i.e.}, $\Tr[\Gamma_m \ \ \rho] \geq 0$, $\forall \rho$. Plugging Eq. (\ref{m_party_li}) into Eq. (\ref{mparty_Svetlichny}), one gets
\begin{equation}\label{smgamma}
    \Tr[\mathscr{S}_m \ \rho] = \sum_{\mu = 1}^{2^{m-1}} \omega_\mu - \Tr[\Gamma_m \ \rho]
\end{equation}
 It is then evident from Eq.~(\ref{smgamma}) that maximization of $\braket{\mathscr{S}_m}$ requires minimization of $\braket{\Gamma_m}$ \textit{i.e.}, $\braket{\Gamma_m} = 0$. Thus, the optimal quantum value of Svetlichny functional is attained by maximising the summation over $\omega_\mu$. Since linear functions are convex, and for convex functions, the maximum of the sum equals the sum of the maxima, then
\begin{equation}\label{m_party_obs_state}
    \begin{aligned}
        \qty(\mathscr{S})^{opt}_Q &= \max\limits_{\{A^k_{i^k_{\mu}}, \ \rho\}} \ \ \qty(\sum_{\mu = 1}^{2^{m-1}} \omega_\mu) \\
        &= \sum_{\mu=1}^{2^{m-1}}\max\limits_{\{A^k_{i^k_{\mu}}, \ \rho\}} \ \ \qty(\sqrt{2 + (-1)^{n+1} \Tr\left[\left\{ A^1_0,A^1_1 \right\} \ \rho\right]})
    \end{aligned}
\end{equation}
The optimal quantum value of the $m-$partite Svetlichny functional is $(\mathscr{S}_m)^{opt}_Q = 2^{m-1}\sqrt{2}$, which is achieved when $\left\{A^1_0,A^1_1\right\} = 0$. This in turn fixes the value $\omega_\mu = \sqrt{2}, \ \ \forall \mu$.


\subsection{Conditions on the state and the measurement settings}

The optimality condition $\Tr[\Gamma_{m}\ \rho] = 0$ implies that $\Tr[(L_\mu)^\dagger L_\mu \ \rho] = 0$ $\forall \mu \in [2^{m-1}]$. Consequently, this yields the following relations from Eq.~(\ref{m_party_li}) as
\begin{equation}\label{obsrm}
    \Tr[\qty(\frac{A^1_0 + (-1)^{n+1} A^1_1}{\sqrt{2}}) \Motimes_{k=2}^{m} A^k_{i^k_{\mu}}\;\rho] = \frac{1}{v^n_{\mu}}  \ \ \forall \mu  \in [2^{m-1}], \rho
\end{equation}
Given that $A^k_{i^k_{\mu}} $ are dichotomic, the aforementioned equation implies that the shared $m-$partite state $\rho$ has to be a pure state, since $v^n_{\mu} = \pm 1$. Furthermore, using the optimality condition of $\Tr[\Gamma_{m}\ \rho] = 0$, and considering $\rho = \ket{\psi}\bra{\psi}$, we have $L_\mu \ket{\psi} = 0$, leading to the following relations.
\begin{equation} \label{allobs}
    \frac{A^1_0 + \qty(-1)^{n+1} A^1_1}{\sqrt{2}}\Motimes_{k=2}^{m}\openone \ket{\psi} = v^n_{\mu} \ \openone \Motimes_{k=2}^{m} A^k_{i^k_{\mu}} \ket{\psi} \ \forall \mu, \ket{\psi}
\end{equation}
To obtain interrelation between the observables of an arbitrary $j^{th}$ party, we proceed as follows. Consider any two relations from  Eq.~(\ref{allobs}), viz., $i_{\mu_1}^k = 0 \ \forall k \in [m]$, and  the one with $i^k_{\mu_2} = 0 \ \forall k\neq j\in [m]$, and,  $i^j_{\mu_2} = 1$. Taking anticommutation between these relations, we straightforwardly obtain $\{A_0^j,A_1^j\} = 0 \ \forall j\in\{2,3,4,\dots,m\}$. Therefore, for the optimal quantum violation of the $m-$partite Svetlichny inequality, the observables of each party must be anti-commuting. Explicit examples of this procedure for $m=3$ and $m=4$ are given in Appendices~\ref{SOS3} and \ref{SOS4} respectively.

Once the anti-commutativity between the observables of each party is established, it is straightforward to show that the joint observables appearing in the left hand side of Eq.~(\ref{obsrm}) are mutually commuting, with $\rho = \ket{\psi}\bra{\psi}$ being the common eigenstate. Thus, one possible representation of $\rho\in\mathscr{L}(\Motimes_{i=1}^m\mathcal{H}_d)$ is
\begin{equation}\label{gen_state}
     \rho = \frac{1}{d^m} \qty[\openone_{d^m} + \sum\limits_{i=1}^{2^{m}-1} \ \alpha_i \ \mathscr{O}_i]
\end{equation}
where $\alpha_i \in\{+1,-1\}$ and $\mathscr{O}_i \in \mathscr{L}(\Motimes_{i=1}^m\mathcal{H}_d)$ are mutually commuting observables with some of them appearing in Eq.~(\ref{obsrm}). Now, since $\rho$ is a pure state, imposing $\Tr[\rho^2] = 1$ on the state given in the Eq.~(\ref{gen_state}), then $\Tr[\mathscr{O}_i] = 0$. This implies that the density matrix of each subsystem is a maximally mixed state, i.e., $\rho_{A^k} = \Tr_{A^i,i\neq k}[\rho] = \frac{\openone}{d}$. Hence, $\rho$ is a $m-$partite maximally entangled state.


\subsection{Self-testing statements}\label{secc}

We can make the following self-testing statements based on the optimal quantum violation of Svetlichny inequality as follows. 

\emph {If the quantum value of the Svetlichny functional corresponding to a behaviour $\mathbb{P}_{*}$ is observed to be $\braket{\mathscr{S}_m(\mathbb{P}_{*})}=2^{m-1}\sqrt{2}$, the following statements hold.}

\emph { i) The shared state between all the parties is the $m-$partite maximally entangled state.}

\emph { ii) All the parties perform projective measurements on their respective subsystems. Each party's observables are anti-commuting, i.e., $\qty{A^k_{0},A^k_{1}}=0 \ \forall k \in [m]$.}

In the following section, we provide an overview of the experimental procedure to test the derived self-testing statements by using the swap-circuit method as illustrated in Fig.~\ref{figswap}.


\section{Result: Self-Testing of $m-$partite GHZ state and observables}

In the previous section, we showed that achieving the optimal quantum violation of the $m-$partite Svetlichny inequality implies the shared state is $m-$partite maximally entangled, with each party's observables being anti-commuting. This inference is predicated on the state and observables satisfying Eq.~(\ref{allobs}). Self-testing using a swap circuit provides a detailed methodology by which the physical experiment in a black-box scenario can be replicated by a reference experiment that assumes the minimum dimension of the system. This approach involves local ancilla systems and local unitary operations, enabling the results obtained from Eq.~(\ref{allobs}) to be swapped to the local ancilla systems. The swap circuit method is particularly elegant, as it facilitates the swapping of characteristics from the physical system to a reference ancillary system, thereby establishing a direct correspondence between the reference state and observables and those of the physical system.

However, practical experimental scenarios inevitably involve noise and imperfections making the self-testing schemes more challenging. Experimental imperfections result in sub-optimal violations of a Bell inequality, and hence it becomes essential to rigorously analyse to what extent these noises affect the self-testing relations. The swap circuit method offers a valuable advantage in this context by allowing for the analysis of the effects of experimental errors on self-testing relations. Notably, when assessing the robustness of these relations, it is not necessary to identify the exact source of errors. Instead, one can attribute all errors to the imperfect implementation of the observables used to construct the swap circuit. This simplification makes the evaluation of the robustness of the self-testing protocol more comprehensible. Rather than computing errors for each self-testing relation in Eq.~(\ref{allobs}), one can measure the extent to which the extracted state in the swap circuit deviates from the ideal scenario. This deviation serves as a quantitative indicator of robustness.

Let the state $\ket{\psi}_m$ and the observables $Z_k$ and $X_k$ (satisfying $\{Z_k,X_k\} = 0$, $\ \forall k\in[m]$) produce the statistics $\mathbb{P}_*$, which maximally violate the $m-$partite Svetlichny inequality, and satisfy the self-testing relations given by Eqs.~(\ref{obsrm}) and (\ref{allobs}). Such a state and measurements are referred to as the \textit{physical state} and \textit{physical measurements}. Given the unknown dimensions of the physical state and measurements, our objective is to establish a \textit{reference state} and \textit{reference measurements} that replicate the same statistics as the physical ones. This entails demonstrating the existence of a local isometry that jointly acts on both the ancillary and physical systems, extracting the reference state and measurements within the ancillary system. Mathematically, this can be expressed as follows
\begin{equation}
        \Phi \ \qty[\mathscr{A}^l \ket{\psi}_{m} \otimes \ket{ancilla}_{m'}] = \ket{\chi}_m \otimes \sigma^l_k \ket{ent}_{m'}
\end{equation}
where $\mathscr{A}^{l=0} = \openone$, $\mathscr{A}^{l=1} = Z_k$ and $\mathscr{A}^{l=2} = X_k$; $\ket{\chi}_{m}$ is called `junk state'. $\ket{ancilla}_{m'}$ is the state of the local ancillary system, $\ket{ent}_{m'}$ is an entangled state between local ancillary systems, $\sigma^l_k$ is a unitary operator acting on the $k^{th}$ subsystem of the entangled state $\ket{ent}_{m'}$. We provide an explicit construction of the local isometry using the states and measurement unitaries to be implemented by each of the parties to reproduce the statistics achieving the optimal quantum value of $\braket{\mathscr{S}_m}$.

  \begin{figure*}[ht] 
		\centering
		\includegraphics[width=0.8\textwidth]{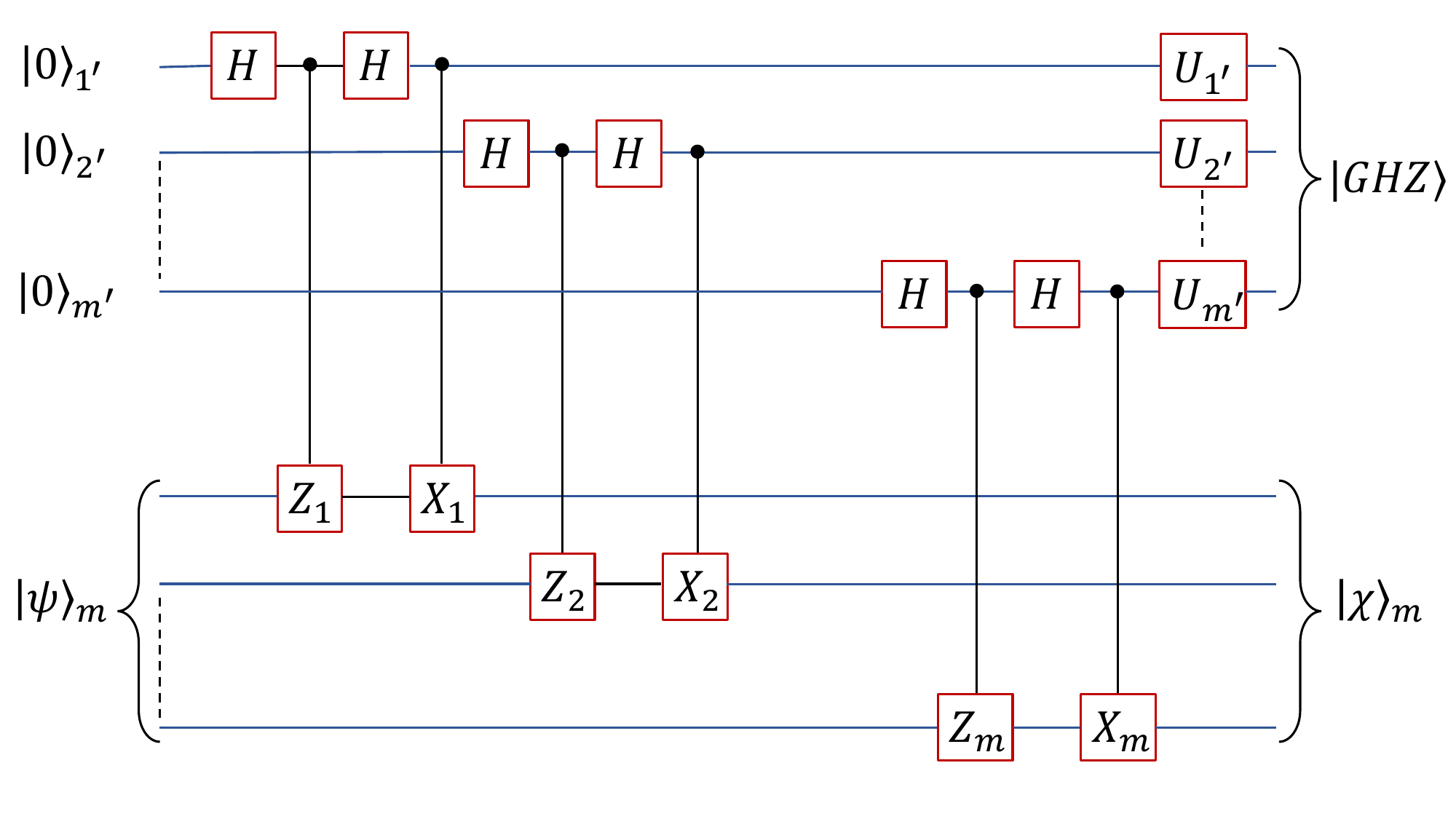} 
		\caption{Example of a swap circuit used for self-testing of the $m-$partite GHZ state and local anti-commuting observables. The circuit takes as input a state $\ket{\psi}_m$ that satisfies Eqs.~(\ref{obsrm}) and (\ref{allobs}), along with $m$-ancillary qubits, each initialised in the state $\ket{0}$. It outputs the state $\ket{GHZ}$ in tensor product with an auxiliary state $\ket{\chi}_m$. Here, $H$ denotes the usual Hadamard gate, and $U$ represents suitably constructed Unitary gates.}
  \label{figswap}
  \end{figure*}

Let us first define the normalised scaled observables of $k^{th}$ party as $X_k$ and $Z_k \ (\forall k\in[m])$ as follows
\begin{equation}\label{XZ_relation}
    \begin{aligned}
        Z_1 &= \frac{A^1_0+A^1_1}{\sqrt{2}} \ ;  & Z_k = A^k_0 \ \ \forall k\neq 1;   \\
        X_1 &= \frac{A^1_0-A^1_1}{\sqrt{2}} \ ; & X_k = A^k_1 \ \ \forall k \neq 1;  
    \end{aligned}
\end{equation}

{\thm If the $m-$partite state $\ket{\psi}_m$ and the observables $Z_k$ and, $X_k \  \forall k \in [m]$ attains the optimal quantum value of the Svetlichny functional $\mathscr{S}_m$ given by Eq.~(\ref{svm}), then the isometry $\Phi$ implemented as the swap circuit illustrated in Fig.~\ref{figswap}, extracts the $m-$partite GHZ state and qubit observables, such that the following relation holds
\begin{equation} \label{stswap}
        \Phi \ \qty[\mathscr{A}^l \ket{\psi}_{m} \otimes \ket{0}^{\otimes m}] = \ket{\chi}_m \otimes \sigma^l_k \ket{GHZ}_{m'}
\end{equation}
where $\ket{0}^{\otimes m'}$ is the local ancillary state; $\mathscr{A}^{l=0} = \openone$, $\mathscr{A}^{l=1} = Z_k$ and $\mathscr{A}^{l=2} = X_k$; $\ket{\chi}_{m}$ is the so called `junk state', given by $\ket{\chi}_{m}=\frac{1}{2^{m-1} \sqrt{2}}\qty[\Motimes_{k=1}^m \qty(\openone+ Z_k)]\ket{\psi}_{m}$, and the state $\ket{GHZ}_{m'}$ is a local-unitary equivalent form of the maximally entangled $m$-qubit state and $\sigma_k^l$ are Pauli spin observables satisfying $\{\sigma^{l=1}_k,\sigma^{l=2}_k\} = 0 $\label{theoswap}} and $\sigma^{l=0}_k = \openone$.

\textit{Sketch of the proof.--} The proof of the Theorem \ref{theoswap} is hugely cumbersome and quite challenging to present in full detail here. We provide only a brief outline. A more intuitive understanding of the proof can be gained by carefully examining the examples for $m=3$ and $m=4$ (see, Appx.~\ref{sss3} and \ref{sss4}, respectively).

Based on the outputs of the isometries for $m=3$ and $m=4$, as evaluated in Appx~\ref{isoo3} and \ref{isoo4} respectively, the isometry represented by the swap circuit in Fig.~\ref{figswap} for the general $m-$partite case produces the following output (prior to the application of the unitaries $U_{i'}$).
\begin{widetext}
\begin{equation}\label{isometry_output}
 \ket{\Psi} =  \Phi \ \qty[\mathscr{A}^l \ket{\psi}_{m} \otimes \ket{0}^{\otimes m}] =\frac{1}{2^m} \sum_{a_{k'} \in \{0,1\}}  \qty(\Motimes_{k=1}^m   \qty(X_k)^{a_{k'=k}}\qty[\openone + (-1)^{a_{k'=k}}Z_k]) \mathscr{A}^l \ket{\psi}_{m}\qty(\Motimes_{k'={1}}^{m} \ket{a_{k'}}_{k'})
\end{equation}
\end{widetext}
By imposing the self-testing relations in Eq.~(\ref{allobs}) to Eq.~(\ref{isometry_output}), we show that ancillary systems for $m$ parties effectively form an $m-$partite GHZ state and that the observables corresponding to each party are qubit observables. This proves that the swap circuit self-tests the state and measurements for $\qty(\mathscr{S})_Q^{opt}$ optimally. 

We begin by demonstrating the self-testing of the state, focusing on the term in Eq.~(\ref{isometry_output}), where all ancillary bits are in the state $\ket{0}$, \textit{i.e.}, $a_{k'} = 0 \ \forall k' \in [m]$ in the output. This term is
\begin{equation}
    \Motimes_{k=1}^{m} \qty(\openone + Z_k)\ket{\psi}_m \ket{0}^{\otimes m}
\end{equation}
Next, by considering the term where the first ancillary bit is in the state $\ket{1}$ and the rest of the ancillary bits are in the state $\ket{0}$, we get
\begin{equation}\label{self_testing_term_mparty}
\begin{aligned}
    &X_1\qty(\openone - Z_1)\Motimes_{k=2}^{m} \qty(\openone + Z_k)\ket{\psi}_m \ket{1}\otimes\ket{0}^{\otimes m-1} \\
    &= \Motimes_{k=1}^{m} \qty(\openone + Z_k)X_1\ket{\psi}_m \ket{1}\otimes\ket{0}^{\otimes m-1}
\end{aligned}
\end{equation}
Recalling the self-testing relations in Eq.~(\ref{allobs}) for $\mu = 1$, \textit{i.e.}, $i_1 = 00\dots0$, we have
\begin{equation}\label{x_igen}
\begin{aligned}
    \frac{A^1_0 - A^1_1}{\sqrt{2}}\Motimes_{k=2}^{m}\openone \ket{\psi} &= \ \openone \Motimes_{k=2}^{m} A^k_{0} \ket{\psi} \implies 
    X_1 \ket{\psi} = \Motimes_{k=2}^m Z_k \ket{\psi}
\end{aligned}
\end{equation}
Using Eq.~(\ref{x_igen}) and $(\openone+Z_k)Z_k = (\openone+Z_k)$, Eq.~(\ref{self_testing_term_mparty}) simplifies to the following form
\begin{equation}
    \Motimes_{k=1}^{m} \qty(\openone + Z_k)\ket{\psi}_m \ket{1}\ket{0}^{\otimes m-1}
\end{equation}
Similarly, considering the output term in which first $(m-2)$ ancillas are in $\ket{0}$ state, and the remaining two are in $\ket{1}$ state, we have
\begin{eqnarray}
    && \Motimes_{k=1}^{m-2} (\openone+Z_k) X_{m-1}(\openone-Z_{m-1})X_m(\openone-Z_{m}) \ket{\psi}\otimes\ket{0}^{\otimes m-2}\ket{1}\ket{1} \nonumber \\
    &=&\Motimes_{k=1}^{m} (\openone+Z_k) X_{m-1}X_m\ket{\psi}\otimes\ket{0}^{\otimes m-2}\ket{1}\ket{1} \label{eq24} 
\end{eqnarray}
To simplify Eq.~(\ref{eq24}), we invoke the self-testing relation in Eq.~(\ref{allobs}) by taking $\mu = 4$, \textit{i.e.}, $i_4 = 00\dots011$ which gives the following.
\begin{equation}
\begin{aligned}\label{eq25}
     &\frac{A^1_0 - A^1_1}{\sqrt{2}}\Motimes_{k=2}^{m}\openone \ket{\psi} = -\ \openone \Motimes_{k=2}^{m-2} A^k_{0}\otimes A^{m-1}_{1}\otimes A^m_{1}\ket{\psi}\\
     &\implies 
    X_1 \ket{\psi} = -\Motimes_{k=2}^{m-2} Z_k X_{m-1}X_m\ket{\psi}  \\
    &\implies \Motimes_{k=2}^m Z_k \ket{\psi} = -\Motimes_{k=2}^{m-2} Z_k X_{m-1}X_m\ket{\psi}  \ \ \ \text{[using Eq.~(\ref{x_igen})]} \\
    &\implies -Z_{m-1}Z_{m} \ket{\psi} = X_{m-1}X_{m}\ket{\psi}
\end{aligned}
\end{equation}
Using Eq.~(\ref{eq25}), Eq.~(\ref{eq24}) can be rewritten as
\begin{eqnarray}
    &=&\Motimes_{k=1}^{m} (\openone+Z_k) X_{m-1}X_m\ket{\psi}\otimes\ket{0}^{\otimes m-2}\ket{1}\ket{1}\\
    &=&-\Motimes_{k=1}^{m} (\openone+Z_k)\ket{\psi}\otimes\ket{0}^{\otimes m-2}\ket{1}\ket{1}
\end{eqnarray}
The above procedure can be repeated for any combination of ancillas $\otimes_{k'=1}^{m}\ket{a_{k'}}$, and by invoking the self-testing relations in Eq.~(\ref{allobs}), we can obtain simplified terms. Finally, the output of the isometry would be a linear combination of the following form
\begin{equation}
    \ket{\Psi} = \frac{1}{2^m}\qty[\Motimes_{k=1}^m \qty(\openone+ Z_k)]\ket{\psi}\otimes\qty[\sum_{a_{k'}\in\{0,1\}} g_1(\boldsymbol{a}) \Motimes_{k' = 1}^m\ket{a_{k'}}]
\end{equation}
where $\boldsymbol{a} = (a_1,a_2,\cdots,a_m)$ with $g_1(\boldsymbol{a})\in\{-1,+1\}$ and is dependent on the function $v^n_\mu$. Since the further evaluation of this general term and showing that the GHZ state (or some rotated form of it) has indeed been extracted in the ancillary system is quite extensive, we take recourse to the cases of $m=3$ and $m=4$ as shown in Appx.~\ref{sss3} and Appx.~\ref{sss4}. 

Similarly, for observables we can show that 
\begin{equation}
    \ket{\Psi} = \frac{1}{2^m}\qty[\Motimes_{k=1}^m \qty(\openone+ Z_k)]\ket{\psi}\otimes\qty[\sum_{a_{k'}\in\{0,1\}}^m g_2(\boldsymbol{a})  \Motimes_{k' = 1}^m\ket{a_{k'}}]
\end{equation}
where $\boldsymbol{a} = (a_1,a_2,\cdots,a_m)$ with $g_2(\boldsymbol{a})\in\{-1,+1\}$. We have presented explicit calculations for the cases of $m=3$ and $m=4$ in Appx.~\ref{sss3} and Appx.~\ref{sss4} to show that the observables are also self-tested using the isometry presented in Fig. \ref{figswap}.


\section{Result: Robust Self-Testing of $m-$partite GHZ state and observables} \label{robust}

The self-testing relations in Eq.~(\ref{allobs}) are derived under ideal conditions. However, in real experimental situations, the Bell test subject to unavoidable noise and imperfections. Consequently, the implemented observables may deviate from those initially intended. These deviated observables may not necessarily be unitary operators and are referred to as unregularised observables, denoted by $\tilde{Z}_k$ and $\tilde{X}_k$ with $k\in[m]$. We assume that these observables are close to the ideal set of observables $Z_k$ and $X_k$, and we define them as follows.
\begin{equation}\label{normdiffobs}
\norm{\qty(\tilde{X}_k - X_k)\ket{\psi}} \leq \xi_k \ ; \ \ \norm{\qty(\tilde{Z}_k - Z_k)\ket{\psi}} \leq \zeta_k \ ;    
\end{equation}
where $ \xi_k$ and $\zeta_k$ are small positive numbers. For perfect or ideal measurement scenarios, $\xi_k=\zeta_k=0$.

Due to the implementation of such unregularised observables (where eigenvalues are not necessarily $\pm1$) instead of the intended regularised ones modifies the construction of $L_{\mu}$ in Eq.(~\ref{m_party_li}) to $\tilde{L}_{\mu}$. This modification introduces an upper-bound on the SOS-derived optimality condition $ \norm{L_\mu\ket{\psi}}=0$, given by
\begin{equation}\label{mpartite_relro}
     0 \leq \norm{\Tilde{L}_\mu\ket{\psi}} \leq \epsilon_\mu \ \ \ ; \forall \mu \in [2^{m-1}] 
\end{equation}
Therefore, the optimality condition $\Tr[\Gamma_m \ \rho]=0$ changes to the following form
\begin{equation}
    \Tr[\tilde{\Gamma}_m \ \rho]=\frac{1}{\sqrt{2}} \sum\limits_{\mu=1}^{2^{m-1}} \bra{\psi} \tilde{L}^\dagger_\mu \tilde{L}_\mu \ket{\psi} \leq \frac{1}{\sqrt{2}}\sum_{\mu = 1}^{2^{m-1}}{\epsilon_\mu}^2
\end{equation}
where $\epsilon_{\mu} \geq 0$ is a small positive number functionally related to $\xi_k$ and $\zeta_k$. This leads to a sub-optimal value of the Svetlichny functional, given by
\begin{equation}
    (\mathscr{S}_m)_Q = 2^{m-1}\sqrt{2} - \epsilon \ \ \text{with} \ \ \epsilon = \sum_{\mu = 1}^{2^{m-1}}{\epsilon_\mu}^2 \geq 0
\end{equation}

The following claim addresses the extent to which the self-testing statements in  Sec.~\ref{secc} hold when there is a sub-optimal violation of the Svetlichny functional due to noise in the experimental implementation of observables.

 {\Claim Assuming that in an experiment the optimal quantum violation of the Svetlichny inequality deviates by a small amount $\epsilon \geq 0$, \textit{i.e.}, $2^{m-1}\sqrt{2}-\epsilon$, the self-testing remains robust if the isometry continues to extract a state close to the $\ket{GHZ}$. This closeness is measured by the trace-norm, which quantifies the distance between the output state of the imperfect isometry, constructed with non-ideal unitaries, and the ideal isometry, constructed with ideal unitaries \label{cl2}
\begin{equation} \label{th2ro}
\norm{ \tilde{\Phi} \ \qty[\tilde{\mathscr{A}} ^l \ket{\psi} \otimes \ket{0}^{\otimes m}] -\Phi \ \qty[\mathscr{A} ^l \ket{\psi} \otimes \ket{0}^{\otimes m}]}   \leq f(\boldsymbol{\xi},\boldsymbol{\zeta})
\end{equation}
where $\Phi \ \qty[\mathscr{A} ^l \ket{\psi} \otimes \ket{0}^{\otimes m}]= \ket{\chi} \otimes  \sigma_l^k \ket{GHZ}$, $\mathscr{\tilde{A}}^l \in \qty{\openone,\tilde{Z}_k, \tilde{X_k}}$, $f(\boldsymbol{\xi},\boldsymbol{\zeta})$ represents a functional depending on $\boldsymbol{\xi},\boldsymbol{\zeta} \in \mathbb{R}^m$ that tends to zero as $\boldsymbol{\xi},\boldsymbol{\zeta} \to \boldsymbol{0}$.

\begin{proof}
    The unregularised observables $\tilde{Z}_k$ and $\tilde{X}_k$ are defined in terms of unregularised hermitian operators $\tilde{A}^k_{i^{k}_{\mu}}$ as follows
\begin{equation}\label{swapobsro1}
        \begin{aligned}
        \tilde{Z}_1 &= \frac{\tilde{A}^1_0+\tilde{A}^1_1}{\sqrt{2}} \ ;  & \tilde{Z}_k = \tilde{A}^k_0 \ \\
        \tilde{X}_1 &=  \frac{\tilde{A}^1_0-\tilde{A}^1_1}{\sqrt{2}} \ ; & \tilde{X}_k = \tilde{A}^k_1 \  
        \end{aligned}
    \end{equation}
In order to make the unregularised operators (denoted by using tilde notations) unitary, we employ the regularisation trick used in  \cite{Bowles2018a}. Initially, by replacing zero eigenvalues with $1$ and denoting such operator as $\qty(\tilde{A}^k_{i^{k}_{\mu}})^{*}$ with $i^k_\mu \in \{0,1\}$, we normalise the eigenvalues by $\hat{\tilde{A}}^k_{i^{k}_{\mu}}=\frac{\qty(\tilde{A}^k_{i^{k}_{\mu}})^{*}}{\norm{(\tilde{A}^k_{0})^{*}}}$, ensuring that the resulting hermitian operator $\hat{\tilde{A}}^k_{i^{k}_{\mu}}$ becomes unitary and satisfies $\qty(\hat{\tilde{A}}^k_{i^{k}_{\mu}})^2=\openone$. This regularisation will not affect the error-bound in the robustness analysis. Hence, from now on, without loss of generality, we treat $\tilde{A}^k_{i^{k}_{\mu}}$ as hermitian and unitary operators.

Furthermore, it is important to note that for optimal quantum violation, while each party's ideal observables anti-commute, i.e., $\{A_0^k,A_1^k\}=0$, the implemented observables may not necessarily exhibit this property, \textit{i.e.,} $\{\tilde{A}^k_{0},\tilde{A}^k_1\}\neq 0$. Additionally, the unitarity of $\tilde{A}^k$ ensures that $\tilde{Z}_k$ and $\tilde{X}_k$ are unitary for $k \geq 2$, the lack of anti-commutativity  between $\tilde{A}^1_0$ and $\tilde{A}^1_1$ means that $\tilde{Z}_1$ and $\Tilde{X}_1$ are not unitary (unregularised) operators.


\subsection{Robust self-testing of state}

The output of the isometry with imperfect implementation of the observables is
\begin{equation*}
\tilde{\Phi}\qty({\ket{\psi}\ket{0}^{\otimes m}})=\frac{1}{2^m} \sum_{a_{k'} \in \{0,1\}}  \qty[\Motimes_{k=1}^m   \qty(\tilde{X}_k)^{a_{k' = k}}\qty(\openone + (-1)^{a_{k'=k}}\tilde{Z}_k)]\ket{\psi} \Motimes_{k'=1}^m \ket{a_{k'}}
\end{equation*}
Now, using the triangle inequality $\norm{C+D}\leq \norm{C}+\norm{D}$, the trace-distance between the outputs of imperfectly implemented $(\tilde{\Phi})$ and ideal $(\Phi)$ isometries is upper-bounded as follows
\begin{widetext}
    \begin{eqnarray}
\norm{\tilde{\Phi}\qty({\ket{\psi}\ket{0}^{\otimes m}}) - \Phi\qty({\ket{\psi}\ket{0}^{\otimes m }})} &\leq& \frac{1}{2^m} \sum_{ a_{k'} \in \{0,1\}} \Bigg\Vert \Motimes_{k=1}^m   \Bigg[\qty(\tilde{X}_k)^{a_{k'=k}}\qty(\openone + (-1)^{a_{k'=k}}\tilde{Z}_k)  -   \qty(X_k)^{a_{k'=k}}\qty(\openone + (-1)^{a_{k'=k}}Z_k)\Bigg]\ket{\psi} \Motimes_{k'=1}^m \ket{a_{k'}} \bigg\Vert \nonumber \\
&\leq& \frac{1}{2^m}  \sum_{a_{k'} \in \{0,1\}} \Biggl[ \big\Vert\qty(\boldsymbol{\mathcal{\tilde{X}}} - \boldsymbol{\mathcal{X}}) \ket{\psi}\Motimes_{k'=1}^m \ket{a_{k'}} \big\Vert + \sum_{i=1}^{m} \norm{\qty(\boldsymbol{\mathcal{\tilde{X}}} \ \tilde{Z}_i - \boldsymbol{\mathcal{X}} \ Z_i)\ket{\psi}\Motimes_{k'=1}^m \ket{a_{k'}}} \nonumber\\
&+& \sum_{i<j} \norm{ \qty(\boldsymbol{\mathcal{\tilde{X}}} \ \tilde{Z}_i\tilde{Z}_j - \boldsymbol{\mathcal{X}} \ Z_iZ_j)\ket{\psi}\Motimes_{k'=1}^m \ket{a_{k'}}} + \cdots +\sum \norm{ \qty(\boldsymbol{\mathcal{\tilde{X}}} \ \boldsymbol{\mathcal{\tilde{Z}}} - \boldsymbol{\mathcal{X}} \ \boldsymbol{\mathcal{Z}})\ket{\psi}\Motimes_{k'=1}^m \ket{a_{k'}}}  \Biggl] \label{ups1}
\end{eqnarray}
\end{widetext}
where $\boldsymbol{\mathcal{X}} = \Motimes_{k=1}^m \qty(X_k)^{a_{k'=k}}$, $\boldsymbol{\mathcal{\tilde{X}}} = \Motimes_{k=1}^m \qty(\tilde{X}_k)^{a_{k'=k}} $, $\boldsymbol{\mathcal{Z}} = \Motimes_{k=1}^m Z_k$ and $\boldsymbol{\mathcal{\tilde{Z}}} = \Motimes_{k=1}^m \tilde{Z}_k $. Note that while $\boldsymbol{\mathcal{X}}$ depends on the term appearing in the ancillary system; $\boldsymbol{\mathcal{Z}}$ does not.

We use the following approximations of Eq.~(\ref{normdiffobs}) in order to evaluate the upper bound
\begin{eqnarray}
        \norm{\qty(\tilde{X}_i - X_i)\ket{\psi}} &\leq& \xi_i \implies \tilde{X}_i \approx \xi_i\openone + X_i \label{approx1} \\
        \norm{\qty(\tilde{Z}_i - Z_i)\ket{\psi}} &\leq& \zeta_i \implies \tilde{Z}_i \approx \zeta_i\openone + Z_i \label{approx2}
\end{eqnarray}
Using Eq.~(\ref{approx1}), we can obtain (see Appx.~\ref{rsca} for more detailed calculation)
\begin{equation}
    \begin{aligned}
        \norm{\qty(\tilde{X}_\gamma \tilde{X}_\delta - X_\gamma X_\delta)\ket{\psi}} &\leq \xi_\gamma + \xi_\delta \norm{\tilde{X}_\gamma \ket{\psi}}  \\
        \norm{\qty(\tilde{X}_\gamma \tilde{X}_\delta\tilde{X}_\eta - X_\gamma X_\delta X_\eta)\ket{\psi}} &\leq  \xi_\gamma + \xi_\delta \norm{\tilde{X}_\gamma \ket{\psi}} + \xi_\eta \norm{\tilde{X}_\gamma \tilde{X}_\delta \ket{\psi}}
    \end{aligned}
\end{equation}
Where $\gamma\neq\delta\neq\eta\in[m]$. Now, it is straightforward to obtain the following relation
\begin{equation} \label{apx1}
    \norm{\qty(\boldsymbol{\mathcal{\tilde{X}}} - \boldsymbol{\mathcal{X}})\ket{\psi}} \leq f_1(\boldsymbol{\xi})
\end{equation}
Following the similar procedure, using Eq.~(\ref{approx2}), we also derive (see Appx.~\ref{rsca} for more detailed calculation)
\begin{equation} \label{apz1}
    \norm{\qty(\boldsymbol{\mathcal{\tilde{Z}}} - \boldsymbol{\mathcal{Z}})\ket{\psi}} \leq f_2(\boldsymbol{\zeta})
\end{equation}
Using these relations given by Eqs.~(\ref{approx1}) and (\ref{approx2}), we get 
\begin{equation} \label{apxz1}
    \begin{aligned}
        \norm{\qty(\tilde{X}_\gamma \tilde{Z}_i - X_\gamma Z_i)\ket{\psi}} &\leq \xi_\gamma + \zeta_j\norm{\tilde{Z}_i\ket{\psi}}  \\
        \norm{\qty(\tilde{X}_\gamma \tilde{X}_\delta \tilde{Z}_i - X_\gamma X_\delta Z_i)\ket{\psi}} &\leq \xi_\gamma + \xi_\delta \norm{\tilde{X}_\gamma \ket{\psi}} + \zeta_i\norm{\tilde{X}_\gamma \tilde{X}_\delta \ket{\psi}} 
    \end{aligned}
\end{equation}
Where $\gamma\neq\delta\neq i\in[m]$. From Eq.~(\ref{apxz1}), straightforward evaluation leads to the following
\begin{equation}
    \norm{\qty(\boldsymbol{\tilde{\mathcal{X}}}\tilde{Z}_i - \boldsymbol{\mathcal{X}} Z_i)\ket{\psi}} \leq f_{3,i}(\boldsymbol{\zeta},\boldsymbol{\xi}) \ \ \forall i \in [m]
\end{equation}
Next using Eqs.~(\ref{approx1}), (\ref{approx2}) and (\ref{apxz1}), we derive (see Appx.~\ref{rsca})
\begin{equation} \label{apxxzz1}
    \begin{aligned}
        \norm{\qty(\tilde{X}_\gamma\tilde{Z}_i\tilde{Z}_j - X_\gamma Z_iZ_j)\ket{\psi}} &\leq \xi_\gamma + \zeta_i\norm{\tilde{Z}_i\ket{\psi}} + \zeta_j \norm{\tilde{X}_\gamma\tilde{Z}_i\ket{\psi}} \\
        \norm{\qty(\tilde{X}_\gamma \tilde{X}_\delta \tilde{Z}_i\tilde{Z}_j - X_\gamma X_\delta Z_iZ_j)\ket{\psi}} &\leq \xi_\gamma + \xi_\delta \norm{\tilde{X}_\gamma \ket{\psi}} \\
        &+ \zeta_i\norm{\tilde{X}_\gamma \tilde{X}_\delta \ket{\psi}} 
        + \zeta_j \norm{\tilde{X}_\gamma\tilde{X}_\delta \tilde{Z}_i\ket{\psi}}
    \end{aligned}
\end{equation}
Where $\gamma\neq\delta\neq i \neq j\in[m]$. Finally, generalising Eq.~(\ref{apxxzz1}), we derive
\begin{equation}
    \norm{\qty(\boldsymbol{\tilde{\mathcal{X}}}\tilde{Z}_i\tilde{Z}_j - \boldsymbol{\mathcal{X}}Z_iZ_j)\ket{\psi}} \leq f_{i,j}(\boldsymbol{\zeta},\boldsymbol{\xi})
\end{equation}
and this further can be generalised using the similar methodology to all intermediate terms (containing $3, 4, \dots, m$ $\tilde{Z}$ and $Z$).  
\begin{equation}\label{apfxfz}
    \norm{\qty(\boldsymbol{\tilde{\mathcal{X}}}\boldsymbol{\mathcal{\tilde{Z}}} - \boldsymbol{\mathcal{X}}\boldsymbol{\mathcal{Z}})\ket{\psi}} \leq f_4(\boldsymbol{\zeta},\boldsymbol{\xi})
\end{equation}
Note that the summation outside the square bracket of Eq.~(\ref{ups1}) acts on the $m-$partite ancillary system $\Motimes_{k'} \ket{a_{k'}}$. Depending on the different values of $a_{k'} \in \{0,1\}$, there are $2^m$ distinct $\Motimes_{k'} \ket{a_{k'}}$. In Appx.~\ref{rsca}, we have shown that each term is function of $\xi_i, \zeta_i$. For instance, if we take $a_{k'}=0 \ \forall k$, then $\boldsymbol{\mathcal{X}}=\boldsymbol{\tilde{\mathcal{X}}}=\openone$ and the right hand side of Eq.~(\ref{ups1}) is upper-bounded as follows
\begin{eqnarray}
        && \frac{1}{2^m}\qty(\sum_{i=1}^m \norm{\qty(\tilde{Z}_i -  Z_i)\ket{\psi}} +  \sum_{i<j} \norm{\qty(\tilde{Z}_i\tilde{Z}_j -Z_iZ_j)\ket{\psi}} + \dots) \nonumber \\
        &\leq&  \frac{1}{2^m}\qty(\sum_{i=1}^m \zeta_i + \sum_{i<j} \qty(\zeta_i + \zeta_j\norm{\tilde{Z}_i\ket{\psi}} ) + \dots + f_4(\boldsymbol{\zeta},\boldsymbol{\xi})) \nonumber \\
        &\leq& \frac{1}{2^m} \ h_1(\boldsymbol{\zeta,\xi})
\end{eqnarray}
Following the similar evaluation in Appx.~\ref{rsca}, by considering other possible combinations of $\Motimes_{k'} \ket{a_{k'}}$, we obtain each term is some function of $(\boldsymbol{\zeta,\xi})$. This then leads to the following conclusion
\begin{equation}\label{gentradis}
\norm{\tilde{\Phi}\qty({\ket{\psi}\ket{0}^{\otimes m}}) - \Phi\qty({\ket{\psi}\ket{0}^{\otimes m }})} \leq F(\boldsymbol{\xi},\boldsymbol{\zeta})
\end{equation}
with $F(\boldsymbol{\xi},\boldsymbol{\zeta}) \rightarrow 0$ when $\boldsymbol{\xi},\boldsymbol{\zeta} \rightarrow 0$.\\


\subsection{Robust self-testing of observables}

We begin by analysing the robustness of the observables $X_{\alpha}$ and $Z_{\alpha}$ associated with the $(k=\alpha)$-th party. Following this, we will demonstrate that, by adhering to the same procedure, all observables across the parties can be collectively and robustly self-tested.

For the robust self-testing of $X_{\alpha}$,using Eq.~(\ref{ups1}), we get
\begin{widetext}
    \begin{eqnarray}\label{robsx}
        \norm{\tilde{\Phi}\qty({\tilde{X}_\alpha\ket{\psi}\ket{0}^{\otimes m}}) - \Phi\qty(X_\alpha{\ket{\psi}\ket{0}^{\otimes m }})} &\leq& \frac{1}{2^m} \sum_{a_{k'} \in \{0,1\}} \Bigg\Vert \Motimes_{k=1}^m   \Bigg[\qty(\tilde{X}_k)^{a_{k'=k}}\qty(\openone + (-1)^{a_{k'=k}}\tilde{Z}_k)\tilde{X}_\alpha  -   \qty(X_k)^{a_{k'=k}}\qty(\openone + (-1)^{a_{k'=k}}Z_k)X_\alpha\Bigg]\ket{\psi} \Motimes_{k'=1}^m \ket{a_{k'}} \bigg\Vert \nonumber \\
        &\leq& \frac{1}{2^m}  \sum_{a_{k'} \in \{0,1\}} \Biggl[ \big\Vert\qty(\boldsymbol{\mathcal{\tilde{X}}}\tilde{X}_\alpha - \boldsymbol{\mathcal{X}}X_\alpha) \ket{\psi}\Motimes_{k'=1}^m \ket{a_{k'}} \big\Vert + \sum_{i=1}^{m} \norm{\qty(\boldsymbol{\mathcal{\tilde{X}}} \ \tilde{Z}_i\tilde{X}_\alpha - \boldsymbol{\mathcal{X}} \ Z_iX_\alpha)\ket{\psi}\Motimes_{k'=1}^m \ket{a_{k'}}} \nonumber\\
        &+& \sum_{i<j} \norm{ \qty(\boldsymbol{\mathcal{\tilde{X}}} \ \tilde{Z}_i\tilde{Z}_j\tilde{X}_\alpha - \boldsymbol{\mathcal{X}} \ Z_iZ_jX_\alpha)\ket{\psi}\Motimes_{k'=1}^m \ket{a_{k'}}} + \cdots +\sum \norm{ \qty(\boldsymbol{\mathcal{\tilde{X}}} \ \boldsymbol{\mathcal{\tilde{Z}}}\tilde{X}_\alpha - \boldsymbol{\mathcal{X}} \ \boldsymbol{\mathcal{Z}}X_\alpha)\ket{\psi}\Motimes_{k'=1}^m \ket{a_{k'}}}  \Biggl] \nonumber \\
         &\leq& \frac{1}{2^m}  \sum_{a_{k'} \in \{0,1\}} \Biggl[ \xi_\alpha \norm{\boldsymbol{\mathcal{\tilde{X}}}\ket{\psi}\Motimes_{k'=1}^m \ket{a_{k'}}} + \big\Vert\qty(\boldsymbol{\mathcal{\tilde{X}}} - \boldsymbol{\mathcal{X}}) \ket{\psi}\Motimes_{k'=1}^m \ket{a_{k'}} \big\Vert \nonumber\\
         &+& \sum_{i=1}^{m} \qty(\xi_\alpha \norm{\boldsymbol{\mathcal{\tilde{X}}} \ \tilde{Z}_i\ket{\psi}\Motimes_{k'=1}^m \ket{a_{k'}}} + \norm{\qty(\boldsymbol{\mathcal{\tilde{X}}} \ \tilde{Z}_i - \boldsymbol{\mathcal{X}} \ Z_i)\ket{\psi}\Motimes_{k'=1}^m \ket{a_{k'}}}) \nonumber\\
         &+& \sum_{i<j} \qty(\xi_\alpha \norm{\boldsymbol{\mathcal{\tilde{X}}} \ \tilde{Z}_i\tilde{Z}_j\ket{\psi}\Motimes_{k'=1}^m \ket{a_{k'}}} +\norm{ \qty(\boldsymbol{\mathcal{\tilde{X}}} \ \tilde{Z}_i\tilde{Z}_j - \boldsymbol{\mathcal{X}} \ Z_iZ_j)\ket{\psi}\Motimes_{k'=1}^m \ket{a_{k'}}} ) + \cdots \nonumber \\
         &+& \sum \qty( \xi_\alpha \norm{\boldsymbol{\mathcal{\tilde{X}}} \ \boldsymbol{\mathcal{\tilde{Z}}}\Motimes_{k=1}^m \ket{a_{k'}}} + 
         \norm{ \qty(\boldsymbol{\mathcal{\tilde{X}}} \ \boldsymbol{\mathcal{\tilde{Z}}} - \boldsymbol{\mathcal{X}} \ \boldsymbol{\mathcal{Z}})\ket{\psi}\Motimes_{k'=1}^m \ket{a_{k'}}})  \Biggl] \nonumber \\
         &\leq& F_{1,\alpha}(\boldsymbol{\xi},\boldsymbol{\zeta}) \ \ \ \ \forall \alpha \in [m]
    \end{eqnarray}
\end{widetext}
Robust self-testing of $Z_\alpha$ can be done by replacing $\tilde{X}_{\alpha}$ with $\tilde{Z}_{\alpha}$ in Eq.~(\ref{robsx}) and using the approximation of $\tilde{Z}_{\alpha}$ as per Eq.~(\ref{approx2}). Such evaluation will lead to the following
\begin{equation}
    \norm{\tilde{\Phi}\qty({\tilde{Z}_\alpha\ket{\psi}\ket{0}^{\otimes m}}) - \Phi\qty(Z_\alpha{\ket{\psi}\ket{0}^{\otimes m }})} \leq F_{2,\alpha}(\boldsymbol{\xi},\boldsymbol{\zeta}) \ \forall \alpha \in [m]
\end{equation}
When observing sub-optimal statistics in the $m-$partite Svetlichny functional, it is essential to determine whether the imperfections stem from the preparation of the state, the implementation of observables, or both. However, it is sufficient to consider that the state preparation is ideal (i.e., the $m-$partite maximally entangled state) and that the errors originate solely from the implementation of observables by the first party \cite{Bowles2018a}.


\subsubsection{Special case: Only first party implements imperfect observables}

Here, we consider the scenario where the sub-optimal violation of the Svetlichny functional arises only from the imperfect implementation of the first party's observables. In this context, 
\begin{equation}
\norm{\qty(\tilde{X}_1 - X_1)\ket{\psi}} \leq \xi_1 \ ; \ \ \norm{\qty(\tilde{Z}_1 - Z_1)\ket{\psi}} \leq \zeta_1 
\end{equation}
The construction of $L_\mu$ would be affected and constructed as $\tilde{L}_\mu$. Now, there are $2^{m-1}$ number of $\tilde{L}_{\mu}$ out of which, $2^{m-2}$ are constructed using $\tilde{X}_1$ and the other $2^{m-2}$ are constructed using $\tilde{Z}_1$. This leads to 
\begin{equation}
    \Tr[\tilde{\Gamma}_m \ \rho]=\frac{1}{\sqrt{2}} \sum\limits_{\mu=1}^{2^{m-1}} \bra{\psi} \tilde{L}^\dagger_\mu \tilde{L}_\mu \ket{\psi} \leq 2^{m-\frac{3}{2}} \qty(\zeta_1^2 + \xi_1^2)
\end{equation}
So the if the maximal violation of Svetlichny functional is $2^{m-1}\sqrt{2} - \epsilon$ then $\epsilon = 2^{m-\frac{3}{2}} \qty(\zeta_1^2 + \xi_1^2)$.
For robust self-testing of state we wish to calculate
\begin{eqnarray}
    \norm{\tilde{\Phi}\qty({\ket{\psi}\ket{0}^{\otimes m}}) - \Phi\qty({\ket{\psi}\ket{0}^{\otimes m }})} \leq \frac{1}{2} \qty[\zeta_1\qty(2+\xi_1) + 2 \xi_1  ]
\end{eqnarray}
To robust self-test the observables, taking the robust self-testing of $Z_1$, we will have
\begin{equation}
    \begin{aligned}
        &\norm{\tilde{\Phi}\qty({\tilde{Z}_1\ket{\psi}\ket{0}^{\otimes m}}) - \Phi\qty({Z_1\ket{\psi}\ket{0}^{\otimes m }})} \\ 
        &\leq \frac{1}{2} \qty(6 \zeta_1 + 2 \zeta_1^2 + \zeta_1^2\xi_1 + 3\zeta_1 \xi_1 + 2\xi_ 1)     
    \end{aligned}
\end{equation}
Similarly self-testing analysis of $X_1$ can be done.
\begin{equation}
\begin{aligned}
        &\norm{\tilde{\Phi}\qty({\tilde{X}_1\ket{\psi}\ket{0}^{\otimes m}}) - \Phi\qty({X _1\ket{\psi}\ket{0}^{\otimes m }})} \\
        &\leq \frac{1}{2} \qty(6\xi_1 + 2\xi_1^2 + \xi_1^2\zeta_1 + 3\zeta_1\xi_1 + 2\zeta_1)
\end{aligned}
\end{equation}
\end{proof}
} 
Further, we analyze the case when  $\xi_1 = \zeta_1 = \epsilon_1$, i.e., the difference from the optimal observables is same for both the observables of the first party.We then have
\begin{equation}
\norm{\qty(\tilde{X}_1 - X_1)\ket{\psi}} \leq \epsilon_1 \ ; \ \ \norm{\qty(\tilde{Z}_1 - Z_1)\ket{\psi}} \leq \epsilon_1 
\end{equation}
This will give the relation between the amount of deviation from the optimal violation of Svetlichny inequality and the error in the implementation of the observables
\begin{equation}
    \epsilon = 2^{m-\frac{1}{2}}\epsilon_1^2
\end{equation}
The output of the isometry then differs from the ideal scenario as follows
\begin{equation}\label{risse}
\begin{aligned}
        \norm{\tilde{\Phi}\qty({\ket{\psi}\ket{0}^{\otimes m}}) - \Phi\qty({\ket{\psi}\ket{0}^{\otimes m }})} &\leq \frac{\epsilon_1}{2} \qty(4+\epsilon_1) \\
        &= 2^{\frac{-2m-3}{4}}\sqrt{\epsilon} \ \qty(2^{\frac{1-2m}{4}}\sqrt{\epsilon}+4)
\end{aligned}
\end{equation}
For observables, we obtain
\begin{equation}\label{riose}
\begin{aligned}
        \norm{\tilde{\Phi}\qty({\tilde{X}_1\ket{\psi}\ket{0}^{\otimes m}}) - \Phi\qty({X _1\ket{\psi}\ket{0}^{\otimes m }})} &\leq \frac{\epsilon_1}{2} \qty(8 + 5 \epsilon_1+\epsilon_1^2) \\
        \norm{\tilde{\Phi}\qty({\tilde{Z}_1\ket{\psi}\ket{0}^{\otimes m}}) - \Phi\qty({Z_1\ket{\psi}\ket{0}^{\otimes m }})} &\leq \frac{\epsilon_1}{2} \qty(8 + 5 \epsilon_1+\epsilon_1^2) \\
\end{aligned}
\end{equation}
\begin{equation} \label{riose1}
       \frac{\epsilon_1}{2} \qty(8 + 5 \epsilon_1+\epsilon_1^2)= 2^{\frac{-2m-3}{4}}\sqrt{\epsilon}\qty(8 + 2^{\frac{-2m+1}{4}}5\sqrt{\epsilon}+2^{\frac{-2m+1}{2}}\epsilon)
\end{equation}

\begin{figure} [h!]
	\centering
	\begin{tikzpicture}
	\begin{axis}[legend pos=south east, legend cell align=left, enlargelimits=false, xlabel={Error ($\epsilon$)}, ylabel style ={align=center}, ylabel= {Error in the extracted isometry \\ (State)}, xticklabel style={ 
		/pgf/number format/fixed, /pgf/number format/fixed zerofill,
		/pgf/number format/precision=2
	}, scaled ticks=false, xtick={1,2,3,4,5,6,7}, yticklabel style={ 
		/pgf/number format/fixed, /pgf/number format/fixed zerofill,
		/pgf/number format/precision=2
	}, scaled ticks=false, ytick={0.00,0.2,0.4,0.6,0.8,1.0,1.2,1.4},xmin=0,xmax=7,ymin=0,ymax=1.3           
	]	

\addplot [domain=0:1.66, samples=200, red, thick, dashdotted] {2^((-3-2*3)/4)*x^(0.5)*(4+2^((1-2*3)/4)*x^(0.5))};
\addlegendentry{\textit{a} : $m = 3$}
\addplot [domain=0:3.31, samples=200, green, thick] {2^((-3-2*4)/4)*x^(0.5)*(4+2^((1-2*4)/4)*x^(0.5))};
\addlegendentry{\textit{b} : $m=4$}
\addplot [domain=0:6.63, samples=200, blue, dashed, thick] {2^((-3-2*5)/4)*x^(0.5)*(4+2^((1-2*5)/4)*x^(0.5))};
\addlegendentry{\textit{c} : $m=5$}

\node[above] at (140,100) {$a$};
\node[above] at (290,100) {$b$};
\node[above] at (570,100) {$c$};

\end{axis}
\end{tikzpicture}
	\caption{Plot of the error in the extracted state given by Eq.~(\ref{risse}) vs the deviation from the optimal quantum value of the Svetlichny inequality for robust self-testing of state.}
	\label{figstate}
\end{figure}
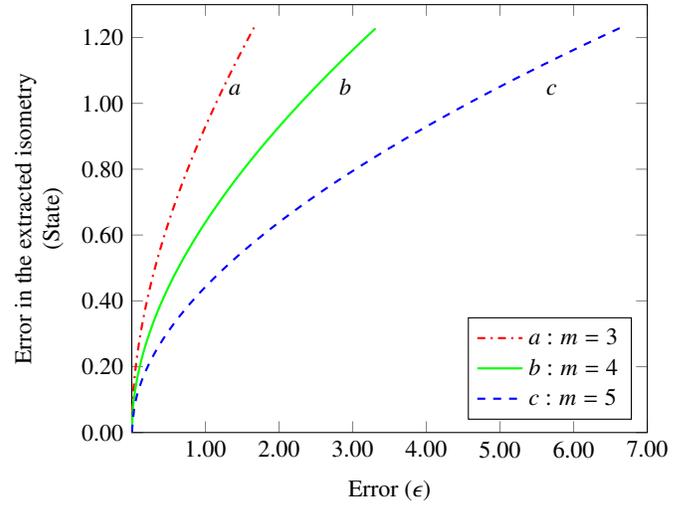

If the observed violation of the $m-$partite Svetlichny inequality is sub-optimal, specifically $2^{m-1}\sqrt{2}-\epsilon$ with $0 \leq \epsilon < 2^{m-1}\qty(\sqrt{2}-1)$, our proposed self-testing scheme remains robust. In particular, when the error $\epsilon$ is attributed to the imperfect implementation of observables by the first party, and the deviations of the imperfect observables from the ideal ones are characterised by $\epsilon_1 = 2^{\frac{1}{2}\qty(-m+\frac{1}{2})}\sqrt{\epsilon}$, the state extracted through the imperfect isometry will differ from the ideal extracted state. This discrepancy allows us to determine the robust self-testing bounds for both the state and observables as functions of $\epsilon$ in Eqs.~(\ref{risse}) and (~\ref{riose1}), respectively. 

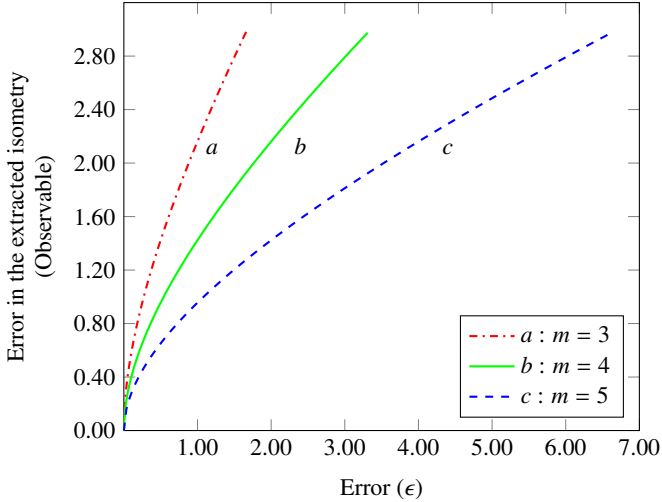
\begin{figure}[h!]
	\centering
	\begin{tikzpicture}
	\begin{axis}[legend pos=south east, legend cell align=left, enlargelimits=false, xlabel={Error ($\epsilon$)}, ylabel style ={align=center}, ylabel= {Error in the extracted isometry \\ (Observable)}, xticklabel style={ 
		/pgf/number format/fixed, /pgf/number format/fixed zerofill,
		/pgf/number format/precision=2
	}, scaled ticks=false, xtick={1,2,3,4,5,6,7}, yticklabel style={ 
		/pgf/number format/fixed, /pgf/number format/fixed zerofill,
		/pgf/number format/precision=2
	}, scaled ticks=false, ytick={0.00,0.4,0.8,1.2,1.6,2.0,2.4,2.8},xmin=0,xmax=7,ymin=0,ymax=3.2
	]	

\addplot [domain=0:1.66, samples=200, red, thick, dashdotted] {2^(-1)*2^((1-2*3)/4)*x^(0.5)*(8+5*2^((1-2*3)/4)*x^(0.5)+2^((1-2*3)/2)*x};
\addlegendentry{\textit{a} : $m = 3$}
\addplot [domain=0:3.31, samples=200, green, thick] {2^(-1)*2^((1-2*4)/4)*x^(0.5)*(8+5*2^((1-2*4)/4)*x^(0.5)+2^((1-2*4)/2)*x};
\addlegendentry{\textit{b} : $m=4$}
\addplot [domain=0:6.63, samples=200, blue, dashed, thick] {2^(-1)*2^((1-2*5)/4)*x^(0.5)*(8+5*2^((1-2*5)/4)*x^(0.5)+2^((1-2*5)/2)*x};
\addlegendentry{\textit{c} : $m=5$}

\node[above] at (120,200) {$a$};
\node[above] at (240,200) {$b$};
\node[above] at (440,200) {$c$};

\end{axis}
 
\end{tikzpicture}

	\caption{Plot of the error in the extracted state given by Eq.~(\ref{riose1}) vs the deviation from the optimal quantum value of the Svetlichny inequality for robust self-testing of observables.}
	\label{figobs}

\end{figure}

It is important to note that as $\epsilon$ increases, so does the error in the isometries, leading to a less precise self-testing outcome. Further, as the number of parties $m$ increases, the overall errors in the self-testing process decrease. These relationships are illustrated in Figs.~\ref{figstate} and \ref{figobs}, which depict the robustness of the state and observables, respectively.

\section{Application in randomness generation}
Our self-testing scheme leads to the generation of device-independent certified randomness. We then demonstrate that whenever the $m-$partite Bell inequality is maximally violated, each party can generate one random bit locally, which is the maximum possible in the two-outcome scenario.

Recently, it has been pointed out that in the bipartite scenario, a violation of the CHSH inequality or Hardy relations implies that both the operational assumptions of `predictability' and `nosignalling' cannot be simultaneously consistent with the quantum theory \cite{Masanes2006, Cavalcanti2012, Sasmal2024}. This indicates that, correlations violating CHSH inequality or Hardy relations result in genuinely random outputs for both parties. Following this argument, it can be shown that any multipartite joint probability distribution at operational level satisfying predictability and nosignalling must satisfy Eq.~(\ref{gn}). Therefore, similar to the CHSH inequality in the bipartite case, violation of $m-$partite Svetlichny inequality implies violations of predictability, thereby certifying genuine randomness. Notably, this method of certifying randomness relies solely on input-output statistics and does not require any knowledge of the inner workings of the devices, a process known as device-independent certification of randomness \cite{Pironio2010, Colbeck2011, Colbeck2012, Acin2012, Wooltorton2022}. 

However, in order to estimate the amount of randomness, we must consider a specific system and account for the presence of an adversary who can control the devices. For unconditional security and privacy against an adversary constrained by quantum theory, we assume that the user knows nothing about the internal workings of the system. The user can only characterise the input-output data produced by the device in terms of the observed behaviour, $\mathbb{P}_{obs}$. Since the adversary can control the devices, they can produce $\mathbb{P}_{obs}$ by using different convex mixtures of extreme points $\mathbb{P}_{ext}$ such that $\mathbb{P}_{obs}=\sum_{ext}q_{ext}\mathbb{P}_{ext}$. The adversary knows $q_{ext}$ perfectly, but the user does not. The adversary can choose the distribution $q_{ext}$ to maximise their probability of guessing the output. Hence, the secure way to quantify the amount of randomness, $\mathcal{R}(\mathbb{P}_{obs})$, present in a correlation is to quantify the maximum guessing probability of an adversary, $G(\mathbb{P}_{obs})=\max_{a_k, x_k} p(\textbf{a}|\textbf{x})$, which can be expressed in terms of min.-Entropy to measure the randomness in bits \cite{Konig2008, Pironio2010, Scarani2019book}. Thus, the amount of device-independent randomness pertaining to an observed nonlocal behaviour is given by the following optimisation problem
\begin{eqnarray}
   && \mathcal{R}(\mathbb{P}_{obs}) = -\log_2 \qty[\max_{q_{ext},\mathbb{P}_{ext}} \sum_{ext} q_{ext}G(\mathbb{P}_{ext})]  \nonumber \\
  && \text{Subject to} \nonumber \\
  && \text{(i)} \ \mathscr{F}.\mathbb{P}_{obs} > \beta_L \nonumber \\
  && \text{(ii)} \ \mathbb{P}_{obs} = \sum_{ext}q_{ext}\mathbb{P}_{ext} 
\end{eqnarray} 
where $G(\mathbb{P}_{ext}) = \max_{a_k, x_k} p_{ext}(\textbf{a}|\textbf{x})$ and $\mathscr{F}$ is a Bell functional with a local bound $\beta_L$. In general, the aforementioned convex optimization problem is challenging to compute in quantum theory because quantum theory, not being a polytope, makes optimization over all possible convex mixtures of extreme points a difficult task \cite{Navascues2007, Navascues2008, Silleras2014, Bancal2014, Silleras2018}. 
However, one can bypass the optimization problem if the observed behavior can be self-tested and thereby is an extremal behavior \cite{Acin2012, Andersson2018, Woodhead2020, Borkala2022, Wooltorton2022}. Hence, for each of these extreme points $\mathbb{P}_{ext}$, evaluation of the guessing probability is straightforward since there is a unique way of preparing this probability distribution. We show that the maximal quantum violation of the Svetlichny inequality self-tests the behavior and thus is an extremal distribution. Such extremal behaviour that leads to the maximal violation $(2^{m-1}\sqrt{2})$ of the $m-$partite Svetlichny inequality is given by
\begin{equation} \label{mpartmax}
    \mathbb{P}_{ext}=\qty{\frac{1}{2^m}\qty(1\pm\frac{1}{\sqrt{2}})} 
\end{equation}
The amount of global randomness in the probability distribution ($\mathbb{P}_{ext}$) maximally violating the Svetlichny inequality is given by 
\begin{equation}
    \mathcal{R}(\mathbb{P}_{ext})_{global} = m - \log_2\qty(1+\frac{1}{\sqrt{2}})
\end{equation}
It is important to note here that $m-$partite Svetlichny inequality is maximised by an $m-$partite maximally entangled state, implying that the subsystem of each party is in a maximally mixed state, thus performing any measurement on it will give one bit of randomness. This amount can also be obtained by marginalising the joint probability distribution given by Eq.~(\ref{mpartmax}).


\section{Summary and Discussion}

The self-testing protocols rely on proving the uniqueness of a quantum behaviour uniquely obtainable by certain state and observables. Uniqueness in this context implies that a self-testable behaviour must be an extremal that cannot be reproduced by a convex mixture of other behaviours within the quantum set, nor should it be realisable through any local realist strategy. Even in the simplest bipartite qubit scenarios, establishing uniqueness remains a formidable challenge because quantum probability space is convex but not a polytope. Nevertheless, one can identify certain extremal behaviours, if not the entire set, using the optimal quantum violations of Bell inequalities \cite{Nuszkiewicz2023}.  In the simplest bipartite scenario, extensive research has shown that by designing various Bell inequalities \cite{Nuszkiewicz2023}, one can characterise a wide array of quantum extremal points and demonstrate the self-testing of pure two-qubit states and measurements. This is typically achieved by finding the maximal quantum values of Bell functionals, often under the assumption of the system's dimensionality.

In the multipartite scenario, deriving an optimal quantum value for a relevant multipartite Bell functional becomes increasingly challenging with a growing number of parties. In this work, we employ a simple and elegant SOS method to derive the optimal quantum value of the Svetlichny functional \cite{Seevinck2002}, crucially without assuming the dimensionality of the system. This method imposes unique constraints on the nature of the state and determines the properties of the observables. Specifically, we find that achieving the optimal value of the Svetlichny functional requires that all parties' observables must be anti-commuting, and the shared state must be an $m-$partite maximally entangled state. Thus, the optimal violation confirms that the behaviour is extremal and therefore unique, effectively self-testing the $m-$partite maximally entangled state and local anti-commuting observables for each party.

It is crucial to note that the maximal quantum violation of a Bell inequality is necessary for self-testing, but not sufficient \cite{Le2023}. Recent studies \cite{Goh2018, Le2023} show that while all the extremal behaviour are self testable in the simplest Bell scenario, the relation between self-testing, optimal violation and the boundary of the quantum set is more nuanced than one might hope for  beyond this scenario. However, if we can demonstrate the existence of a local isometry through a swap circuit for an extremal behavior, mapping the observed correlations to a specific quantum state and measurements (up to local unitaries), then we can assert that the extremal behavior corresponds to a unique state and set of observables.  This approach involves using local ancillary systems and local unitary operations, enabling us to swap the characteristics from the physical system (results obtained from Eq.~\ref{allobs}) to a reference ancillary system. Thus, the swap circuit method establishes a direct correspondence between the reference state and observables to those of the physical system. 
 
Furthermore, practical experimental scenarios are often affected by noise and imperfections, resulting in sub-optimal violations of a Bell inequality. The swap circuit method allows us to analyze the effect of experimental errors on self-testing relations without pinpointing the exact source of errors. Instead, we can attribute all errors to the imperfect implementation of observables and evaluate the robustness of the self-testing relation in Eq.~(\ref{allobs}). Rather than computing errors for each self-testing relation, one can measure the extent to which the extracted state in the swap circuit deviates from the ideal one. This deviation serves as a quantitative indicator of robustness.

Here, by developing a swap circuit isometry, we first demonstrate how multipartite entanglement in the physical system is swapped to the qubit ancillary systems to produce the $m-$partite GHZ state. Subsequently, we have quantified the robustness of the self-testing scheme, by demonstrating how the reference state and measurements are close to the ideal ones in the presence of noise.

In a recent work \cite{Sarkar2022}, the self-testing of $m-$partite GHZ state in arbitrary local dimensions was presented, based on the tilted Bell inequality proposed in \cite{Augusiak2019}. Although they have shown the existence of a local isometry, neither robustness is addressed, nor the swap circuit implementation is provided. Another recent study \cite{Panwar2023} has shown the self-testing of the $m-$partite GHZ state and observables, using two classes of multipartite inequalities \cite{Werenr2001,Weinfurter2001,Zukowski2002,Uffink2002}. They derived the optimal quantum violations of these multipartite inequalities employing Jordan's lemma. While they provided a brief outline of the swap circuit method for the state, they did not offer a thorough analysis of the robustness of their scheme.

In contrast, our approach utilizes the SOS technique that enabled a dimension-independent derivation of the optimal quantum violation of the Svetlichny inequality. Note that the Jordan lemma becomes ineffective when each party performs more than two measurements. Our method, therefore, paves a way for deriving the optimal quantum value for any multipartite Bell functional involving more than two inputs. Furthermore, our detailed robustness analysis makes our self-testing scheme more adoptable to the experimental scenario.

We wish to point out that the well-known Mermin's inequality \cite{Mermin1990} does not have the power to robustly self-test the GHZ state and observables, as the behavior violating Mermin's inequality does not exhibit genuine nonlocality. Additionally, we have found that the generalization of the Svetlichny functional \cite{Svetlichny1987}, as proposed in \cite{Bancal2011}, also fails to capture genuine nonlocality. In Appx. \ref{bancalgen}, we provide a dimension-independent derivation of the optimal quantum value of the inequality in \cite{Bancal2011}. We show that for $m\geq 4$, to achieve maximum quantum violation, all but any three parties in the network must perform commuting measurements. 

Finally, we have also shown that the optimal violation of the Svetlichny inequality leads to device-independent generation of certified randomness. The behaviour that maximally violates the Svetlichny inequality yields one bit of local randomness for each party, and the global randomness of $m - \log_2\qty(1+\frac{1}{\sqrt{2}})$ bit.

The self-testing protocols we proposed using SOS decomposition suggest a few immediate future directions. Firstly,  this SOS decomposition can be generalised to find maximal violations for higher setting multi-partite Bell inequalities. Combined with robustness analysis, this approach holds potential for developing robust delegated quantum computing protocols involving a single verifier multiple provers. Additionally, this technique can be applied to revisit the proposed multi-partite key distributions protocols \cite{Epping2017, Das2021} in a dimension independent manner.


\section{Acknowledgement}
RKS acknowledges the financial support from the Council of Scientific and Industrial Research (CSIR, 09/1001(17051)/2023-EMR-I), Government of India. SS acknowledges the support from the National Natural Science Fund of China (Grant No. G0512250610191) and the local hospitality from the research grant SG160 of IIT Hyderabad, India. A.K.P. acknowledges the support
from Research Grant No. SERB/CRG/2021/004258, Government of India.



\begin{thebibliography}{88}%
	\makeatletter
	\providecommand \@ifxundefined [1]{%
		\@ifx{#1\undefined}
	}%
	\providecommand \@ifnum [1]{%
		\ifnum #1\expandafter \@firstoftwo
		\else \expandafter \@secondoftwo
		\fi
	}%
	\providecommand \@ifx [1]{%
		\ifx #1\expandafter \@firstoftwo
		\else \expandafter \@secondoftwo
		\fi
	}%
	\providecommand \natexlab [1]{#1}%
	\providecommand \enquote  [1]{``#1''}%
	\providecommand \bibnamefont  [1]{#1}%
	\providecommand \bibfnamefont [1]{#1}%
	\providecommand \citenamefont [1]{#1}%
	\providecommand \href@noop [0]{\@secondoftwo}%
	\providecommand \href [0]{\begingroup \@sanitize@url \@href}%
	\providecommand \@href[1]{\@@startlink{#1}\@@href}%
	\providecommand \@@href[1]{\endgroup#1\@@endlink}%
	\providecommand \@sanitize@url [0]{\catcode `\\12\catcode `\$12\catcode
		`\&12\catcode `\#12\catcode `\^12\catcode `\_12\catcode `\%12\relax}%
	\providecommand \@@startlink[1]{}%
	\providecommand \@@endlink[0]{}%
	\providecommand \url  [0]{\begingroup\@sanitize@url \@url }%
	\providecommand \@url [1]{\endgroup\@href {#1}{\urlprefix }}%
	\providecommand \urlprefix  [0]{URL }%
	\providecommand \Eprint [0]{\href }%
	\providecommand \doibase [0]{https://doi.org/}%
	\providecommand \selectlanguage [0]{\@gobble}%
	\providecommand \bibinfo  [0]{\@secondoftwo}%
	\providecommand \bibfield  [0]{\@secondoftwo}%
	\providecommand \translation [1]{[#1]}%
	\providecommand \BibitemOpen [0]{}%
	\providecommand \bibitemStop [0]{}%
	\providecommand \bibitemNoStop [0]{.\EOS\space}%
	\providecommand \EOS [0]{\spacefactor3000\relax}%
	\providecommand \BibitemShut  [1]{\csname bibitem#1\endcsname}%
	\let\auto@bib@innerbib\@empty
	\bibitem [{\citenamefont {Ekert}(1991)}]{Ekert1991}%
	\BibitemOpen
	\bibfield  {author} {\bibinfo {author} {\bibfnamefont {A.~K.}\ \bibnamefont
			{Ekert}},\ }\bibfield  {title} {\bibinfo {title} {Quantum cryptography based
			on bell's theorem},\ }\href {https://doi.org/10.1103/PhysRevLett.67.661}
	{\bibfield  {journal} {\bibinfo  {journal} {Phys. Rev. Lett.}\ }\textbf
		{\bibinfo {volume} {67}},\ \bibinfo {pages} {661} (\bibinfo {year}
		{1991})}\BibitemShut {NoStop}%
	\bibitem [{\citenamefont {Pironio}\ \emph {et~al.}(2009)\citenamefont
		{Pironio}, \citenamefont {Acín}, \citenamefont {Brunner}, \citenamefont
		{Gisin}, \citenamefont {Massar},\ and\ \citenamefont
		{Scarani}}]{Pironio2009}%
	\BibitemOpen
	\bibfield  {author} {\bibinfo {author} {\bibfnamefont {S.}~\bibnamefont
			{Pironio}}, \bibinfo {author} {\bibfnamefont {A.}~\bibnamefont {Acín}},
		\bibinfo {author} {\bibfnamefont {N.}~\bibnamefont {Brunner}}, \bibinfo
		{author} {\bibfnamefont {N.}~\bibnamefont {Gisin}}, \bibinfo {author}
		{\bibfnamefont {S.}~\bibnamefont {Massar}},\ and\ \bibinfo {author}
		{\bibfnamefont {V.}~\bibnamefont {Scarani}},\ }\bibfield  {title} {\bibinfo
		{title} {Device-independent quantum key distribution secure against
			collective attacks},\ }\href {https://doi.org/10.1088/1367-2630/11/4/045021}
	{\bibfield  {journal} {\bibinfo  {journal} {New Journal of Physics}\ }\textbf
		{\bibinfo {volume} {11}},\ \bibinfo {pages} {045021} (\bibinfo {year}
		{2009})}\BibitemShut {NoStop}%
	\bibitem [{\citenamefont {Wooltorton}\ \emph {et~al.}(2024)\citenamefont
		{Wooltorton}, \citenamefont {Brown},\ and\ \citenamefont
		{Colbeck}}]{Wooltorton2024}%
	\BibitemOpen
	\bibfield  {author} {\bibinfo {author} {\bibfnamefont {L.}~\bibnamefont
			{Wooltorton}}, \bibinfo {author} {\bibfnamefont {P.}~\bibnamefont {Brown}},\
		and\ \bibinfo {author} {\bibfnamefont {R.}~\bibnamefont {Colbeck}},\
	}\bibfield  {title} {\bibinfo {title} {Device-independent quantum key
			distribution with arbitrarily small nonlocality},\ }\href
	{https://doi.org/10.1103/PhysRevLett.132.210802} {\bibfield  {journal}
		{\bibinfo  {journal} {Phys. Rev. Lett.}\ }\textbf {\bibinfo {volume} {132}},\
		\bibinfo {pages} {210802} (\bibinfo {year} {2024})}\BibitemShut {NoStop}%
	\bibitem [{\citenamefont {Pironio}\ \emph {et~al.}(2010)\citenamefont
		{Pironio}, \citenamefont {Ac{\'i}n}, \citenamefont {Massar}, \citenamefont
		{de~la Giroday}, \citenamefont {Matsukevich}, \citenamefont {Maunz},
		\citenamefont {Olmschenk}, \citenamefont {Hayes}, \citenamefont {Luo},
		\citenamefont {Manning},\ and\ \citenamefont {Monroe}}]{Pironio2010}%
	\BibitemOpen
	\bibfield  {author} {\bibinfo {author} {\bibfnamefont {S.}~\bibnamefont
			{Pironio}}, \bibinfo {author} {\bibfnamefont {A.}~\bibnamefont {Ac{\'i}n}},
		\bibinfo {author} {\bibfnamefont {S.}~\bibnamefont {Massar}}, \bibinfo
		{author} {\bibfnamefont {A.~B.}\ \bibnamefont {de~la Giroday}}, \bibinfo
		{author} {\bibfnamefont {D.~N.}\ \bibnamefont {Matsukevich}}, \bibinfo
		{author} {\bibfnamefont {P.}~\bibnamefont {Maunz}}, \bibinfo {author}
		{\bibfnamefont {S.}~\bibnamefont {Olmschenk}}, \bibinfo {author}
		{\bibfnamefont {D.}~\bibnamefont {Hayes}}, \bibinfo {author} {\bibfnamefont
			{L.}~\bibnamefont {Luo}}, \bibinfo {author} {\bibfnamefont {T.~A.}\
			\bibnamefont {Manning}},\ and\ \bibinfo {author} {\bibfnamefont
			{C.}~\bibnamefont {Monroe}},\ }\bibfield  {title} {\bibinfo {title} {Random
			numbers certified by bell's theorem},\ }\href
	{https://doi.org/10.1038/nature09008} {\bibfield  {journal} {\bibinfo
			{journal} {Nature}\ }\textbf {\bibinfo {volume} {464}},\ \bibinfo {pages}
		{1021} (\bibinfo {year} {2010})}\BibitemShut {NoStop}%
	\bibitem [{\citenamefont {Colbeck}\ and\ \citenamefont
		{Renner}(2012)}]{Colbeck2012}%
	\BibitemOpen
	\bibfield  {author} {\bibinfo {author} {\bibfnamefont {R.}~\bibnamefont
			{Colbeck}}\ and\ \bibinfo {author} {\bibfnamefont {R.}~\bibnamefont
			{Renner}},\ }\bibfield  {title} {\bibinfo {title} {Free randomness can be
			amplified},\ }\href {https://doi.org/10.1038/nphys2300} {\bibfield  {journal}
		{\bibinfo  {journal} {Nature Physics}\ }\textbf {\bibinfo {volume} {8}},\
		\bibinfo {pages} {450} (\bibinfo {year} {2012})}\BibitemShut {NoStop}%
	\bibitem [{\citenamefont {Liu}\ \emph {et~al.}(2021)\citenamefont {Liu},
		\citenamefont {Li}, \citenamefont {Ragy}, \citenamefont {Zhao}, \citenamefont
		{Bai}, \citenamefont {Liu}, \citenamefont {Brown}, \citenamefont {Zhang},
		\citenamefont {Colbeck}, \citenamefont {Fan}, \citenamefont {Zhang},\ and\
		\citenamefont {Pan}}]{Liu2021}%
	\BibitemOpen
	\bibfield  {author} {\bibinfo {author} {\bibfnamefont {W.-Z.}\ \bibnamefont
			{Liu}}, \bibinfo {author} {\bibfnamefont {M.-H.}\ \bibnamefont {Li}},
		\bibinfo {author} {\bibfnamefont {S.}~\bibnamefont {Ragy}}, \bibinfo {author}
		{\bibfnamefont {S.-R.}\ \bibnamefont {Zhao}}, \bibinfo {author}
		{\bibfnamefont {B.}~\bibnamefont {Bai}}, \bibinfo {author} {\bibfnamefont
			{Y.}~\bibnamefont {Liu}}, \bibinfo {author} {\bibfnamefont {P.~J.}\
			\bibnamefont {Brown}}, \bibinfo {author} {\bibfnamefont {J.}~\bibnamefont
			{Zhang}}, \bibinfo {author} {\bibfnamefont {R.}~\bibnamefont {Colbeck}},
		\bibinfo {author} {\bibfnamefont {J.}~\bibnamefont {Fan}}, \bibinfo {author}
		{\bibfnamefont {Q.}~\bibnamefont {Zhang}},\ and\ \bibinfo {author}
		{\bibfnamefont {J.-W.}\ \bibnamefont {Pan}},\ }\bibfield  {title} {\bibinfo
		{title} {Device-independent randomness expansion against quantum side
			information},\ }\href {https://doi.org/10.1038/s41567-020-01147-2} {\bibfield
		{journal} {\bibinfo  {journal} {Nature Physics}\ }\textbf {\bibinfo {volume}
			{17}},\ \bibinfo {pages} {448} (\bibinfo {year} {2021})}\BibitemShut
	{NoStop}%
	\bibitem [{\citenamefont {Wooltorton}\ \emph {et~al.}(2022)\citenamefont
		{Wooltorton}, \citenamefont {Brown},\ and\ \citenamefont
		{Colbeck}}]{Wooltorton2022}%
	\BibitemOpen
	\bibfield  {author} {\bibinfo {author} {\bibfnamefont {L.}~\bibnamefont
			{Wooltorton}}, \bibinfo {author} {\bibfnamefont {P.}~\bibnamefont {Brown}},\
		and\ \bibinfo {author} {\bibfnamefont {R.}~\bibnamefont {Colbeck}},\
	}\bibfield  {title} {\bibinfo {title} {Tight analytic bound on the trade-off
			between device-independent randomness and nonlocality},\ }\href
	{https://doi.org/10.1103/PhysRevLett.129.150403} {\bibfield  {journal}
		{\bibinfo  {journal} {Phys. Rev. Lett.}\ }\textbf {\bibinfo {volume} {129}},\
		\bibinfo {pages} {150403} (\bibinfo {year} {2022})}\BibitemShut {NoStop}%
	\bibitem [{\citenamefont {Fitzsimons}(2017)}]{Fitzsimons2017}%
	\BibitemOpen
	\bibfield  {author} {\bibinfo {author} {\bibfnamefont {J.~F.}\ \bibnamefont
			{Fitzsimons}},\ }\bibfield  {title} {\bibinfo {title} {Private quantum
			computation: an introduction to blind quantum computing and related
			protocols},\ }\href {https://doi.org/10.1038/s41534-017-0025-3} {\bibfield
		{journal} {\bibinfo  {journal} {npj Quantum Information}\ }\textbf {\bibinfo
			{volume} {3}},\ \bibinfo {pages} {23} (\bibinfo {year} {2017})}\BibitemShut
	{NoStop}%
	\bibitem [{\citenamefont {Mahadev}(2018)}]{Urmila2018}%
	\BibitemOpen
	\bibfield  {author} {\bibinfo {author} {\bibfnamefont {U.}~\bibnamefont
			{Mahadev}},\ }\bibfield  {title} {\bibinfo {title} {Classical verification of
			quantum computations},\ }in\ \href {https://doi.org/10.1109/FOCS.2018.00033}
	{\emph {\bibinfo {booktitle} {2018 IEEE 59th Annual Symposium on Foundations
				of Computer Science (FOCS)}}}\ (\bibinfo {year} {2018})\ pp.\ \bibinfo
	{pages} {259--267}\BibitemShut {NoStop}%
	\bibitem [{\citenamefont {Gheorghiu}\ \emph {et~al.}(2019)\citenamefont
		{Gheorghiu}, \citenamefont {Kapourniotis},\ and\ \citenamefont
		{Kashefi}}]{Gheorghiu2019}%
	\BibitemOpen
	\bibfield  {author} {\bibinfo {author} {\bibfnamefont {A.}~\bibnamefont
			{Gheorghiu}}, \bibinfo {author} {\bibfnamefont {T.}~\bibnamefont
			{Kapourniotis}},\ and\ \bibinfo {author} {\bibfnamefont {E.}~\bibnamefont
			{Kashefi}},\ }\bibfield  {title} {\bibinfo {title} {Verification of quantum
			computation: An overview of existing approaches},\ }\href
	{https://doi.org/10.1007/s00224-018-9872-3} {\bibfield  {journal} {\bibinfo
			{journal} {Theory of Computing Systems}\ }\textbf {\bibinfo {volume} {63}},\
		\bibinfo {pages} {715} (\bibinfo {year} {2019})}\BibitemShut {NoStop}%
	\bibitem [{\citenamefont {Banaszek}\ \emph {et~al.}(2013)\citenamefont
		{Banaszek}, \citenamefont {Cramer},\ and\ \citenamefont
		{Gross}}]{Banaszek2013}%
	\BibitemOpen
	\bibfield  {author} {\bibinfo {author} {\bibfnamefont {K.}~\bibnamefont
			{Banaszek}}, \bibinfo {author} {\bibfnamefont {M.}~\bibnamefont {Cramer}},\
		and\ \bibinfo {author} {\bibfnamefont {D.}~\bibnamefont {Gross}},\ }\bibfield
	{title} {\bibinfo {title} {Focus on quantum tomography},\ }\href
	{https://doi.org/10.1088/1367-2630/15/12/125020} {\bibfield  {journal}
		{\bibinfo  {journal} {New Journal of Physics}\ }\textbf {\bibinfo {volume}
			{15}},\ \bibinfo {pages} {125020} (\bibinfo {year} {2013})}\BibitemShut
	{NoStop}%
	\bibitem [{\citenamefont {O'Donnell}\ and\ \citenamefont
		{Wright}(2016)}]{Donnel2016}%
	\BibitemOpen
	\bibfield  {author} {\bibinfo {author} {\bibfnamefont {R.}~\bibnamefont
			{O'Donnell}}\ and\ \bibinfo {author} {\bibfnamefont {J.}~\bibnamefont
			{Wright}},\ }\bibfield  {title} {\bibinfo {title} {Efficient quantum
			tomography},\ }in\ \href {https://doi.org/10.1145/2897518.2897544} {\emph
		{\bibinfo {booktitle} {Proceedings of the Forty-Eighth Annual ACM Symposium
				on Theory of Computing}}},\ \bibinfo {series and number} {STOC '16}\
	(\bibinfo  {publisher} {Association for Computing Machinery},\ \bibinfo
	{address} {New York, NY, USA},\ \bibinfo {year} {2016})\ p.\ \bibinfo {pages}
	{899–912}\BibitemShut {NoStop}%
	\bibitem [{\citenamefont {Goh}\ \emph {et~al.}(2019)\citenamefont {Goh},
		\citenamefont {Perumangatt}, \citenamefont {Lee}, \citenamefont {Ling},\ and\
		\citenamefont {Scarani}}]{Goh2019}%
	\BibitemOpen
	\bibfield  {author} {\bibinfo {author} {\bibfnamefont {K.~T.}\ \bibnamefont
			{Goh}}, \bibinfo {author} {\bibfnamefont {C.}~\bibnamefont {Perumangatt}},
		\bibinfo {author} {\bibfnamefont {Z.~X.}\ \bibnamefont {Lee}}, \bibinfo
		{author} {\bibfnamefont {A.}~\bibnamefont {Ling}},\ and\ \bibinfo {author}
		{\bibfnamefont {V.}~\bibnamefont {Scarani}},\ }\bibfield  {title} {\bibinfo
		{title} {Experimental comparison of tomography and self-testing in certifying
			entanglement},\ }\href {https://doi.org/10.1103/PhysRevA.100.022305}
	{\bibfield  {journal} {\bibinfo  {journal} {Phys. Rev. A}\ }\textbf {\bibinfo
			{volume} {100}},\ \bibinfo {pages} {022305} (\bibinfo {year}
		{2019})}\BibitemShut {NoStop}%
	\bibitem [{\citenamefont {Sahoo}\ \emph {et~al.}(2020)\citenamefont {Sahoo},
		\citenamefont {Chakraborti}, \citenamefont {Pati},\ and\ \citenamefont
		{Sinha}}]{Sahoo2020}%
	\BibitemOpen
	\bibfield  {author} {\bibinfo {author} {\bibfnamefont {S.~N.}\ \bibnamefont
			{Sahoo}}, \bibinfo {author} {\bibfnamefont {S.}~\bibnamefont {Chakraborti}},
		\bibinfo {author} {\bibfnamefont {A.~K.}\ \bibnamefont {Pati}},\ and\
		\bibinfo {author} {\bibfnamefont {U.}~\bibnamefont {Sinha}},\ }\bibfield
	{title} {\bibinfo {title} {Quantum state interferography},\ }\href
	{https://doi.org/10.1103/PhysRevLett.125.123601} {\bibfield  {journal}
		{\bibinfo  {journal} {Phys. Rev. Lett.}\ }\textbf {\bibinfo {volume} {125}},\
		\bibinfo {pages} {123601} (\bibinfo {year} {2020})}\BibitemShut {NoStop}%
	\bibitem [{\citenamefont {P\'al}\ \emph {et~al.}(2014)\citenamefont {P\'al},
		\citenamefont {V\'ertesi},\ and\ \citenamefont {Navascu\'es}}]{Pal2014}%
	\BibitemOpen
	\bibfield  {author} {\bibinfo {author} {\bibfnamefont {K.~F.}\ \bibnamefont
			{P\'al}}, \bibinfo {author} {\bibfnamefont {T.}~\bibnamefont {V\'ertesi}},\
		and\ \bibinfo {author} {\bibfnamefont {M.}~\bibnamefont {Navascu\'es}},\
	}\bibfield  {title} {\bibinfo {title} {Device-independent tomography of
			multipartite quantum states},\ }\href
	{https://doi.org/10.1103/PhysRevA.90.042340} {\bibfield  {journal} {\bibinfo
			{journal} {Phys. Rev. A}\ }\textbf {\bibinfo {volume} {90}},\ \bibinfo
		{pages} {042340} (\bibinfo {year} {2014})}\BibitemShut {NoStop}%
	\bibitem [{\citenamefont {Mayers}\ and\ \citenamefont
		{Yao}(2004)}]{Mayers2004}%
	\BibitemOpen
	\bibfield  {author} {\bibinfo {author} {\bibfnamefont {D.}~\bibnamefont
			{Mayers}}\ and\ \bibinfo {author} {\bibfnamefont {A.}~\bibnamefont {Yao}},\
	}\bibfield  {title} {\bibinfo {title} {Self testing quantum apparatus},\
	}\href {https://dl.acm.org/doi/10.5555/2011827.2011830} {\bibfield  {journal}
		{\bibinfo  {journal} {Quantum Info. Comput.}\ }\textbf {\bibinfo {volume}
			{4}},\ \bibinfo {pages} {273–286} (\bibinfo {year} {2004})}\BibitemShut
	{NoStop}%
	\bibitem [{\citenamefont {Bell}(1964)}]{Bell1964}%
	\BibitemOpen
	\bibfield  {author} {\bibinfo {author} {\bibfnamefont {J.~S.}\ \bibnamefont
			{Bell}},\ }\bibfield  {title} {\bibinfo {title} {On the einstein podolsky
			rosen paradox},\ }\href {https://doi.org/10.1103/PhysicsPhysiqueFizika.1.195}
	{\bibfield  {journal} {\bibinfo  {journal} {Physics Physique Fizika}\
		}\textbf {\bibinfo {volume} {1}},\ \bibinfo {pages} {195} (\bibinfo {year}
		{1964})}\BibitemShut {NoStop}%
	\bibitem [{\citenamefont {Brunner}\ \emph {et~al.}(2014)\citenamefont
		{Brunner}, \citenamefont {Cavalcanti}, \citenamefont {Pironio}, \citenamefont
		{Scarani},\ and\ \citenamefont {Wehner}}]{Brunner2014rev}%
	\BibitemOpen
	\bibfield  {author} {\bibinfo {author} {\bibfnamefont {N.}~\bibnamefont
			{Brunner}}, \bibinfo {author} {\bibfnamefont {D.}~\bibnamefont {Cavalcanti}},
		\bibinfo {author} {\bibfnamefont {S.}~\bibnamefont {Pironio}}, \bibinfo
		{author} {\bibfnamefont {V.}~\bibnamefont {Scarani}},\ and\ \bibinfo {author}
		{\bibfnamefont {S.}~\bibnamefont {Wehner}},\ }\bibfield  {title} {\bibinfo
		{title} {Bell nonlocality},\ }\href
	{https://doi.org/10.1103/RevModPhys.86.419} {\bibfield  {journal} {\bibinfo
			{journal} {Rev. Mod. Phys.}\ }\textbf {\bibinfo {volume} {86}},\ \bibinfo
		{pages} {419} (\bibinfo {year} {2014})}\BibitemShut {NoStop}%
	\bibitem [{\citenamefont {{\v{S}}upi{\'{c}}}\ and\ \citenamefont
		{Bowles}(2020)}]{Supic2020}%
	\BibitemOpen
	\bibfield  {author} {\bibinfo {author} {\bibfnamefont {I.}~\bibnamefont
			{{\v{S}}upi{\'{c}}}}\ and\ \bibinfo {author} {\bibfnamefont {J.}~\bibnamefont
			{Bowles}},\ }\bibfield  {title} {\bibinfo {title} {Self-testing of quantum
			systems: a review},\ }\href {https://doi.org/10.22331/q-2020-09-30-337}
	{\bibfield  {journal} {\bibinfo  {journal} {Quantum}\ }\textbf {\bibinfo
			{volume} {4}},\ \bibinfo {pages} {337} (\bibinfo {year} {2020})}\BibitemShut
	{NoStop}%
	\bibitem [{\citenamefont {Clauser}\ \emph {et~al.}(1969)\citenamefont
		{Clauser}, \citenamefont {Horne}, \citenamefont {Shimony},\ and\
		\citenamefont {Holt}}]{Clauser1969}%
	\BibitemOpen
	\bibfield  {author} {\bibinfo {author} {\bibfnamefont {J.~F.}\ \bibnamefont
			{Clauser}}, \bibinfo {author} {\bibfnamefont {M.~A.}\ \bibnamefont {Horne}},
		\bibinfo {author} {\bibfnamefont {A.}~\bibnamefont {Shimony}},\ and\ \bibinfo
		{author} {\bibfnamefont {R.~A.}\ \bibnamefont {Holt}},\ }\bibfield  {title}
	{\bibinfo {title} {Proposed experiment to test local hidden-variable
			theories},\ }\href {https://doi.org/10.1103/PhysRevLett.23.880} {\bibfield
		{journal} {\bibinfo  {journal} {Phys. Rev. Lett.}\ }\textbf {\bibinfo
			{volume} {23}},\ \bibinfo {pages} {880} (\bibinfo {year} {1969})}\BibitemShut
	{NoStop}%
	\bibitem [{\citenamefont {McKague}\ \emph {et~al.}(2012)\citenamefont
		{McKague}, \citenamefont {Yang},\ and\ \citenamefont
		{Scarani}}]{McKague2012}%
	\BibitemOpen
	\bibfield  {author} {\bibinfo {author} {\bibfnamefont {M.}~\bibnamefont
			{McKague}}, \bibinfo {author} {\bibfnamefont {T.~H.}\ \bibnamefont {Yang}},\
		and\ \bibinfo {author} {\bibfnamefont {V.}~\bibnamefont {Scarani}},\
	}\bibfield  {title} {\bibinfo {title} {Robust self-testing of the singlet},\
	}\href {https://doi.org/10.1088/1751-8113/45/45/455304} {\bibfield  {journal}
		{\bibinfo  {journal} {Journal of Physics A: Mathematical and Theoretical}\
		}\textbf {\bibinfo {volume} {45}},\ \bibinfo {pages} {455304} (\bibinfo
		{year} {2012})}\BibitemShut {NoStop}%
	\bibitem [{\citenamefont {Wu}\ \emph {et~al.}(2016)\citenamefont {Wu},
		\citenamefont {Bancal}, \citenamefont {McKague},\ and\ \citenamefont
		{Scarani}}]{Wu2016}%
	\BibitemOpen
	\bibfield  {author} {\bibinfo {author} {\bibfnamefont {X.}~\bibnamefont
			{Wu}}, \bibinfo {author} {\bibfnamefont {J.-D.}\ \bibnamefont {Bancal}},
		\bibinfo {author} {\bibfnamefont {M.}~\bibnamefont {McKague}},\ and\ \bibinfo
		{author} {\bibfnamefont {V.}~\bibnamefont {Scarani}},\ }\bibfield  {title}
	{\bibinfo {title} {Device-independent parallel self-testing of two
			singlets},\ }\href {https://doi.org/10.1103/PhysRevA.93.062121} {\bibfield
		{journal} {\bibinfo  {journal} {Phys. Rev. A}\ }\textbf {\bibinfo {volume}
			{93}},\ \bibinfo {pages} {062121} (\bibinfo {year} {2016})}\BibitemShut
	{NoStop}%
	\bibitem [{\citenamefont {Sarkar}(2023)}]{Sarkar2023a}%
	\BibitemOpen
	\bibfield  {author} {\bibinfo {author} {\bibfnamefont {S.}~\bibnamefont
			{Sarkar}},\ }\bibfield  {title} {\bibinfo {title} {Certification of the
			maximally entangled state using nonprojective measurements},\ }\href
	{https://doi.org/10.1103/PhysRevA.107.032408} {\bibfield  {journal} {\bibinfo
			{journal} {Phys. Rev. A}\ }\textbf {\bibinfo {volume} {107}},\ \bibinfo
		{pages} {032408} (\bibinfo {year} {2023})}\BibitemShut {NoStop}%
	\bibitem [{\citenamefont {Coladangelo}\ \emph {et~al.}(2017)\citenamefont
		{Coladangelo}, \citenamefont {Goh},\ and\ \citenamefont
		{Scarani}}]{Coladangelo_ncomms_2017}%
	\BibitemOpen
	\bibfield  {author} {\bibinfo {author} {\bibfnamefont {A.}~\bibnamefont
			{Coladangelo}}, \bibinfo {author} {\bibfnamefont {K.~T.}\ \bibnamefont
			{Goh}},\ and\ \bibinfo {author} {\bibfnamefont {V.}~\bibnamefont {Scarani}},\
	}\bibfield  {title} {\bibinfo {title} {All pure bipartite entangled states
			can be self-tested},\ }\href {https://doi.org/10.1038/ncomms15485} {\bibfield
		{journal} {\bibinfo  {journal} {Nature Communications}\ }\textbf {\bibinfo
			{volume} {8}},\ \bibinfo {pages} {15485} (\bibinfo {year}
		{2017})}\BibitemShut {NoStop}%
	\bibitem [{\citenamefont {Prabuddha~Roy}(2023)}]{Roy_Pan2023}%
	\BibitemOpen
	\bibfield  {author} {\bibinfo {author} {\bibfnamefont {A.~K.~P.}\
			\bibnamefont {Prabuddha~Roy}},\ }\bibfield  {title} {\bibinfo {title}
		{Device-independent self-testing of unsharp measurements},\ }\bibfield
	{journal} {\bibinfo  {journal} {New J. Phys}\ }\href
	{https://doi.org/10.1088/1367-2630/acb4b5} {10.1088/1367-2630/acb4b5}
	(\bibinfo {year} {2023})\BibitemShut {NoStop}%
	\bibitem [{\citenamefont {Paul}\ \emph {et~al.}(2024)\citenamefont {Paul},
		\citenamefont {Sasmal},\ and\ \citenamefont {Pan}}]{Rajdeep2024}%
	\BibitemOpen
	\bibfield  {author} {\bibinfo {author} {\bibfnamefont {R.}~\bibnamefont
			{Paul}}, \bibinfo {author} {\bibfnamefont {S.}~\bibnamefont {Sasmal}},\ and\
		\bibinfo {author} {\bibfnamefont {A.~K.}\ \bibnamefont {Pan}},\ }\bibfield
	{title} {\bibinfo {title} {Self-testing of multiple unsharpness parameters
			through sequential violations of a noncontextual inequality},\ }\href
	{https://doi.org/10.1103/PhysRevA.110.012444} {\bibfield  {journal} {\bibinfo
			{journal} {Phys. Rev. A}\ }\textbf {\bibinfo {volume} {110}},\ \bibinfo
		{pages} {012444} (\bibinfo {year} {2024})}\BibitemShut {NoStop}%
	\bibitem [{\citenamefont {Sekatski}\ \emph {et~al.}(2018)\citenamefont
		{Sekatski}, \citenamefont {Bancal}, \citenamefont {Wagner},\ and\
		\citenamefont {Sangouard}}]{pawel2018}%
	\BibitemOpen
	\bibfield  {author} {\bibinfo {author} {\bibfnamefont {P.}~\bibnamefont
			{Sekatski}}, \bibinfo {author} {\bibfnamefont {J.-D.}\ \bibnamefont
			{Bancal}}, \bibinfo {author} {\bibfnamefont {S.}~\bibnamefont {Wagner}},\
		and\ \bibinfo {author} {\bibfnamefont {N.}~\bibnamefont {Sangouard}},\
	}\bibfield  {title} {\bibinfo {title} {Certifying the building blocks of
			quantum computers from bell's theorem},\ }\href
	{https://doi.org/10.1103/PhysRevLett.121.180505} {\bibfield  {journal}
		{\bibinfo  {journal} {Phys. Rev. Lett.}\ }\textbf {\bibinfo {volume} {121}},\
		\bibinfo {pages} {180505} (\bibinfo {year} {2018})}\BibitemShut {NoStop}%
	\bibitem [{\citenamefont {Sekatski}\ \emph {et~al.}(2023)\citenamefont
		{Sekatski}, \citenamefont {Bancal}, \citenamefont {Ioannou}, \citenamefont
		{Afzelius},\ and\ \citenamefont {Brunner}}]{Sekatski2023}%
	\BibitemOpen
	\bibfield  {author} {\bibinfo {author} {\bibfnamefont {P.}~\bibnamefont
			{Sekatski}}, \bibinfo {author} {\bibfnamefont {J.-D.}\ \bibnamefont
			{Bancal}}, \bibinfo {author} {\bibfnamefont {M.}~\bibnamefont {Ioannou}},
		\bibinfo {author} {\bibfnamefont {M.}~\bibnamefont {Afzelius}},\ and\
		\bibinfo {author} {\bibfnamefont {N.}~\bibnamefont {Brunner}},\ }\bibfield
	{title} {\bibinfo {title} {Toward the device-independent certification of a
			quantum memory},\ }\href {https://doi.org/10.1103/PhysRevLett.131.170802}
	{\bibfield  {journal} {\bibinfo  {journal} {Phys. Rev. Lett.}\ }\textbf
		{\bibinfo {volume} {131}},\ \bibinfo {pages} {170802} (\bibinfo {year}
		{2023})}\BibitemShut {NoStop}%
	\bibitem [{\citenamefont {Wagner}\ \emph {et~al.}(2020)\citenamefont {Wagner},
		\citenamefont {Bancal}, \citenamefont {Sangouard},\ and\ \citenamefont
		{Sekatski}}]{Wagner2020}%
	\BibitemOpen
	\bibfield  {author} {\bibinfo {author} {\bibfnamefont {S.}~\bibnamefont
			{Wagner}}, \bibinfo {author} {\bibfnamefont {J.-D.}\ \bibnamefont {Bancal}},
		\bibinfo {author} {\bibfnamefont {N.}~\bibnamefont {Sangouard}},\ and\
		\bibinfo {author} {\bibfnamefont {P.}~\bibnamefont {Sekatski}},\ }\bibfield
	{title} {\bibinfo {title} {Device-independent characterization of quantum
			instruments},\ }\href {https://doi.org/10.22331/q-2020-03-19-243} {\bibfield
		{journal} {\bibinfo  {journal} {{Quantum}}\ }\textbf {\bibinfo {volume}
			{4}},\ \bibinfo {pages} {243} (\bibinfo {year} {2020})}\BibitemShut {NoStop}%
	\bibitem [{\citenamefont {Bowles}\ \emph
		{et~al.}(2018{\natexlab{a}})\citenamefont {Bowles}, \citenamefont {\ifmmode
			\check{S}\else \v{S}\fi{}upi\ifmmode~\acute{c}\else \'{c}\fi{}},
		\citenamefont {Cavalcanti},\ and\ \citenamefont {Ac\'{\i}n}}]{Bowles2018a}%
	\BibitemOpen
	\bibfield  {author} {\bibinfo {author} {\bibfnamefont {J.}~\bibnamefont
			{Bowles}}, \bibinfo {author} {\bibfnamefont {I.}~\bibnamefont {\ifmmode
				\check{S}\else \v{S}\fi{}upi\ifmmode~\acute{c}\else \'{c}\fi{}}}, \bibinfo
		{author} {\bibfnamefont {D.}~\bibnamefont {Cavalcanti}},\ and\ \bibinfo
		{author} {\bibfnamefont {A.}~\bibnamefont {Ac\'{\i}n}},\ }\bibfield  {title}
	{\bibinfo {title} {Self-testing of pauli observables for device-independent
			entanglement certification},\ }\href
	{https://doi.org/10.1103/PhysRevA.98.042336} {\bibfield  {journal} {\bibinfo
			{journal} {Phys. Rev. A}\ }\textbf {\bibinfo {volume} {98}},\ \bibinfo
		{pages} {042336} (\bibinfo {year} {2018}{\natexlab{a}})}\BibitemShut
	{NoStop}%
	\bibitem [{\citenamefont {Bowles}\ \emph
		{et~al.}(2018{\natexlab{b}})\citenamefont {Bowles}, \citenamefont {\ifmmode
			\check{S}\else \v{S}\fi{}upi\ifmmode~\acute{c}\else \'{c}\fi{}},
		\citenamefont {Cavalcanti},\ and\ \citenamefont {Ac\'{\i}n}}]{Bowles2018}%
	\BibitemOpen
	\bibfield  {author} {\bibinfo {author} {\bibfnamefont {J.}~\bibnamefont
			{Bowles}}, \bibinfo {author} {\bibfnamefont {I.}~\bibnamefont {\ifmmode
				\check{S}\else \v{S}\fi{}upi\ifmmode~\acute{c}\else \'{c}\fi{}}}, \bibinfo
		{author} {\bibfnamefont {D.}~\bibnamefont {Cavalcanti}},\ and\ \bibinfo
		{author} {\bibfnamefont {A.}~\bibnamefont {Ac\'{\i}n}},\ }\bibfield  {title}
	{\bibinfo {title} {Device-independent entanglement certification of all
			entangled states},\ }\href {https://doi.org/10.1103/PhysRevLett.121.180503}
	{\bibfield  {journal} {\bibinfo  {journal} {Phys. Rev. Lett.}\ }\textbf
		{\bibinfo {volume} {121}},\ \bibinfo {pages} {180503} (\bibinfo {year}
		{2018}{\natexlab{b}})}\BibitemShut {NoStop}%
	\bibitem [{\citenamefont {{\v{S}}upi{\'{c}}}\ \emph {et~al.}(2023)\citenamefont
		{{\v{S}}upi{\'{c}}}, \citenamefont {Bowles}, \citenamefont {Renou},
		\citenamefont {Ac{\'{i}}n},\ and\ \citenamefont {Hoban}}]{Šupić2023}%
	\BibitemOpen
	\bibfield  {author} {\bibinfo {author} {\bibfnamefont {I.}~\bibnamefont
			{{\v{S}}upi{\'{c}}}}, \bibinfo {author} {\bibfnamefont {J.}~\bibnamefont
			{Bowles}}, \bibinfo {author} {\bibfnamefont {M.-O.}\ \bibnamefont {Renou}},
		\bibinfo {author} {\bibfnamefont {A.}~\bibnamefont {Ac{\'{i}}n}},\ and\
		\bibinfo {author} {\bibfnamefont {M.~J.}\ \bibnamefont {Hoban}},\ }\bibfield
	{title} {\bibinfo {title} {Quantum networks self-test all entangled states},\
	}\href {https://doi.org/10.1038/s41567-023-01945-4} {\bibfield  {journal}
		{\bibinfo  {journal} {Nature Physics}\ }\textbf {\bibinfo {volume} {19}},\
		\bibinfo {pages} {670} (\bibinfo {year} {2023})}\BibitemShut {NoStop}%
	\bibitem [{\citenamefont {Munshi}\ and\ \citenamefont {Pan}(2023)}]{Sneha2023}%
	\BibitemOpen
	\bibfield  {author} {\bibinfo {author} {\bibfnamefont {S.}~\bibnamefont
			{Munshi}}\ and\ \bibinfo {author} {\bibfnamefont {A.~K.}\ \bibnamefont
			{Pan}},\ }\bibfield  {title} {\bibinfo {title} {Self-testing of an unbounded
			number of mutually commuting local observables},\ }\href
	{https://doi.org/10.1103/PhysRevA.108.062607} {\bibfield  {journal} {\bibinfo
			{journal} {Phys. Rev. A}\ }\textbf {\bibinfo {volume} {108}},\ \bibinfo
		{pages} {062607} (\bibinfo {year} {2023})}\BibitemShut {NoStop}%
	\bibitem [{\citenamefont {Renou}\ \emph {et~al.}(2021)\citenamefont {Renou},
		\citenamefont {Trillo}, \citenamefont {Weilenmann}, \citenamefont {Le},
		\citenamefont {Tavakoli}, \citenamefont {Gisin}, \citenamefont {Ac\'{i}n},\
		and\ \citenamefont {Navascu{\'e}s}}]{Renou2021}%
	\BibitemOpen
	\bibfield  {author} {\bibinfo {author} {\bibfnamefont {M.-O.}\ \bibnamefont
			{Renou}}, \bibinfo {author} {\bibfnamefont {D.}~\bibnamefont {Trillo}},
		\bibinfo {author} {\bibfnamefont {M.}~\bibnamefont {Weilenmann}}, \bibinfo
		{author} {\bibfnamefont {T.~P.}\ \bibnamefont {Le}}, \bibinfo {author}
		{\bibfnamefont {A.}~\bibnamefont {Tavakoli}}, \bibinfo {author}
		{\bibfnamefont {N.}~\bibnamefont {Gisin}}, \bibinfo {author} {\bibfnamefont
			{A.}~\bibnamefont {Ac\'{i}n}},\ and\ \bibinfo {author} {\bibfnamefont
			{M.}~\bibnamefont {Navascu{\'e}s}},\ }\bibfield  {title} {\bibinfo {title}
		{Quantum theory based on real numbers can be experimentally falsified},\
	}\href {https://doi.org/10.1038/s41586-021-04160-4} {\bibfield  {journal}
		{\bibinfo  {journal} {Nature}\ }\textbf {\bibinfo {volume} {600}},\ \bibinfo
		{pages} {625} (\bibinfo {year} {2021})}\BibitemShut {NoStop}%
	\bibitem [{\citenamefont {Hu}\ \emph {et~al.}(2023)\citenamefont {Hu},
		\citenamefont {Xie}, \citenamefont {Arora}, \citenamefont {Ai}, \citenamefont
		{Bharti}, \citenamefont {Zhang}, \citenamefont {Wu}, \citenamefont {Chen},
		\citenamefont {Cui}, \citenamefont {Liu}, \citenamefont {Huang},
		\citenamefont {Li}, \citenamefont {Guo}, \citenamefont {Roland},
		\citenamefont {Cabello},\ and\ \citenamefont {Kwek}}]{Hu2023}%
	\BibitemOpen
	\bibfield  {author} {\bibinfo {author} {\bibfnamefont {X.-M.}\ \bibnamefont
			{Hu}}, \bibinfo {author} {\bibfnamefont {Y.}~\bibnamefont {Xie}}, \bibinfo
		{author} {\bibfnamefont {A.~S.}\ \bibnamefont {Arora}}, \bibinfo {author}
		{\bibfnamefont {M.-Z.}\ \bibnamefont {Ai}}, \bibinfo {author} {\bibfnamefont
			{K.}~\bibnamefont {Bharti}}, \bibinfo {author} {\bibfnamefont
			{J.}~\bibnamefont {Zhang}}, \bibinfo {author} {\bibfnamefont
			{W.}~\bibnamefont {Wu}}, \bibinfo {author} {\bibfnamefont {P.-X.}\
			\bibnamefont {Chen}}, \bibinfo {author} {\bibfnamefont {J.-M.}\ \bibnamefont
			{Cui}}, \bibinfo {author} {\bibfnamefont {B.-H.}\ \bibnamefont {Liu}},
		\bibinfo {author} {\bibfnamefont {Y.-F.}\ \bibnamefont {Huang}}, \bibinfo
		{author} {\bibfnamefont {C.-F.}\ \bibnamefont {Li}}, \bibinfo {author}
		{\bibfnamefont {G.-C.}\ \bibnamefont {Guo}}, \bibinfo {author} {\bibfnamefont
			{J.}~\bibnamefont {Roland}}, \bibinfo {author} {\bibfnamefont
			{A.}~\bibnamefont {Cabello}},\ and\ \bibinfo {author} {\bibfnamefont {L.-C.}\
			\bibnamefont {Kwek}},\ }\bibfield  {title} {\bibinfo {title} {Self-testing of
			a single quantum system from theory to experiment},\ }\href
	{https://doi.org/10.1038/s41534-023-00769-7} {\bibfield  {journal} {\bibinfo
			{journal} {npj Quantum Information}\ }\textbf {\bibinfo {volume} {9}},\
		\bibinfo {pages} {103} (\bibinfo {year} {2023})}\BibitemShut {NoStop}%
	\bibitem [{\citenamefont {Zhang}\ \emph {et~al.}(2019)\citenamefont {Zhang},
		\citenamefont {Chen}, \citenamefont {Yin}, \citenamefont {Peng},
		\citenamefont {Hu}, \citenamefont {Hou}, \citenamefont {Zhou}, \citenamefont
		{Yu}, \citenamefont {Ye}, \citenamefont {Zhou}, \citenamefont {Xu},
		\citenamefont {Tang}, \citenamefont {Xu}, \citenamefont {Han}, \citenamefont
		{Liu}, \citenamefont {Li},\ and\ \citenamefont {Guo}}]{Zhang2019}%
	\BibitemOpen
	\bibfield  {author} {\bibinfo {author} {\bibfnamefont {W.-H.}\ \bibnamefont
			{Zhang}}, \bibinfo {author} {\bibfnamefont {G.}~\bibnamefont {Chen}},
		\bibinfo {author} {\bibfnamefont {P.}~\bibnamefont {Yin}}, \bibinfo {author}
		{\bibfnamefont {X.-X.}\ \bibnamefont {Peng}}, \bibinfo {author}
		{\bibfnamefont {X.-M.}\ \bibnamefont {Hu}}, \bibinfo {author} {\bibfnamefont
			{Z.-B.}\ \bibnamefont {Hou}}, \bibinfo {author} {\bibfnamefont {Z.-Y.}\
			\bibnamefont {Zhou}}, \bibinfo {author} {\bibfnamefont {S.}~\bibnamefont
			{Yu}}, \bibinfo {author} {\bibfnamefont {X.-J.}\ \bibnamefont {Ye}}, \bibinfo
		{author} {\bibfnamefont {Z.-Q.}\ \bibnamefont {Zhou}}, \bibinfo {author}
		{\bibfnamefont {X.-Y.}\ \bibnamefont {Xu}}, \bibinfo {author} {\bibfnamefont
			{J.-S.}\ \bibnamefont {Tang}}, \bibinfo {author} {\bibfnamefont {J.-S.}\
			\bibnamefont {Xu}}, \bibinfo {author} {\bibfnamefont {Y.-J.}\ \bibnamefont
			{Han}}, \bibinfo {author} {\bibfnamefont {B.-H.}\ \bibnamefont {Liu}},
		\bibinfo {author} {\bibfnamefont {C.-F.}\ \bibnamefont {Li}},\ and\ \bibinfo
		{author} {\bibfnamefont {G.-C.}\ \bibnamefont {Guo}},\ }\bibfield  {title}
	{\bibinfo {title} {Experimental demonstration of robust self-testing for
			bipartite entangled states},\ }\href
	{https://doi.org/10.1038/s41534-018-0120-0} {\bibfield  {journal} {\bibinfo
			{journal} {npj Quantum Information}\ }\textbf {\bibinfo {volume} {5}},\
		\bibinfo {pages} {4} (\bibinfo {year} {2019})}\BibitemShut {NoStop}%
	\bibitem [{\citenamefont {Zhang}\ \emph {et~al.}(2018)\citenamefont {Zhang},
		\citenamefont {Chen}, \citenamefont {Peng}, \citenamefont {Ye}, \citenamefont
		{Yin}, \citenamefont {Xiao}, \citenamefont {Hou}, \citenamefont {Cheng},
		\citenamefont {Wu}, \citenamefont {Xu}, \citenamefont {Li},\ and\
		\citenamefont {Guo}}]{Zhang2018}%
	\BibitemOpen
	\bibfield  {author} {\bibinfo {author} {\bibfnamefont {W.-H.}\ \bibnamefont
			{Zhang}}, \bibinfo {author} {\bibfnamefont {G.}~\bibnamefont {Chen}},
		\bibinfo {author} {\bibfnamefont {X.-X.}\ \bibnamefont {Peng}}, \bibinfo
		{author} {\bibfnamefont {X.-J.}\ \bibnamefont {Ye}}, \bibinfo {author}
		{\bibfnamefont {P.}~\bibnamefont {Yin}}, \bibinfo {author} {\bibfnamefont
			{Y.}~\bibnamefont {Xiao}}, \bibinfo {author} {\bibfnamefont {Z.-B.}\
			\bibnamefont {Hou}}, \bibinfo {author} {\bibfnamefont {Z.-D.}\ \bibnamefont
			{Cheng}}, \bibinfo {author} {\bibfnamefont {Y.-C.}\ \bibnamefont {Wu}},
		\bibinfo {author} {\bibfnamefont {J.-S.}\ \bibnamefont {Xu}}, \bibinfo
		{author} {\bibfnamefont {C.-F.}\ \bibnamefont {Li}},\ and\ \bibinfo {author}
		{\bibfnamefont {G.-C.}\ \bibnamefont {Guo}},\ }\bibfield  {title} {\bibinfo
		{title} {Experimentally robust self-testing for bipartite and tripartite
			entangled states},\ }\href {https://doi.org/10.1103/PhysRevLett.121.240402}
	{\bibfield  {journal} {\bibinfo  {journal} {Phys. Rev. Lett.}\ }\textbf
		{\bibinfo {volume} {121}},\ \bibinfo {pages} {240402} (\bibinfo {year}
		{2018})}\BibitemShut {NoStop}%
	\bibitem [{\citenamefont {Wu}\ \emph {et~al.}(2021)\citenamefont {Wu},
		\citenamefont {Zhao}, \citenamefont {Gu}, \citenamefont {Zhong},
		\citenamefont {Zhou}, \citenamefont {Peng}, \citenamefont {Qin},
		\citenamefont {Luo}, \citenamefont {Chen}, \citenamefont {Li}, \citenamefont
		{Liu}, \citenamefont {Lu},\ and\ \citenamefont {Pan}}]{Dian2021}%
	\BibitemOpen
	\bibfield  {author} {\bibinfo {author} {\bibfnamefont {D.}~\bibnamefont
			{Wu}}, \bibinfo {author} {\bibfnamefont {Q.}~\bibnamefont {Zhao}}, \bibinfo
		{author} {\bibfnamefont {X.-M.}\ \bibnamefont {Gu}}, \bibinfo {author}
		{\bibfnamefont {H.-S.}\ \bibnamefont {Zhong}}, \bibinfo {author}
		{\bibfnamefont {Y.}~\bibnamefont {Zhou}}, \bibinfo {author} {\bibfnamefont
			{L.-C.}\ \bibnamefont {Peng}}, \bibinfo {author} {\bibfnamefont
			{J.}~\bibnamefont {Qin}}, \bibinfo {author} {\bibfnamefont {Y.-H.}\
			\bibnamefont {Luo}}, \bibinfo {author} {\bibfnamefont {K.}~\bibnamefont
			{Chen}}, \bibinfo {author} {\bibfnamefont {L.}~\bibnamefont {Li}}, \bibinfo
		{author} {\bibfnamefont {N.-L.}\ \bibnamefont {Liu}}, \bibinfo {author}
		{\bibfnamefont {C.-Y.}\ \bibnamefont {Lu}},\ and\ \bibinfo {author}
		{\bibfnamefont {J.-W.}\ \bibnamefont {Pan}},\ }\bibfield  {title} {\bibinfo
		{title} {Robust self-testing of multiparticle entanglement},\ }\href
	{https://doi.org/10.1103/PhysRevLett.127.230503} {\bibfield  {journal}
		{\bibinfo  {journal} {Phys. Rev. Lett.}\ }\textbf {\bibinfo {volume} {127}},\
		\bibinfo {pages} {230503} (\bibinfo {year} {2021})}\BibitemShut {NoStop}%
	\bibitem [{\citenamefont {Agresti}\ \emph {et~al.}(2021)\citenamefont
		{Agresti}, \citenamefont {Polacchi}, \citenamefont {Poderini}, \citenamefont
		{Polino}, \citenamefont {Suprano}, \citenamefont {\ifmmode \check{S}\else
			\v{S}\fi{}upi\ifmmode~\acute{c}\else \'{c}\fi{}}, \citenamefont {Bowles},
		\citenamefont {Carvacho}, \citenamefont {Cavalcanti},\ and\ \citenamefont
		{Sciarrino}}]{Agresti2021}%
	\BibitemOpen
	\bibfield  {author} {\bibinfo {author} {\bibfnamefont {I.}~\bibnamefont
			{Agresti}}, \bibinfo {author} {\bibfnamefont {B.}~\bibnamefont {Polacchi}},
		\bibinfo {author} {\bibfnamefont {D.}~\bibnamefont {Poderini}}, \bibinfo
		{author} {\bibfnamefont {E.}~\bibnamefont {Polino}}, \bibinfo {author}
		{\bibfnamefont {A.}~\bibnamefont {Suprano}}, \bibinfo {author} {\bibfnamefont
			{I.}~\bibnamefont {\ifmmode \check{S}\else
				\v{S}\fi{}upi\ifmmode~\acute{c}\else \'{c}\fi{}}}, \bibinfo {author}
		{\bibfnamefont {J.}~\bibnamefont {Bowles}}, \bibinfo {author} {\bibfnamefont
			{G.}~\bibnamefont {Carvacho}}, \bibinfo {author} {\bibfnamefont
			{D.}~\bibnamefont {Cavalcanti}},\ and\ \bibinfo {author} {\bibfnamefont
			{F.}~\bibnamefont {Sciarrino}},\ }\bibfield  {title} {\bibinfo {title}
		{Experimental robust self-testing of the state generated by a quantum
			network},\ }\href {https://doi.org/10.1103/PRXQuantum.2.020346} {\bibfield
		{journal} {\bibinfo  {journal} {PRX Quantum}\ }\textbf {\bibinfo {volume}
			{2}},\ \bibinfo {pages} {020346} (\bibinfo {year} {2021})}\BibitemShut
	{NoStop}%
	\bibitem [{\citenamefont {McKague}(2016)}]{McKague2016a}%
	\BibitemOpen
	\bibfield  {author} {\bibinfo {author} {\bibfnamefont {M.}~\bibnamefont
			{McKague}},\ }\bibfield  {title} {\bibinfo {title} {Interactive proofs for
			$\mathsf{BQP}$ via self-tested graph states},\ }\href
	{https://doi.org/10.4086/toc.2016.v012a003} {\bibfield  {journal} {\bibinfo
			{journal} {Theory of Computing}\ }\textbf {\bibinfo {volume} {12}},\ \bibinfo
		{pages} {1} (\bibinfo {year} {2016})}\BibitemShut {NoStop}%
	\bibitem [{\citenamefont {Tavakoli}\ \emph {et~al.}(2018)\citenamefont
		{Tavakoli}, \citenamefont {Kaniewski}, \citenamefont {V\'ertesi},
		\citenamefont {Rosset},\ and\ \citenamefont {Brunner}}]{Tavakoli2018}%
	\BibitemOpen
	\bibfield  {author} {\bibinfo {author} {\bibfnamefont {A.}~\bibnamefont
			{Tavakoli}}, \bibinfo {author} {\bibfnamefont {J.}~\bibnamefont {Kaniewski}},
		\bibinfo {author} {\bibfnamefont {T.}~\bibnamefont {V\'ertesi}}, \bibinfo
		{author} {\bibfnamefont {D.}~\bibnamefont {Rosset}},\ and\ \bibinfo {author}
		{\bibfnamefont {N.}~\bibnamefont {Brunner}},\ }\bibfield  {title} {\bibinfo
		{title} {Self-testing quantum states and measurements in the
			prepare-and-measure scenario},\ }\href
	{https://doi.org/10.1103/PhysRevA.98.062307} {\bibfield  {journal} {\bibinfo
			{journal} {Phys. Rev. A}\ }\textbf {\bibinfo {volume} {98}},\ \bibinfo
		{pages} {062307} (\bibinfo {year} {2018})}\BibitemShut {NoStop}%
	\bibitem [{\citenamefont {Farkas}\ and\ \citenamefont
		{Kaniewski}(2019)}]{Farkas2019}%
	\BibitemOpen
	\bibfield  {author} {\bibinfo {author} {\bibfnamefont {M.}~\bibnamefont
			{Farkas}}\ and\ \bibinfo {author} {\bibfnamefont {J.~m.~k.}\ \bibnamefont
			{Kaniewski}},\ }\bibfield  {title} {\bibinfo {title} {Self-testing mutually
			unbiased bases in the prepare-and-measure scenario},\ }\href
	{https://doi.org/10.1103/PhysRevA.99.032316} {\bibfield  {journal} {\bibinfo
			{journal} {Phys. Rev. A}\ }\textbf {\bibinfo {volume} {99}},\ \bibinfo
		{pages} {032316} (\bibinfo {year} {2019})}\BibitemShut {NoStop}%
	\bibitem [{\citenamefont {Mohan}\ \emph {et~al.}(2019)\citenamefont {Mohan},
		\citenamefont {Tavakoli},\ and\ \citenamefont {Brunner}}]{Mohan2019}%
	\BibitemOpen
	\bibfield  {author} {\bibinfo {author} {\bibfnamefont {K.}~\bibnamefont
			{Mohan}}, \bibinfo {author} {\bibfnamefont {A.}~\bibnamefont {Tavakoli}},\
		and\ \bibinfo {author} {\bibfnamefont {N.}~\bibnamefont {Brunner}},\
	}\bibfield  {title} {\bibinfo {title} {Sequential random access codes and
			self-testing of quantum measurement instruments},\ }\href
	{https://doi.org/10.1088/1367-2630/ab3773} {\bibfield  {journal} {\bibinfo
			{journal} {New Journal of Physics}\ }\textbf {\bibinfo {volume} {21}},\
		\bibinfo {pages} {083034} (\bibinfo {year} {2019})}\BibitemShut {NoStop}%
	\bibitem [{\citenamefont {Uola}\ \emph {et~al.}(2020)\citenamefont {Uola},
		\citenamefont {Costa}, \citenamefont {Nguyen},\ and\ \citenamefont
		{G\"uhne}}]{Uola2020}%
	\BibitemOpen
	\bibfield  {author} {\bibinfo {author} {\bibfnamefont {R.}~\bibnamefont
			{Uola}}, \bibinfo {author} {\bibfnamefont {A.~C.~S.}\ \bibnamefont {Costa}},
		\bibinfo {author} {\bibfnamefont {H.~C.}\ \bibnamefont {Nguyen}},\ and\
		\bibinfo {author} {\bibfnamefont {O.}~\bibnamefont {G\"uhne}},\ }\bibfield
	{title} {\bibinfo {title} {Quantum steering},\ }\href
	{https://doi.org/10.1103/RevModPhys.92.015001} {\bibfield  {journal}
		{\bibinfo  {journal} {Rev. Mod. Phys.}\ }\textbf {\bibinfo {volume} {92}},\
		\bibinfo {pages} {015001} (\bibinfo {year} {2020})}\BibitemShut {NoStop}%
	\bibitem [{\citenamefont {\ifmmode \check{S}\else
			\v{S}\fi{}upi\ifmmode~\acute{c}\else \'{c}\fi{}}\ and\ \citenamefont
		{Matty}(2016)}]{supic2016}%
	\BibitemOpen
	\bibfield  {author} {\bibinfo {author} {\bibfnamefont {I.}~\bibnamefont
			{\ifmmode \check{S}\else \v{S}\fi{}upi\ifmmode~\acute{c}\else \'{c}\fi{}}}\
		and\ \bibinfo {author} {\bibfnamefont {J.~H.}\ \bibnamefont {Matty}},\
	}\bibfield  {title} {\bibinfo {title} {Self-testing through epr-steering},\
	}\href {https://doi.org/10.1088/1367-2630/18/7/075006} {\bibfield  {journal}
		{\bibinfo  {journal} {New J. Phys.}\ }\textbf {\bibinfo {volume} {18}}
		(\bibinfo {year} {2016})}\BibitemShut {NoStop}%
	\bibitem [{\citenamefont {Cavalcanti}\ and\ \citenamefont
		{Skrzypczyk}(2016)}]{Cavalcanti_2017}%
	\BibitemOpen
	\bibfield  {author} {\bibinfo {author} {\bibfnamefont {D.}~\bibnamefont
			{Cavalcanti}}\ and\ \bibinfo {author} {\bibfnamefont {P.}~\bibnamefont
			{Skrzypczyk}},\ }\bibfield  {title} {\bibinfo {title} {Quantum steering: a
			review with focus on semidefinite programming},\ }\href
	{https://doi.org/10.1088/1361-6633/80/2/024001} {\bibfield  {journal}
		{\bibinfo  {journal} {Reports on Progress in Physics}\ }\textbf {\bibinfo
			{volume} {80}},\ \bibinfo {pages} {024001} (\bibinfo {year}
		{2016})}\BibitemShut {NoStop}%
	\bibitem [{\citenamefont {Šupić}\ \emph {et~al.}(2020)\citenamefont
		{Šupić}, \citenamefont {Hoban}, \citenamefont {Colomer},\ and\
		\citenamefont {Acín}}]{Šupić_2020}%
	\BibitemOpen
	\bibfield  {author} {\bibinfo {author} {\bibfnamefont {I.}~\bibnamefont
			{Šupić}}, \bibinfo {author} {\bibfnamefont {M.~J.}\ \bibnamefont {Hoban}},
		\bibinfo {author} {\bibfnamefont {L.~D.}\ \bibnamefont {Colomer}},\ and\
		\bibinfo {author} {\bibfnamefont {A.}~\bibnamefont {Acín}},\ }\bibfield
	{title} {\bibinfo {title} {Self-testing and certification using trusted
			quantum inputs},\ }\href {https://doi.org/10.1088/1367-2630/ab90d1}
	{\bibfield  {journal} {\bibinfo  {journal} {New Journal of Physics}\ }\textbf
		{\bibinfo {volume} {22}},\ \bibinfo {pages} {073006} (\bibinfo {year}
		{2020})}\BibitemShut {NoStop}%
	\bibitem [{\citenamefont {Mukherjee}\ and\ \citenamefont
		{Pan}(2021)}]{Sumit2021}%
	\BibitemOpen
	\bibfield  {author} {\bibinfo {author} {\bibfnamefont {S.}~\bibnamefont
			{Mukherjee}}\ and\ \bibinfo {author} {\bibfnamefont {A.~K.}\ \bibnamefont
			{Pan}},\ }\bibfield  {title} {\bibinfo {title} {Semi-device-independent
			certification of multiple unsharpness parameters through sequential
			measurements},\ }\href {https://doi.org/10.1103/PhysRevA.104.062214}
	{\bibfield  {journal} {\bibinfo  {journal} {Phys. Rev. A}\ }\textbf {\bibinfo
			{volume} {104}},\ \bibinfo {pages} {062214} (\bibinfo {year}
		{2021})}\BibitemShut {NoStop}%
	\bibitem [{\citenamefont {Pan}(2021)}]{Pan2021}%
	\BibitemOpen
	\bibfield  {author} {\bibinfo {author} {\bibfnamefont {A.~K.}\ \bibnamefont
			{Pan}},\ }\bibfield  {title} {\bibinfo {title} {Oblivious communication game,
			self-testing of projective and nonprojective measurements, and certification
			of randomness},\ }\href {https://doi.org/10.1103/PhysRevA.104.022212}
	{\bibfield  {journal} {\bibinfo  {journal} {Phys. Rev. A}\ }\textbf {\bibinfo
			{volume} {104}},\ \bibinfo {pages} {022212} (\bibinfo {year}
		{2021})}\BibitemShut {NoStop}%
	\bibitem [{\citenamefont {Svetlichny}(1987)}]{Svetlichny1987}%
	\BibitemOpen
	\bibfield  {author} {\bibinfo {author} {\bibfnamefont {G.}~\bibnamefont
			{Svetlichny}},\ }\bibfield  {title} {\bibinfo {title} {Distinguishing
			three-body from two-body nonseparability by a bell-type inequality},\ }\href
	{https://doi.org/10.1103/PhysRevD.35.3066} {\bibfield  {journal} {\bibinfo
			{journal} {Phys. Rev. D}\ }\textbf {\bibinfo {volume} {35}},\ \bibinfo
		{pages} {3066} (\bibinfo {year} {1987})}\BibitemShut {NoStop}%
	\bibitem [{\citenamefont {Roy}\ and\ \citenamefont {Singh}(1989)}]{Roy1989}%
	\BibitemOpen
	\bibfield  {author} {\bibinfo {author} {\bibfnamefont {S.}~\bibnamefont
			{Roy}}\ and\ \bibinfo {author} {\bibfnamefont {V.}~\bibnamefont {Singh}},\
	}\bibfield  {title} {\bibinfo {title} {Hidden variable theories without
			non-local signalling and their experimental tests},\ }\href
	{https://doi.org/https://doi.org/10.1016/0375-9601(89)90940-7} {\bibfield
		{journal} {\bibinfo  {journal} {Physics Letters A}\ }\textbf {\bibinfo
			{volume} {139}},\ \bibinfo {pages} {437} (\bibinfo {year}
		{1989})}\BibitemShut {NoStop}%
	\bibitem [{\citenamefont {Roy}\ and\ \citenamefont {Singh}(1991)}]{Roy1991}%
	\BibitemOpen
	\bibfield  {author} {\bibinfo {author} {\bibfnamefont {S.~M.}\ \bibnamefont
			{Roy}}\ and\ \bibinfo {author} {\bibfnamefont {V.}~\bibnamefont {Singh}},\
	}\bibfield  {title} {\bibinfo {title} {Tests of signal locality and
			einstein-bell locality for multiparticle systems},\ }\href
	{https://doi.org/10.1103/PhysRevLett.67.2761} {\bibfield  {journal} {\bibinfo
			{journal} {Phys. Rev. Lett.}\ }\textbf {\bibinfo {volume} {67}},\ \bibinfo
		{pages} {2761} (\bibinfo {year} {1991})}\BibitemShut {NoStop}%
	\bibitem [{\citenamefont {Mermin}(1990)}]{Mermin1990}%
	\BibitemOpen
	\bibfield  {author} {\bibinfo {author} {\bibfnamefont {N.~D.}\ \bibnamefont
			{Mermin}},\ }\bibfield  {title} {\bibinfo {title} {Extreme quantum
			entanglement in a superposition of macroscopically distinct states},\ }\href
	{https://doi.org/10.1103/PhysRevLett.65.1838} {\bibfield  {journal} {\bibinfo
			{journal} {Phys. Rev. Lett.}\ }\textbf {\bibinfo {volume} {65}},\ \bibinfo
		{pages} {1838} (\bibinfo {year} {1990})}\BibitemShut {NoStop}%
	\bibitem [{\citenamefont {Uffink}(2002)}]{Uffink2002}%
	\BibitemOpen
	\bibfield  {author} {\bibinfo {author} {\bibfnamefont {J.}~\bibnamefont
			{Uffink}},\ }\bibfield  {title} {\bibinfo {title} {Quadratic bell
			inequalities as tests for multipartite entanglement},\ }\href
	{https://doi.org/10.1103/PhysRevLett.88.230406} {\bibfield  {journal}
		{\bibinfo  {journal} {Phys. Rev. Lett.}\ }\textbf {\bibinfo {volume} {88}},\
		\bibinfo {pages} {230406} (\bibinfo {year} {2002})}\BibitemShut {NoStop}%
	\bibitem [{\citenamefont {Seevinck}\ and\ \citenamefont
		{Svetlichny}(2002)}]{Seevinck2002}%
	\BibitemOpen
	\bibfield  {author} {\bibinfo {author} {\bibfnamefont {M.}~\bibnamefont
			{Seevinck}}\ and\ \bibinfo {author} {\bibfnamefont {G.}~\bibnamefont
			{Svetlichny}},\ }\bibfield  {title} {\bibinfo {title} {Bell-type inequalities
			for partial separability in $n$-particle systems and quantum mechanical
			violations},\ }\href {https://doi.org/10.1103/PhysRevLett.89.060401}
	{\bibfield  {journal} {\bibinfo  {journal} {Phys. Rev. Lett.}\ }\textbf
		{\bibinfo {volume} {89}},\ \bibinfo {pages} {060401} (\bibinfo {year}
		{2002})}\BibitemShut {NoStop}%
	\bibitem [{\citenamefont {Collins}\ \emph {et~al.}(2002)\citenamefont
		{Collins}, \citenamefont {Gisin}, \citenamefont {Popescu}, \citenamefont
		{Roberts},\ and\ \citenamefont {Scarani}}]{Collins2002}%
	\BibitemOpen
	\bibfield  {author} {\bibinfo {author} {\bibfnamefont {D.}~\bibnamefont
			{Collins}}, \bibinfo {author} {\bibfnamefont {N.}~\bibnamefont {Gisin}},
		\bibinfo {author} {\bibfnamefont {S.}~\bibnamefont {Popescu}}, \bibinfo
		{author} {\bibfnamefont {D.}~\bibnamefont {Roberts}},\ and\ \bibinfo {author}
		{\bibfnamefont {V.}~\bibnamefont {Scarani}},\ }\bibfield  {title} {\bibinfo
		{title} {Bell-type inequalities to detect true $\mathit{n}$-body
			nonseparability},\ }\href {https://doi.org/10.1103/PhysRevLett.88.170405}
	{\bibfield  {journal} {\bibinfo  {journal} {Phys. Rev. Lett.}\ }\textbf
		{\bibinfo {volume} {88}},\ \bibinfo {pages} {170405} (\bibinfo {year}
		{2002})}\BibitemShut {NoStop}%
	\bibitem [{\citenamefont {Gallego}\ \emph {et~al.}(2012)\citenamefont
		{Gallego}, \citenamefont {W\"urflinger}, \citenamefont {Ac\'{\i}n},\ and\
		\citenamefont {Navascu\'es}}]{Gallego2012}%
	\BibitemOpen
	\bibfield  {author} {\bibinfo {author} {\bibfnamefont {R.}~\bibnamefont
			{Gallego}}, \bibinfo {author} {\bibfnamefont {L.~E.}\ \bibnamefont
			{W\"urflinger}}, \bibinfo {author} {\bibfnamefont {A.}~\bibnamefont
			{Ac\'{\i}n}},\ and\ \bibinfo {author} {\bibfnamefont {M.}~\bibnamefont
			{Navascu\'es}},\ }\bibfield  {title} {\bibinfo {title} {Operational framework
			for nonlocality},\ }\href {https://doi.org/10.1103/PhysRevLett.109.070401}
	{\bibfield  {journal} {\bibinfo  {journal} {Phys. Rev. Lett.}\ }\textbf
		{\bibinfo {volume} {109}},\ \bibinfo {pages} {070401} (\bibinfo {year}
		{2012})}\BibitemShut {NoStop}%
	\bibitem [{\citenamefont {Dutta}\ \emph {et~al.}(2020)\citenamefont {Dutta},
		\citenamefont {Mukherjee},\ and\ \citenamefont {Banik}}]{Sagnik2020}%
	\BibitemOpen
	\bibfield  {author} {\bibinfo {author} {\bibfnamefont {S.}~\bibnamefont
			{Dutta}}, \bibinfo {author} {\bibfnamefont {A.}~\bibnamefont {Mukherjee}},\
		and\ \bibinfo {author} {\bibfnamefont {M.}~\bibnamefont {Banik}},\ }\bibfield
	{title} {\bibinfo {title} {Operational characterization of multipartite
			nonlocal correlations},\ }\href {https://doi.org/10.1103/PhysRevA.102.052218}
	{\bibfield  {journal} {\bibinfo  {journal} {Phys. Rev. A}\ }\textbf {\bibinfo
			{volume} {102}},\ \bibinfo {pages} {052218} (\bibinfo {year}
		{2020})}\BibitemShut {NoStop}%
	\bibitem [{\citenamefont {Bancal}\ \emph {et~al.}(2013)\citenamefont {Bancal},
		\citenamefont {Barrett}, \citenamefont {Gisin},\ and\ \citenamefont
		{Pironio}}]{Bancal2013}%
	\BibitemOpen
	\bibfield  {author} {\bibinfo {author} {\bibfnamefont {J.-D.}\ \bibnamefont
			{Bancal}}, \bibinfo {author} {\bibfnamefont {J.}~\bibnamefont {Barrett}},
		\bibinfo {author} {\bibfnamefont {N.}~\bibnamefont {Gisin}},\ and\ \bibinfo
		{author} {\bibfnamefont {S.}~\bibnamefont {Pironio}},\ }\bibfield  {title}
	{\bibinfo {title} {Definitions of multipartite nonlocality},\ }\href
	{https://doi.org/10.1103/PhysRevA.88.014102} {\bibfield  {journal} {\bibinfo
			{journal} {Phys. Rev. A}\ }\textbf {\bibinfo {volume} {88}},\ \bibinfo
		{pages} {014102} (\bibinfo {year} {2013})}\BibitemShut {NoStop}%
	\bibitem [{\citenamefont {Wu}\ \emph {et~al.}(2014)\citenamefont {Wu},
		\citenamefont {Cai}, \citenamefont {Yang}, \citenamefont {Le}, \citenamefont
		{Bancal},\ and\ \citenamefont {Scarani}}]{wu2014}%
	\BibitemOpen
	\bibfield  {author} {\bibinfo {author} {\bibfnamefont {X.}~\bibnamefont
			{Wu}}, \bibinfo {author} {\bibfnamefont {Y.}~\bibnamefont {Cai}}, \bibinfo
		{author} {\bibfnamefont {T.~H.}\ \bibnamefont {Yang}}, \bibinfo {author}
		{\bibfnamefont {H.~N.}\ \bibnamefont {Le}}, \bibinfo {author} {\bibfnamefont
			{J.-D.}\ \bibnamefont {Bancal}},\ and\ \bibinfo {author} {\bibfnamefont
			{V.}~\bibnamefont {Scarani}},\ }\bibfield  {title} {\bibinfo {title} {Robust
			self-testing of the three-qubit {$W$} state},\ }\href
	{https://doi.org/10.1103/PhysRevA.90.042339} {\bibfield  {journal} {\bibinfo
			{journal} {Phys. Rev. A}\ }\textbf {\bibinfo {volume} {90}},\ \bibinfo
		{pages} {042339} (\bibinfo {year} {2014})}\BibitemShut {NoStop}%
	\bibitem [{\citenamefont {\ifmmode \check{S}\else
			\v{S}\fi{}upi\ifmmode~\acute{c}\else \'{c}\fi{}}\ \emph
		{et~al.}(2018)\citenamefont {\ifmmode \check{S}\else
			\v{S}\fi{}upi\ifmmode~\acute{c}\else \'{c}\fi{}}, \citenamefont
		{Coladangelo}, \citenamefont {Augusiak},\ and\ \citenamefont
		{Acín}}]{Supic2018}%
	\BibitemOpen
	\bibfield  {author} {\bibinfo {author} {\bibfnamefont {I.}~\bibnamefont
			{\ifmmode \check{S}\else \v{S}\fi{}upi\ifmmode~\acute{c}\else \'{c}\fi{}}},
		\bibinfo {author} {\bibfnamefont {A.}~\bibnamefont {Coladangelo}}, \bibinfo
		{author} {\bibfnamefont {R.}~\bibnamefont {Augusiak}},\ and\ \bibinfo
		{author} {\bibfnamefont {A.}~\bibnamefont {Acín}},\ }\bibfield  {title}
	{\bibinfo {title} {Self-testing multipartite entangled states through
			projections onto two systems},\ }\href
	{https://doi.org/10.1088/1367-2630/aad89b} {\bibfield  {journal} {\bibinfo
			{journal} {New Journal of Physics}\ }\textbf {\bibinfo {volume} {20}},\
		\bibinfo {pages} {083041} (\bibinfo {year} {2018})}\BibitemShut {NoStop}%
	\bibitem [{\citenamefont {Sarkar}\ and\ \citenamefont
		{Augusiak}(2022)}]{Sarkar2022}%
	\BibitemOpen
	\bibfield  {author} {\bibinfo {author} {\bibfnamefont {S.}~\bibnamefont
			{Sarkar}}\ and\ \bibinfo {author} {\bibfnamefont {R.}~\bibnamefont
			{Augusiak}},\ }\bibfield  {title} {\bibinfo {title} {Self-testing of
			multipartite greenberger-horne-zeilinger states of arbitrary local dimension
			with arbitrary number of measurements per party},\ }\href
	{https://doi.org/10.1103/PhysRevA.105.032416} {\bibfield  {journal} {\bibinfo
			{journal} {Phys. Rev. A}\ }\textbf {\bibinfo {volume} {105}},\ \bibinfo
		{pages} {032416} (\bibinfo {year} {2022})}\BibitemShut {NoStop}%
	\bibitem [{\citenamefont {Augusiak}\ \emph {et~al.}(2019)\citenamefont
		{Augusiak}, \citenamefont {Salavrakos}, \citenamefont {Tura},\ and\
		\citenamefont {Acín}}]{Augusiak2019}%
	\BibitemOpen
	\bibfield  {author} {\bibinfo {author} {\bibfnamefont {R.}~\bibnamefont
			{Augusiak}}, \bibinfo {author} {\bibfnamefont {A.}~\bibnamefont
			{Salavrakos}}, \bibinfo {author} {\bibfnamefont {J.}~\bibnamefont {Tura}},\
		and\ \bibinfo {author} {\bibfnamefont {A.}~\bibnamefont {Acín}},\ }\bibfield
	{title} {\bibinfo {title} {Bell inequalities tailored to the
			greenberger–horne–zeilinger states of arbitrary local dimension},\ }\href
	{https://doi.org/10.1088/1367-2630/ab4d9f} {\bibfield  {journal} {\bibinfo
			{journal} {New Journal of Physics}\ }\textbf {\bibinfo {volume} {21}},\
		\bibinfo {pages} {113001} (\bibinfo {year} {2019})}\BibitemShut {NoStop}%
	\bibitem [{\citenamefont {Panwar}\ \emph {et~al.}(2023)\citenamefont {Panwar},
		\citenamefont {Pandya},\ and\ \citenamefont {Wie{\'{s}}niak}}]{Panwar2023}%
	\BibitemOpen
	\bibfield  {author} {\bibinfo {author} {\bibfnamefont {E.}~\bibnamefont
			{Panwar}}, \bibinfo {author} {\bibfnamefont {P.}~\bibnamefont {Pandya}},\
		and\ \bibinfo {author} {\bibfnamefont {M.}~\bibnamefont {Wie{\'{s}}niak}},\
	}\bibfield  {title} {\bibinfo {title} {An elegant scheme of self-testing for
			multipartite bell inequalities},\ }\href
	{https://doi.org/10.1038/s41534-023-00735-3} {\bibfield  {journal} {\bibinfo
			{journal} {npj Quantum Information}\ }\textbf {\bibinfo {volume} {9}},\
		\bibinfo {pages} {71} (\bibinfo {year} {2023})}\BibitemShut {NoStop}%
	\bibitem [{\citenamefont {Werner}\ and\ \citenamefont
		{Wolf}(2001)}]{Werenr2001}%
	\BibitemOpen
	\bibfield  {author} {\bibinfo {author} {\bibfnamefont {R.~F.}\ \bibnamefont
			{Werner}}\ and\ \bibinfo {author} {\bibfnamefont {M.~M.}\ \bibnamefont
			{Wolf}},\ }\bibfield  {title} {\bibinfo {title} {All-multipartite
			bell-correlation inequalities for two dichotomic observables per site},\
	}\href {https://doi.org/10.1103/PhysRevA.64.032112} {\bibfield  {journal}
		{\bibinfo  {journal} {Phys. Rev. A}\ }\textbf {\bibinfo {volume} {64}},\
		\bibinfo {pages} {032112} (\bibinfo {year} {2001})}\BibitemShut {NoStop}%
	\bibitem [{\citenamefont {Weinfurter}\ and\ \citenamefont
		{\ifmmode~\dot{Z}\else \.{Z}\fi{}ukowski}(2001)}]{Weinfurter2001}%
	\BibitemOpen
	\bibfield  {author} {\bibinfo {author} {\bibfnamefont {H.}~\bibnamefont
			{Weinfurter}}\ and\ \bibinfo {author} {\bibfnamefont {M.}~\bibnamefont
			{\ifmmode~\dot{Z}\else \.{Z}\fi{}ukowski}},\ }\bibfield  {title} {\bibinfo
		{title} {Four-photon entanglement from down-conversion},\ }\href
	{https://doi.org/10.1103/PhysRevA.64.010102} {\bibfield  {journal} {\bibinfo
			{journal} {Phys. Rev. A}\ }\textbf {\bibinfo {volume} {64}},\ \bibinfo
		{pages} {010102} (\bibinfo {year} {2001})}\BibitemShut {NoStop}%
	\bibitem [{\citenamefont {\ifmmode~\dot{Z}\else \.{Z}\fi{}ukowski}\ and\
		\citenamefont {Brukner}(2002)}]{Zukowski2002}%
	\BibitemOpen
	\bibfield  {author} {\bibinfo {author} {\bibfnamefont {M.}~\bibnamefont
			{\ifmmode~\dot{Z}\else \.{Z}\fi{}ukowski}}\ and\ \bibinfo {author}
		{\bibfnamefont {i.~c.~v.}\ \bibnamefont {Brukner}},\ }\bibfield  {title}
	{\bibinfo {title} {Bell's theorem for general n-qubit states},\ }\href
	{https://doi.org/10.1103/PhysRevLett.88.210401} {\bibfield  {journal}
		{\bibinfo  {journal} {Phys. Rev. Lett.}\ }\textbf {\bibinfo {volume} {88}},\
		\bibinfo {pages} {210401} (\bibinfo {year} {2002})}\BibitemShut {NoStop}%
	\bibitem [{\citenamefont {Masanes}\ \emph {et~al.}(2006)\citenamefont
		{Masanes}, \citenamefont {Ac\'{i}n},\ and\ \citenamefont
		{Gisin}}]{Masanes2006}%
	\BibitemOpen
	\bibfield  {author} {\bibinfo {author} {\bibfnamefont {L.}~\bibnamefont
			{Masanes}}, \bibinfo {author} {\bibfnamefont {A.}~\bibnamefont {Ac\'{i}n}},\
		and\ \bibinfo {author} {\bibfnamefont {N.}~\bibnamefont {Gisin}},\ }\bibfield
	{title} {\bibinfo {title} {General properties of nonsignaling theories},\
	}\href {https://doi.org/10.1103/PhysRevA.73.012112} {\bibfield  {journal}
		{\bibinfo  {journal} {Phys. Rev. A}\ }\textbf {\bibinfo {volume} {73}},\
		\bibinfo {pages} {012112} (\bibinfo {year} {2006})}\BibitemShut {NoStop}%
	\bibitem [{\citenamefont {Cavalcanti}\ and\ \citenamefont
		{Wiseman}(2012)}]{Cavalcanti2012}%
	\BibitemOpen
	\bibfield  {author} {\bibinfo {author} {\bibfnamefont {E.~G.}\ \bibnamefont
			{Cavalcanti}}\ and\ \bibinfo {author} {\bibfnamefont {H.~M.}\ \bibnamefont
			{Wiseman}},\ }\bibfield  {title} {\bibinfo {title} {Bell nonlocality, signal
			locality and unpredictability (or what bohr could have told einstein at
			solvay had he known about bell experiments)},\ }\href
	{https://doi.org/10.1007/s10701-012-9669-1} {\bibfield  {journal} {\bibinfo
			{journal} {Foundations of Physics}\ }\textbf {\bibinfo {volume} {42}},\
		\bibinfo {pages} {1329} (\bibinfo {year} {2012})}\BibitemShut {NoStop}%
	\bibitem [{\citenamefont {Sasmal}\ \emph {et~al.}(2024)\citenamefont {Sasmal},
		\citenamefont {Rai}, \citenamefont {Gangopadhyay}, \citenamefont {Home},\
		and\ \citenamefont {Sinha}}]{Sasmal2024}%
	\BibitemOpen
	\bibfield  {author} {\bibinfo {author} {\bibfnamefont {S.}~\bibnamefont
			{Sasmal}}, \bibinfo {author} {\bibfnamefont {A.}~\bibnamefont {Rai}},
		\bibinfo {author} {\bibfnamefont {S.}~\bibnamefont {Gangopadhyay}}, \bibinfo
		{author} {\bibfnamefont {D.}~\bibnamefont {Home}},\ and\ \bibinfo {author}
		{\bibfnamefont {U.}~\bibnamefont {Sinha}},\ }\bibfield  {title} {\bibinfo
		{title} {Revealing incommensurability between device-independent randomness,
			nonlocality, and entanglement using hardy and hardy-type relations},\ }\href
	{https://doi.org/10.1088/1402-4896/ad24a5} {\bibfield  {journal} {\bibinfo
			{journal} {Physica Scripta}\ }\textbf {\bibinfo {volume} {99}},\ \bibinfo
		{pages} {035012} (\bibinfo {year} {2024})}\BibitemShut {NoStop}%
	\bibitem [{\citenamefont {Colbeck}\ and\ \citenamefont
		{Kent}(2011)}]{Colbeck2011}%
	\BibitemOpen
	\bibfield  {author} {\bibinfo {author} {\bibfnamefont {R.}~\bibnamefont
			{Colbeck}}\ and\ \bibinfo {author} {\bibfnamefont {A.}~\bibnamefont {Kent}},\
	}\bibfield  {title} {\bibinfo {title} {Private randomness expansion with
			untrusted devices},\ }\href {https://doi.org/10.1088/1751-8113/44/9/095305}
	{\bibfield  {journal} {\bibinfo  {journal} {Journal of Physics A:
				Mathematical and Theoretical}\ }\textbf {\bibinfo {volume} {44}},\ \bibinfo
		{pages} {095305} (\bibinfo {year} {2011})}\BibitemShut {NoStop}%
	\bibitem [{\citenamefont {Ac\ifmmode~\acute{i}\else \'{i}\fi{}n}\ \emph
		{et~al.}(2012)\citenamefont {Ac\ifmmode~\acute{i}\else \'{i}\fi{}n},
		\citenamefont {Massar},\ and\ \citenamefont {Pironio}}]{Acin2012}%
	\BibitemOpen
	\bibfield  {author} {\bibinfo {author} {\bibfnamefont {A.}~\bibnamefont
			{Ac\ifmmode~\acute{i}\else \'{i}\fi{}n}}, \bibinfo {author} {\bibfnamefont
			{S.}~\bibnamefont {Massar}},\ and\ \bibinfo {author} {\bibfnamefont
			{S.}~\bibnamefont {Pironio}},\ }\bibfield  {title} {\bibinfo {title}
		{Randomness versus nonlocality and entanglement},\ }\href
	{https://doi.org/10.1103/PhysRevLett.108.100402} {\bibfield  {journal}
		{\bibinfo  {journal} {Phys. Rev. Lett.}\ }\textbf {\bibinfo {volume} {108}},\
		\bibinfo {pages} {100402} (\bibinfo {year} {2012})}\BibitemShut {NoStop}%
	\bibitem [{\citenamefont {Konig}\ \emph {et~al.}(2009)\citenamefont {Konig},
		\citenamefont {Renner},\ and\ \citenamefont {Schaffner}}]{Konig2008}%
	\BibitemOpen
	\bibfield  {author} {\bibinfo {author} {\bibfnamefont {R.}~\bibnamefont
			{Konig}}, \bibinfo {author} {\bibfnamefont {R.}~\bibnamefont {Renner}},\ and\
		\bibinfo {author} {\bibfnamefont {C.}~\bibnamefont {Schaffner}},\ }\bibfield
	{title} {\bibinfo {title} {The operational meaning of min- and max-entropy},\
	}\href {https://doi.org/10.1109/TIT.2009.2025545} {\bibfield  {journal}
		{\bibinfo  {journal} {IEEE Transactions on Information Theory}\ }\textbf
		{\bibinfo {volume} {55}},\ \bibinfo {pages} {4337} (\bibinfo {year}
		{2009})}\BibitemShut {NoStop}%
	\bibitem [{\citenamefont {Scarani}(2019)}]{Scarani2019book}%
	\BibitemOpen
	\bibfield  {author} {\bibinfo {author} {\bibfnamefont {V.}~\bibnamefont
			{Scarani}},\ }\href
	{https://fdslive.oup.com/www.oup.com/academic/pdf/openaccess/9780198788416.pdf}
	{\emph {\bibinfo {title} {Bell Nonlocality}}},\ Oxford Graduate Texts\
	(\bibinfo  {publisher} {Oxford University Press},\ \bibinfo {year}
	{2019})\BibitemShut {NoStop}%
	\bibitem [{\citenamefont {Navascu\'es}\ \emph {et~al.}(2007)\citenamefont
		{Navascu\'es}, \citenamefont {Pironio},\ and\ \citenamefont
		{Ac\'{\i}n}}]{Navascues2007}%
	\BibitemOpen
	\bibfield  {author} {\bibinfo {author} {\bibfnamefont {M.}~\bibnamefont
			{Navascu\'es}}, \bibinfo {author} {\bibfnamefont {S.}~\bibnamefont
			{Pironio}},\ and\ \bibinfo {author} {\bibfnamefont {A.}~\bibnamefont
			{Ac\'{\i}n}},\ }\bibfield  {title} {\bibinfo {title} {Bounding the set of
			quantum correlations},\ }\href
	{https://doi.org/10.1103/PhysRevLett.98.010401} {\bibfield  {journal}
		{\bibinfo  {journal} {Phys. Rev. Lett.}\ }\textbf {\bibinfo {volume} {98}},\
		\bibinfo {pages} {010401} (\bibinfo {year} {2007})}\BibitemShut {NoStop}%
	\bibitem [{\citenamefont {Navascués}\ \emph {et~al.}(2008)\citenamefont
		{Navascués}, \citenamefont {Pironio},\ and\ \citenamefont
		{Acín}}]{Navascues2008}%
	\BibitemOpen
	\bibfield  {author} {\bibinfo {author} {\bibfnamefont {M.}~\bibnamefont
			{Navascués}}, \bibinfo {author} {\bibfnamefont {S.}~\bibnamefont
			{Pironio}},\ and\ \bibinfo {author} {\bibfnamefont {A.}~\bibnamefont
			{Acín}},\ }\bibfield  {title} {\bibinfo {title} {A convergent hierarchy of
			semidefinite programs characterizing the set of quantum correlations},\
	}\href {https://doi.org/10.1088/1367-2630/10/7/073013} {\bibfield  {journal}
		{\bibinfo  {journal} {New Journal of Physics}\ }\textbf {\bibinfo {volume}
			{10}},\ \bibinfo {pages} {073013} (\bibinfo {year} {2008})}\BibitemShut
	{NoStop}%
	\bibitem [{\citenamefont {Nieto-Silleras}\ \emph {et~al.}(2014)\citenamefont
		{Nieto-Silleras}, \citenamefont {Pironio},\ and\ \citenamefont
		{Silman}}]{Silleras2014}%
	\BibitemOpen
	\bibfield  {author} {\bibinfo {author} {\bibfnamefont {O.}~\bibnamefont
			{Nieto-Silleras}}, \bibinfo {author} {\bibfnamefont {S.}~\bibnamefont
			{Pironio}},\ and\ \bibinfo {author} {\bibfnamefont {J.}~\bibnamefont
			{Silman}},\ }\bibfield  {title} {\bibinfo {title} {Using complete measurement
			statistics for optimal device-independent randomness evaluation},\ }\href
	{https://doi.org/10.1088/1367-2630/16/1/013035} {\bibfield  {journal}
		{\bibinfo  {journal} {New Journal of Physics}\ }\textbf {\bibinfo {volume}
			{16}},\ \bibinfo {pages} {013035} (\bibinfo {year} {2014})}\BibitemShut
	{NoStop}%
	\bibitem [{\citenamefont {Bancal}\ \emph {et~al.}(2014)\citenamefont {Bancal},
		\citenamefont {Sheridan},\ and\ \citenamefont {Scarani}}]{Bancal2014}%
	\BibitemOpen
	\bibfield  {author} {\bibinfo {author} {\bibfnamefont {J.-D.}\ \bibnamefont
			{Bancal}}, \bibinfo {author} {\bibfnamefont {L.}~\bibnamefont {Sheridan}},\
		and\ \bibinfo {author} {\bibfnamefont {V.}~\bibnamefont {Scarani}},\
	}\bibfield  {title} {\bibinfo {title} {More randomness from the same data},\
	}\href {https://doi.org/10.1088/1367-2630/16/3/033011} {\bibfield  {journal}
		{\bibinfo  {journal} {New Journal of Physics}\ }\textbf {\bibinfo {volume}
			{16}},\ \bibinfo {pages} {033011} (\bibinfo {year} {2014})}\BibitemShut
	{NoStop}%
	\bibitem [{\citenamefont {Nieto-Silleras}\ \emph {et~al.}(2018)\citenamefont
		{Nieto-Silleras}, \citenamefont {Bamps}, \citenamefont {Silman},\ and\
		\citenamefont {Pironio}}]{Silleras2018}%
	\BibitemOpen
	\bibfield  {author} {\bibinfo {author} {\bibfnamefont {O.}~\bibnamefont
			{Nieto-Silleras}}, \bibinfo {author} {\bibfnamefont {C.}~\bibnamefont
			{Bamps}}, \bibinfo {author} {\bibfnamefont {J.}~\bibnamefont {Silman}},\ and\
		\bibinfo {author} {\bibfnamefont {S.}~\bibnamefont {Pironio}},\ }\bibfield
	{title} {\bibinfo {title} {Device-independent randomness generation from
			several bell estimators},\ }\href {https://doi.org/10.1088/1367-2630/aaaa06}
	{\bibfield  {journal} {\bibinfo  {journal} {New Journal of Physics}\ }\textbf
		{\bibinfo {volume} {20}},\ \bibinfo {pages} {023049} (\bibinfo {year}
		{2018})}\BibitemShut {NoStop}%
	\bibitem [{\citenamefont {Andersson}\ \emph {et~al.}(2018)\citenamefont
		{Andersson}, \citenamefont {Badziag}, \citenamefont {Dumitru},\ and\
		\citenamefont {Cabello}}]{Andersson2018}%
	\BibitemOpen
	\bibfield  {author} {\bibinfo {author} {\bibfnamefont {O.}~\bibnamefont
			{Andersson}}, \bibinfo {author} {\bibfnamefont {P.}~\bibnamefont {Badziag}},
		\bibinfo {author} {\bibfnamefont {I.}~\bibnamefont {Dumitru}},\ and\ \bibinfo
		{author} {\bibfnamefont {A.}~\bibnamefont {Cabello}},\ }\bibfield  {title}
	{\bibinfo {title} {Device-independent certification of two bits of randomness
			from one entangled bit and gisin's elegant bell inequality},\ }\href
	{https://doi.org/10.1103/PhysRevA.97.012314} {\bibfield  {journal} {\bibinfo
			{journal} {Phys. Rev. A}\ }\textbf {\bibinfo {volume} {97}},\ \bibinfo
		{pages} {012314} (\bibinfo {year} {2018})}\BibitemShut {NoStop}%
	\bibitem [{\citenamefont {Woodhead}\ \emph {et~al.}(2020)\citenamefont
		{Woodhead}, \citenamefont {Kaniewski}, \citenamefont {Bourdoncle},
		\citenamefont {Salavrakos}, \citenamefont {Bowles}, \citenamefont
		{Ac\'{\i}n},\ and\ \citenamefont {Augusiak}}]{Woodhead2020}%
	\BibitemOpen
	\bibfield  {author} {\bibinfo {author} {\bibfnamefont {E.}~\bibnamefont
			{Woodhead}}, \bibinfo {author} {\bibfnamefont {J.}~\bibnamefont {Kaniewski}},
		\bibinfo {author} {\bibfnamefont {B.}~\bibnamefont {Bourdoncle}}, \bibinfo
		{author} {\bibfnamefont {A.}~\bibnamefont {Salavrakos}}, \bibinfo {author}
		{\bibfnamefont {J.}~\bibnamefont {Bowles}}, \bibinfo {author} {\bibfnamefont
			{A.}~\bibnamefont {Ac\'{\i}n}},\ and\ \bibinfo {author} {\bibfnamefont
			{R.}~\bibnamefont {Augusiak}},\ }\bibfield  {title} {\bibinfo {title}
		{Maximal randomness from partially entangled states},\ }\href
	{https://doi.org/10.1103/PhysRevResearch.2.042028} {\bibfield  {journal}
		{\bibinfo  {journal} {Phys. Rev. Res.}\ }\textbf {\bibinfo {volume} {2}},\
		\bibinfo {pages} {042028} (\bibinfo {year} {2020})}\BibitemShut {NoStop}%
	\bibitem [{\citenamefont {Borkala}\ \emph {et~al.}(2022)\citenamefont
		{Borkala}, \citenamefont {Jebarathinam}, \citenamefont {Sarkar},\ and\
		\citenamefont {Augusiak}}]{Borkala2022}%
	\BibitemOpen
	\bibfield  {author} {\bibinfo {author} {\bibfnamefont {J.~J.}\ \bibnamefont
			{Borkala}}, \bibinfo {author} {\bibfnamefont {C.}~\bibnamefont
			{Jebarathinam}}, \bibinfo {author} {\bibfnamefont {S.}~\bibnamefont
			{Sarkar}},\ and\ \bibinfo {author} {\bibfnamefont {R.}~\bibnamefont
			{Augusiak}},\ }\bibfield  {title} {\bibinfo {title} {Device-independent
			certification of maximal randomness from pure entangled two-qutrit states
			using non-projective measurements},\ }\bibfield  {journal} {\bibinfo
		{journal} {Entropy}\ }\textbf {\bibinfo {volume} {24}},\ \href
	{https://doi.org/10.3390/e24030350} {10.3390/e24030350} (\bibinfo {year}
	{2022})\BibitemShut {NoStop}%
	\bibitem [{\citenamefont {Mikos-Nuszkiewicz}\ and\ \citenamefont
		{Kaniewski}(2023)}]{Nuszkiewicz2023}%
	\BibitemOpen
	\bibfield  {author} {\bibinfo {author} {\bibfnamefont {A.}~\bibnamefont
			{Mikos-Nuszkiewicz}}\ and\ \bibinfo {author} {\bibfnamefont {J.~m.~k.}\
			\bibnamefont {Kaniewski}},\ }\bibfield  {title} {\bibinfo {title} {Extremal
			points of the quantum set in the clauser-horne-shimony-holt scenario:
			Conjectured analytical solution},\ }\href
	{https://doi.org/10.1103/PhysRevA.108.012212} {\bibfield  {journal} {\bibinfo
			{journal} {Phys. Rev. A}\ }\textbf {\bibinfo {volume} {108}},\ \bibinfo
		{pages} {012212} (\bibinfo {year} {2023})}\BibitemShut {NoStop}%
	\bibitem [{\citenamefont {Le}\ \emph {et~al.}(2023)\citenamefont {Le},
		\citenamefont {Meroni}, \citenamefont {Sturmfels}, \citenamefont {Werner},\
		and\ \citenamefont {Ziegler}}]{Le2023}%
	\BibitemOpen
	\bibfield  {author} {\bibinfo {author} {\bibfnamefont {T.~P.}\ \bibnamefont
			{Le}}, \bibinfo {author} {\bibfnamefont {C.}~\bibnamefont {Meroni}}, \bibinfo
		{author} {\bibfnamefont {B.}~\bibnamefont {Sturmfels}}, \bibinfo {author}
		{\bibfnamefont {R.~F.}\ \bibnamefont {Werner}},\ and\ \bibinfo {author}
		{\bibfnamefont {T.}~\bibnamefont {Ziegler}},\ }\bibfield  {title} {\bibinfo
		{title} {Quantum {C}orrelations in the {M}inimal {S}cenario},\ }\href
	{https://doi.org/10.22331/q-2023-03-16-947} {\bibfield  {journal} {\bibinfo
			{journal} {{Quantum}}\ }\textbf {\bibinfo {volume} {7}},\ \bibinfo {pages}
		{947} (\bibinfo {year} {2023})}\BibitemShut {NoStop}%
	\bibitem [{\citenamefont {Goh}\ \emph {et~al.}(2018)\citenamefont {Goh},
		\citenamefont {Kaniewski}, \citenamefont {Wolfe}, \citenamefont {V\'ertesi},
		\citenamefont {Wu}, \citenamefont {Cai}, \citenamefont {Liang},\ and\
		\citenamefont {Scarani}}]{Goh2018}%
	\BibitemOpen
	\bibfield  {author} {\bibinfo {author} {\bibfnamefont {K.~T.}\ \bibnamefont
			{Goh}}, \bibinfo {author} {\bibfnamefont {J.~m.~k.}\ \bibnamefont
			{Kaniewski}}, \bibinfo {author} {\bibfnamefont {E.}~\bibnamefont {Wolfe}},
		\bibinfo {author} {\bibfnamefont {T.}~\bibnamefont {V\'ertesi}}, \bibinfo
		{author} {\bibfnamefont {X.}~\bibnamefont {Wu}}, \bibinfo {author}
		{\bibfnamefont {Y.}~\bibnamefont {Cai}}, \bibinfo {author} {\bibfnamefont
			{Y.-C.}\ \bibnamefont {Liang}},\ and\ \bibinfo {author} {\bibfnamefont
			{V.}~\bibnamefont {Scarani}},\ }\bibfield  {title} {\bibinfo {title}
		{Geometry of the set of quantum correlations},\ }\href
	{https://doi.org/10.1103/PhysRevA.97.022104} {\bibfield  {journal} {\bibinfo
			{journal} {Phys. Rev. A}\ }\textbf {\bibinfo {volume} {97}},\ \bibinfo
		{pages} {022104} (\bibinfo {year} {2018})}\BibitemShut {NoStop}%
	\bibitem [{\citenamefont {Bancal}\ \emph {et~al.}(2011)\citenamefont {Bancal},
		\citenamefont {Brunner}, \citenamefont {Gisin},\ and\ \citenamefont
		{Liang}}]{Bancal2011}%
	\BibitemOpen
	\bibfield  {author} {\bibinfo {author} {\bibfnamefont {J.-D.}\ \bibnamefont
			{Bancal}}, \bibinfo {author} {\bibfnamefont {N.}~\bibnamefont {Brunner}},
		\bibinfo {author} {\bibfnamefont {N.}~\bibnamefont {Gisin}},\ and\ \bibinfo
		{author} {\bibfnamefont {Y.-C.}\ \bibnamefont {Liang}},\ }\bibfield  {title}
	{\bibinfo {title} {Detecting genuine multipartite quantum nonlocality: A
			simple approach and generalization to arbitrary dimensions},\ }\href
	{https://doi.org/10.1103/PhysRevLett.106.020405} {\bibfield  {journal}
		{\bibinfo  {journal} {Phys. Rev. Lett.}\ }\textbf {\bibinfo {volume} {106}},\
		\bibinfo {pages} {020405} (\bibinfo {year} {2011})}\BibitemShut {NoStop}%
	\bibitem [{\citenamefont {Epping}\ \emph {et~al.}(2017)\citenamefont {Epping},
		\citenamefont {Kampermann}, \citenamefont {macchiavello},\ and\ \citenamefont
		{Bruß}}]{Epping2017}%
	\BibitemOpen
	\bibfield  {author} {\bibinfo {author} {\bibfnamefont {M.}~\bibnamefont
			{Epping}}, \bibinfo {author} {\bibfnamefont {H.}~\bibnamefont {Kampermann}},
		\bibinfo {author} {\bibfnamefont {C.}~\bibnamefont {macchiavello}},\ and\
		\bibinfo {author} {\bibfnamefont {D.}~\bibnamefont {Bruß}},\ }\bibfield
	{title} {\bibinfo {title} {Multi-partite entanglement can speed up quantum
			key distribution in networks},\ }\href
	{https://doi.org/10.1088/1367-2630/aa8487} {\bibfield  {journal} {\bibinfo
			{journal} {New Journal of Physics}\ }\textbf {\bibinfo {volume} {19}},\
		\bibinfo {pages} {093012} (\bibinfo {year} {2017})}\BibitemShut {NoStop}%
	\bibitem [{\citenamefont {Das}\ \emph {et~al.}(2021)\citenamefont {Das},
		\citenamefont {B\"auml}, \citenamefont {Winczewski},\ and\ \citenamefont
		{Horodecki}}]{Das2021}%
	\BibitemOpen
	\bibfield  {author} {\bibinfo {author} {\bibfnamefont {S.}~\bibnamefont
			{Das}}, \bibinfo {author} {\bibfnamefont {S.}~\bibnamefont {B\"auml}},
		\bibinfo {author} {\bibfnamefont {M.}~\bibnamefont {Winczewski}},\ and\
		\bibinfo {author} {\bibfnamefont {K.}~\bibnamefont {Horodecki}},\ }\bibfield
	{title} {\bibinfo {title} {Universal limitations on quantum key distribution
			over a network},\ }\href {https://doi.org/10.1103/PhysRevX.11.041016}
	{\bibfield  {journal} {\bibinfo  {journal} {Phys. Rev. X}\ }\textbf {\bibinfo
			{volume} {11}},\ \bibinfo {pages} {041016} (\bibinfo {year}
		{2021})}\BibitemShut {NoStop}%
\end{thebibliography}

%


\appendix

\onecolumngrid

\section{Evaluation of the optimal quantum bound of the $3$-partite Svetlichny inequality, using SOS technique}\label{SOS3}

For $m=3$, the strings $i_{\mu}$ are explicitly given as $i_1 \equiv 000$, $i_2\equiv 001$, $i_3\equiv010$ and $i_4 \equiv 111$. The parameter $v^n_{\mu}$ are given as $v^0_1 = 1$ for $i_1$, $v^1_2 = -1$ for $i_2$, $v^1_3 = -1$ for $i_3$ and $v^2_4 = -1$ for $i_4$. From Eq.~(\ref{svm_mod}), using these parameters, the $3-$party  Svetlichny functional is given by
\begin{equation}
    \mathscr{S}_3 = \qty(A^1_0 - A^1_1)\otimes A^2_0 \otimes A^3_0 - \qty(A^1_0 + A^1_1)\otimes A^2_0 \otimes A^3_1 - \qty(A^1_0 + A^1_1)\otimes A^2_1 \otimes A^3_0 - \qty(A^1_0 - A^1_1)\otimes A^2_1 \otimes A^3_1
\end{equation}
Similar to Eq.~(\ref{mparty_Svetlichny}), we define the positive semidefinite operator as $\Gamma_3=\frac{1}{2}\sum_{\mu=1}^4 \omega_\mu L^{\dagger}_\mu L_\mu$, where the hermitian operators $L_\mu$ are defined as follows
\begin{equation}
    \begin{aligned}
        L_{1} &= \frac{1}{\omega_{1}}\qty[A_0^1 - A_1^1]\otimes\openone\otimes\openone - \openone\otimes A_0^2\otimes A_0^3 \ \ \ \ \ \ \ \ \ \ \ \ &L_{2} &= \frac{1}{\omega_{2}}\qty[A_0^1 + A_1^1]\otimes\openone\otimes\openone + \openone\otimes A_0^2\otimes A_1^3 \\
        L_{3} &= \frac{1}{\omega_{3}}\qty[A_0^1 + A_1^1]\otimes\openone\otimes\openone + \openone\otimes A_1^2\otimes A_0^3 \ \ \ \ \ \ \ \ \ \ \ \ &L_{4} &= \frac{1}{\omega_{4}}\qty[A_0^1 - A_1^1]\otimes\openone\otimes\openone + \openone\otimes A_1^2\otimes A_1^3 \\
    \end{aligned}   
\end{equation}
with $\omega_{1}=\omega_{4
}=\norm{A_0^1-A_1^1}_\rho = \sqrt{1-\Tr[\{A_0^1,A_1^1\}\ \rho]}$ and $\omega_{2}=\omega_{3}=\norm{A_0^1+A_1^1}_\rho= \sqrt{1+\Tr[\{A_0^1,A_1^1\}\ \rho]}$. Now, following the arguments of optimality discussed in the main text (see Eq.~(\ref{smgamma}) and the preceding discussions), the optimal value is given by
\begin{equation}
      (\mathscr{S}_3)_Q^{opt} = 2\qty(\sqrt{1-\Tr[\{A_0^1,A_1^1\}\ \rho]}+\sqrt{1+\Tr[\{A_0^1,A_1^1\}\ \rho]}) = 4\sqrt{2} \ \ \ \  \text{\textit{iff} $\{A_0^1,A_1^1\} = 0$}
\end{equation}
The optimality condition $\Tr[\Gamma_3 \ \rho] = 0$ implies that $\Tr[L^{\dagger}_\mu L_\mu \ \rho]=0$, leading to the following
\begin{equation}\label{a4}
    \begin{aligned}
        \Tr[\frac{A_0^1-A_1^1}{\sqrt{2}} \otimes A_0^2 \otimes A_0^3\ \rho] &= 1 \ \ (\text{for} \ \mu=1, \ i_1 = 000) \ ; \ \  
        \Tr[\frac{A_0^1+A_1^1}{\sqrt{2}} \otimes A_0^2 \otimes A_1^3\ \rho] &= -1\ \  (\text{for} \ \mu=2, \ i_2 = 001) \ ; \\
        \Tr[\frac{A_0^1+A_1^1}{\sqrt{2}} \otimes A_1^2 \otimes A_0^3\ \rho] &= -1\ \ (\text{for} \ \mu=3, \ i_3 = 010) \ ; \ \
        \Tr[\frac{A_0^1-A_1^1}{\sqrt{2}} \otimes A_1^2 \otimes A_1^3\ \rho] &= -1\ \  (\text{for} \ \mu=4, \ i_4 = 011) \ ;
    \end{aligned}
\end{equation}
From the argument presented for the generalised case in the main text, we infer that $\rho = \ket{\psi}\bra{\psi}$. Furthermore, the optimality condition $\Tr[L^{\dagger}_\mu L_\mu \ \rho]$ implies $L_\mu\ket{\psi} = 0$, which gives
\begin{equation}\label{three_party_L}
    \begin{aligned}
        \frac{A_0^1-A_1^1}{\sqrt{2}} \otimes \openone \otimes \openone  \ket{\psi} &= \openone \otimes A_0^2 \otimes A_0^3 \ket{\psi}\  \ (\text{for} \ \mu=1, \ i_1 = 000) \\
        \frac{A_0^1+A_1^1}{\sqrt{2}} \otimes \openone \otimes \openone  \ket{\psi} &= -\openone \otimes A_0^2 \otimes A_1^3 \ket{\psi}\  \ (\text{for} \ \mu=2, \ i_2 = 001) \\
        \frac{A_0^1+A_1^1}{\sqrt{2}} \otimes \openone \otimes \openone  \ket{\psi} &= -\openone \otimes A_1^2 \otimes A_0^3 \ket{\psi} \ \ (\text{for} \ \mu=3, \ i_3 = 010)\\
        \frac{A_0^1-A_1^1}{\sqrt{2}} \otimes \openone \otimes \openone  \ket{\psi} &= -\openone \otimes A_1^2 \otimes A_1^3 \ket{\psi} \ \ (\text{for} \ \mu=4, \ i_4 = 011)
    \end{aligned}
\end{equation}
Taking anticommutation between first and second equations of Eq.~(\ref{three_party_L}), which correspond to $i_1=000$ and $i_2=001$ we get $\{A_0^3,A_1^3\}= 0$. Similarly taking the anticommutation between first and third equations of Eq.~(\ref{three_party_L}), corresponding to $i_1=000$ and $i_3=010$ we get we get $\{A_0^2,A_1^2\}= 0$. Note that, we could have chosen any pair of $\mu$ which only differ only at some bit. For example $i_2 = 001$ and $i_4=011$ will also give the anticommutation relation $\{A_0^2,A_1^2\}= 0$. 

Hence, each of the joint observables appearing in the left hand side of Eq.~(\ref{a4}) are mutually commuting, with $\rho = \ket{\psi}\bra{\psi}$ being the common eigenstate. Thus, one possible representation of $\rho\in\mathscr{L}(\Motimes_{i=1}^3\mathcal{H}_d)$ can be given by Eq.~(\ref{gen_state}).


\section{Evaluation of the optimal quantum bound of the $4$-partite Svetlichny inequality, using SOS technique}\label{SOS4}

For $m=4$, there are eight 4-bit string given by $i_\mu = \{0000,0001,0010,0011,0100,0101,0110,0111\}$ corresponding to $v_\mu^n = \{1,-1,-1,-1,-1,-1,-1,1\}$. From Eq.~(\ref{svm_mod}), The four-partite Svetlichny functional is given by 
\begin{equation}
    \begin{aligned}
        \mathscr{S}_4&= \qty(A^1_0 - A^1_1)A^2_0A^3_0A^4_0 - \qty(A^1_0 + A^1_1)A^2_0A^3_0A^4_1 -\qty(A^1_0 + A^1_1)A^2_0A^3_1A^4_0 - \qty(A^1_0 - A^1_1)A^2_0A^3_1A^4_1 \\
        & -\qty(A^1_0 + A^1_1)A^2_1A^3_0A^4_0 - \qty(A^1_0 - A^1_1)A^2_1A^3_0A^4_1 - \qty(A^1_0 - A^1_1)A^2_1A^3_1A^4_0 + \qty(A^1_0 + A^1_1)A^2_1A^3_1A^4_1 \\
    \end{aligned}
\end{equation}
In order to evaluate the optimal quantum value of the four-partite Svetlichny functional, we follow the procedure similar to that has been introduced in preceding section, we define a positive semi-definite operator as  $\Gamma_{4}=\frac{1}{2} \sum^{8}_{\mu=1} \ \omega_\mu \ L^\dagger_\mu L_\mu$, where $L_\mu$'s are arbitrary hermitian operators constructed as follows
\begin{equation}\label{li4}
    \begin{aligned}
        L_1 &=&\frac{\qty(A^1_0 - A^1_1)\otimes\openone\otimes\openone\otimes\openone}{\omega_1} - \openone\otimes A^2_0\otimes A^3_0 \otimes A^4_0 ; \ \ 
        L_2 &=& \frac{\qty(A^1_0 + A^1_1)\otimes\openone\otimes\openone\otimes\openone}{\omega_2} + \openone\otimes A^2_0\otimes A^3_0 \otimes A^4_1\\ 
        L_3 &=& \frac{\qty(A^1_0 + A^1_1)\otimes\openone\otimes\openone\otimes\openone}{\omega_3} + \openone\otimes A^2_0\otimes A^3_1 \otimes A^4_0; \ \
        L_4 &=& \frac{\qty(A^1_0 - A^1_1)\otimes\openone\otimes\openone\otimes\openone}{\omega_4} + \openone\otimes A^2_0\otimes A^3_1 \otimes A^4_1 \\
        L_5 &=&\frac{\qty(A^1_0 + A^1_1)\otimes\openone\otimes\openone\otimes\openone}{\omega_5} + \openone\otimes A^2_1\otimes A^3_0 \otimes A^4_0 ; \ \ 
        L_6 &=& \frac{\qty(A^1_0 - A^1_1)\otimes\openone\otimes\openone\otimes\openone}{\omega_6} + \openone\otimes A^2_1\otimes A^3_0 \otimes A^4_1\\ 
        L_7 &=& \frac{\qty(A^1_0 - A^1_1)\otimes\openone\otimes\openone\otimes\openone}{\omega_7} + \openone\otimes A^2_1\otimes A^3_1 \otimes A^4_0; \ \ 
        L_8 &=& \frac{\qty(A^1_0 + A^1_1)\otimes\openone\otimes\openone\otimes\openone}{\omega_8} - \openone\otimes A^2_1\otimes A^3_1 \otimes A^4_1 
    \end{aligned}
\end{equation}
and we define each $\omega_i$ as follows
\begin{equation}\label{omegai}
  \begin{aligned}
    \omega_1 = \omega_4 = \omega_6 = \omega_7 &=  \sqrt{\Tr[ \openone \otimes \left( A^1_0 - A^1_1 \right)\left( A^1_0 - A^1_1 \right) \otimes\openone \ \rho ]}= \sqrt{2 - \Tr\left[\left\{ A^1_0,A^1_1 \right\} \ \rho\right]}\\
    \omega_2 = \omega_3 = \omega_5 = \omega_8 &= \sqrt{\Tr[ \openone \otimes \left( A^1_0 + A^1_1 \right)\left( A^1_0 + A^1_1 \right) \otimes\openone \ \rho ]}=\sqrt{2 + \Tr\left[\left\{ A^1_0,A^1_1 \right\} \ \rho\right]}
  \end{aligned}
\end{equation}
The optimal quantum value of the four-partite Svetlichny functional given by
\begin{equation}
    (\mathscr{S}_4)^{opt}_Q \leq 4 \ \qty[ \max \qty(\omega_1+\omega_2)] \nonumber = 4 \ \max \Bigg(\sqrt{2 + \Tr[\qty{ A^1_0,A^1_1 } \ \rho]} \nonumber + \sqrt{2 - \Tr[\qty{ A^1_0,A^1_1 } \ \rho]}\Bigg) = 8\sqrt{2} \ \ \ \ \text{\textit{iff} $\qty{A_0^1,A_1^1}=0$}
\end{equation}
The optimality condition $\Tr[\Gamma_{4} \; \rho] = 0$ implies that $\Tr[L^\dagger_\mu L_\mu\; \rho] = 0 \; \forall \mu$, implying $\rho$ is a pure state and this further leads to the following relations
\begin{equation}\label{four_party_L}
    \begin{aligned}
        \frac{\qty(A^1_0 - A^1_1)\otimes\openone\otimes\openone\otimes\openone}{\sqrt{2}} \ket{\psi} &= \openone\otimes A^2_0\otimes A^3_0 \otimes A^4_0 \ket{\psi} \ \ (\text{for} \ \mu = 1, \ i_1 = 0000) \\
        \frac{\qty(A^1_0 + A^1_1)\otimes\openone\otimes\openone\otimes\openone}{\sqrt{2}}\ket{\psi} &= - \openone\otimes A^2_0\otimes A^3_0 \otimes A^4_1\ket{\psi}\ \  (\text{for} \ \mu = 2, \ i_2 = 0001)\\ 
        \frac{\qty(A^1_0 + A^1_1)\otimes\openone\otimes\openone\otimes\openone}{\sqrt{2}} \ket{\psi}&= -\openone\otimes A^2_0\otimes A^3_1 \otimes A^4_0\ket{\psi} \ \ (\text{for} \ \mu = 3, \ i_3 = 0010)\\
        \frac{\qty(A^1_0 - A^1_1)\otimes\openone\otimes\openone\otimes\openone}{\sqrt{2}}\ket{\psi}&= -\openone\otimes A^2_0\otimes A^3_1 \otimes A^4_1 \ket{\psi}\ \ (\text{for} \ \mu = 4, \ i_4 = 0011)\\
        \frac{\qty(A^1_0 + A^1_1)\otimes\openone\otimes\openone\otimes\openone}{\sqrt{2}} \ket{\psi}&= -\openone\otimes A^2_1\otimes A^3_0 \otimes A^4_0 \ket{\psi} \ \ (\text{for} \ \mu = 5, \ i_5 = 0100)\\
        \frac{\qty(A^1_0 - A^1_1)\otimes\openone\otimes\openone\otimes\openone}{\sqrt{2}}\ket{\psi}&= -\openone\otimes A^2_1\otimes A^3_0 \otimes A^4_1\ket{\psi}\ \ \ \ (\text{for} \ \mu = 6, \ i_6 = 0101)\\ 
        \frac{\qty(A^1_0 - A^1_1)\otimes\openone\otimes\openone\otimes\openone}{\sqrt{2}}\ket{\psi}&= - \openone\otimes A^2_1\otimes A^3_1 \otimes A^4_0\ket{\psi}\ \ (\text{for} \ \mu = 7, \ i_7 = 0110)\\
        \frac{\qty(A^1_0 + A^1_1)\otimes\openone\otimes\openone\otimes\openone}{\sqrt{2}} \ket{\psi} &= \openone\otimes A^2_1\otimes A^3_1 \otimes A^4_1 \ket{\psi} \ \ (\text{for} \ \mu = 8, \ i_8 = 0111)
    \end{aligned}
\end{equation}
By taking the anticommutation of relations corresponding to $i_1 = 0000$ and $i_2=0001$, one can obtain $\{A_0^4,A_1^4\} = 0$. Taking the relations corresponding to $i_1 = 0000$ and $i_3=0010$, we obtain $\{A_0^3,A_1^3\} = 0$ and for obtaining $\{A_0^2,A_1^2\} = 0$, we use the anticommutation between the relations corresponding to $i_1 = 0000$ and $i_5=0100$.


\section{Quantum optimal of the $m-$partite inequality proposed by Bancal \textit{et al.} \cite{Bancal2011}}\label{bancalgen}

A generalised Svetlichny-type inequality was proposed in \cite{Bancal2011}. This inequality was claimed to detect genuine multipartite nonlocality. Note that for $m=3$, this inequality corresponds to the tripartite Svetlichny inequality. However, we will show that for $m\geq3$ this inequality does not capture genuine multipartite nonlocality. This is because the proposed inequality with $m\geq 4$ can be optimised even if one of the parties measures commuting observables, implying that the correlations pertaining to the optimal value of this inequality will have a local ontic model of the form given in Eq.~(\ref{gn}). The proposed $m-$partite inequality is given by \cite{Bancal2011}
\begin{equation}\label{eqbancal}
    \braket{\mathscr{S}_m} = \braket{\mathscr{S}_{m-1}A^m_1} + \braket{\mathscr{S}'_{m-1}A^m_0} \leq 2^m
\end{equation}
The quantum optimal bound is found to be $2^{m-1}\sqrt{2}$ corresponding to $m$-qubit entangled state and qubit observables. Here, we will derive the optimal quantum bound without assuming the dimension of the system and consequently determine the conditions on the state and observables for achieving this optimal bound. The inequality given by Eq.~(\ref{eqbancal}) can be re-expressed as follows
\begin{equation}
    \mathscr{S}_m = \qty[\qty(A^1_0 \otimes (A^2_0 + A^2_1) + A^1_1 \otimes (A^2_0 - A^2_1))\otimes A^3_1 + \qty(A^1_1 \otimes (A^2_0+A^2_1) - A^1_0 \otimes (A^2_0-A^2_1)) \otimes A^3_0] \Motimes_{k=1}^{m} (A^k_0 + A^k_1)
\end{equation}
To achieve the optimal quantum bound, without loss of generality, we first construct an operator $\Gamma_{m}$ as follows
\begin{equation}
    \Gamma_{m} = \sum_{\mu = 1}^{2^{m-3}} \qty[ \frac{\omega}{2}\qty{ \qty(T_\mu)^\dagger T_\mu + \qty(U_\mu)^\dagger U_\mu} +  \frac{\omega'}{2}\qty{\qty(V_\mu)^\dagger V_\mu +  \qty(W_\mu)^\dagger W_\mu}]
\end{equation}
where $T_\mu$, $U_\mu$, $V_\mu$ and $W_\mu$ are given as follows
\begin{align}
    T_\mu &= \frac{1}{\omega}\openone\otimes\left(A^2_0 + A^2_1\right)\otimes\openone\Motimes_{k=4}^{m}\openone - A^1_0\otimes\openone\otimes A^3_1 \Motimes_{k=4}^{m} A^k_{i^{k-3}_\mu} \  \ \ \nonumber \\
    U_\mu &= \frac{1}{\omega}\openone\otimes\left(A^2_0 + A^2_1\right)\otimes\openone\Motimes_{k=4}^{m}\openone - A^1_1\otimes\openone\otimes A^3_0 \Motimes_{k=4}^{m} A^k_{i^{k-3}_\mu} \  \ \ \omega= \norm{A^2_0 + A^2_1} \nonumber \\
    V_\mu &= \frac{1}{\omega'}\openone\otimes\left(A^2_0 - A^2_1\right)\otimes\openone\Motimes_{k=4}^{m}\openone - A^1_1\otimes\openone\otimes A^3_1 \Motimes_{k=4}^{m} A^k_{i^{k-3}_\mu} \  \ \ \\
    W_\mu &= \frac{1}{\omega'}\openone\otimes\left(A^2_1 - A^2_0\right)\otimes\openone\Motimes_{k=4}^{m}\openone - A^1_0\otimes\openone\otimes A^3_0 \Motimes_{k=4}^{m} A^k_{i^{k-3}_\mu}\  \ \ \omega'= \norm{A^2_1-A^2_0}\nonumber 
\end{align}
where $i_{\mu}$ is a $(m-3)$-bit string with $m \geq 4$, hence $i^{k-3}_\mu$ represents the bit value at the $(k-3)$ index for a chosen $\mu$. Taking the shared state to be $\rho$, and evaluating $\omega$, we obtain $\omega = \sqrt{2 + \Tr\left[\left\{ A^2_0,A^2_1 \right\} \ \rho\right]}$ and $\omega'= \sqrt{2 - \Tr\left[\left\{ A^2_0,A^2_1 \right\} \ \rho\right]}$. 

This will make $\braket{\Gamma_{m}} = 2^{m-2}\qty(\omega+\omega') - \braket{\mathscr{S}_m}$. Following the argument presented earlier (refer to Appx.~\ref{SOS3}), the maximum value occurs when $\braket{\Gamma_{m}} = 0$. Consequently, the optimal value of the $m-$partite inequality becomes $(\mathscr{S}_m)_Q = 2^{m-1}\sqrt{2}$ with all $\omega = \omega' = \sqrt{2}$ and $\left\{A^2_0,A^2_1\right\} = 0$. The optimality condition $\Tr[\Gamma_{m} \ \rho] = 0$ implies $\Tr[(T_\mu)^\dagger T_\mu \ \rho]=\Tr[ (U_\mu)^\dagger U_\mu \ \rho] = \Tr[(V_\mu)^\dagger V_\mu \ \rho] = \Tr[(W_\mu)^\dagger W_\mu \ \rho] = 0$ leading to the following set of relations 
\begin{equation}\label{obsrm1}
\begin{aligned}
&& \Tr[\frac{A^1_0\otimes\left(A^2_0 + A^2_1\right)\otimes A^3_1 \Motimes_{k=4}^{m} A^k_{i_\mu^{k-3}}}{\sqrt{2}} \ \rho] =\Tr[\frac{A^1_1\otimes\left(A^2_0 - A^2_1\right)\otimes A^3_0 \Motimes_{k=4}^{m} A^k_{i_\mu^{k-3}}}{\sqrt{2}} \ \rho] =1 \\
&&  \Tr[\frac{A^1_1\otimes\left(A^2_1 - A^2_0\right)\otimes A^3_1 \Motimes_{k=4}^{m} A^k_{i_\mu^{k-3}}}{\sqrt{2}} \ \rho] = \Tr[\frac{A^1_0\otimes\left(A^2_1 - A^2_0\right)\otimes A^3_0 \Motimes_{k=4}^{m} A^k_{i_\mu^{k-3}}}{\sqrt{2}} \ \rho]=1
\end{aligned}
\end{equation}
The above optimality conditions given by Eq.~(\ref{obsrm1}) imply that the shared state $\rho$ is a pure state since all the expectation values of the dichotomic observables are extremal. Hence, we write

\begin{align}
    \openone\otimes\frac{A^2_0 + A^2_1}{\sqrt{2}}\otimes\openone\Motimes_{k=4}^{m}\openone\ket{\psi} &= A^1_0\otimes\openone\otimes A^3_1 \Motimes_{k=4}^{m} A^k_{i_\mu^{k-3}} \ket{\psi} \label{obsrm11} \\
    \openone\otimes\frac{A^2_0 + A^2_1}{\sqrt{2}}\otimes\openone\Motimes_{k=4}^{m}\openone\ket{\psi} &= A^1_1\otimes\openone\otimes A^3_0 \Motimes_{k=4}^{m} A^k_{i_\mu^{k-3}} \ket{\psi} \label{obsrm12} \\
    \openone\otimes\frac{A^2_0 -A^2_1}{\sqrt{2}}\otimes\openone\Motimes_{k=4}^{m}\openone\ket{\psi} &=  A^1_1\otimes\openone\otimes A^3_1 \Motimes_{k=4}^{m} A^k_{i_\mu^{k-3}} \ket{\psi} \label{obsrm13}\\
    \openone\otimes\frac{A^2_1 - A^2_0}{\sqrt{2}}\otimes\openone\Motimes_{k=4}^{m}\openone \ket{\psi} &= A^1_0\otimes\openone\otimes A^3_0 \Motimes_{k=4}^{m} A^k_{i_\mu^{k-3}} \ket{\psi} \label{obsrm14}
\end{align}
Now to derive the condition on the observables, we first fix a particular $i_\mu^{k-3}$ and take the anticommutation between Eq.~(\ref{obsrm11}) and Eq.~(\ref{obsrm13}). It is the straightforward to obtain first party's observables are anti-commuting, i.e., $\{A^1_0,A^1_1\} = 0$. Taking the anticommutation between Eq.~(\ref{obsrm11}) and Eq.~(\ref{obsrm14}), we get $\{A^3_0,A^3_1\} = 0$. Let us now find the observables conditions for $k(\geq 4)$-th party observables. We take recourse to the $m=4$ case. Then, relations in Eqs.~(\ref{obsrm11})-(\ref{obsrm14}) are re-expressed as follows
\begin{align}
\openone\otimes\frac{A^2_0 + A^2_1}{\sqrt{2}}\otimes\openone\otimes\openone\ket{\psi} &= A^1_0\otimes\openone\otimes A^3_1 \otimes A^4_0 \ket{\psi}; \ \ \ \ &\openone\otimes\frac{A^2_0 + A^2_1}{\sqrt{2}}\otimes\openone\otimes \openone\ket{\psi} &= A^1_0\otimes\openone\otimes A^3_1 \otimes A^4_1 \ket{\psi} \label{c10}\\
\openone\otimes\frac{A^2_0 + A^2_1}{\sqrt{2}}\otimes\openone\otimes\openone\ket{\psi} &= A^1_1\otimes\openone\otimes A^3_0 \otimes A^4_0 \ket{\psi}; \ \ \ \ &\openone\otimes\frac{A^2_0 + A^2_1}{\sqrt{2}}\otimes\openone\otimes \openone\ket{\psi} &= A^1_1\otimes\openone\otimes A^3_0 \otimes A^4_1 \ket{\psi} \\
\openone\otimes\frac{A^2_0 - A^2_1}{\sqrt{2}}\otimes\openone\otimes\openone\ket{\psi} &= A^1_1\otimes\openone\otimes A^3_1 \otimes A^4_0 \ket{\psi}; \ \ \ \ &\openone\otimes\frac{A^2_0 - A^2_1}{\sqrt{2}}\otimes\openone\otimes \openone\ket{\psi} &= A^1_1\otimes\openone\otimes A^3_1 \otimes A^4_1 \ket{\psi} \\
\openone\otimes\frac{A^2_1 - A^2_0}{\sqrt{2}}\otimes\openone\otimes\openone\ket{\psi} &= A^1_0\otimes\openone\otimes A^3_0 \otimes A^4_0 \ket{\psi}; \ \ \ \ &\openone\otimes\frac{A^2_1 - A^2_0}{\sqrt{2}}\otimes\openone\otimes \openone\ket{\psi} &= A^1_0\otimes\openone\otimes A^3_0 \otimes A^4_1 \ket{\psi}
\end{align}
Now it is evident from the above equations that the fourth party must implement commuting observables so that the functional in Eq.~(\ref{eqbancal}) obtains the optimal quantum bound. This is readily generalised to arbitrary number of parties leading us to the conclusion that any party beyond three must implement commuting observables to obtain optimal quantum bound. We present an explicit example with local dimension $2$ for $m=4$ case. For $m=4$, the functional in Eq.~(\ref{eqbancal}) becomes
\begin{equation}
    \mathscr{S}_4= \qty[\qty{A^1_0 \otimes \qty(A^2_0 + A^2_1) + A^1_1 \otimes \qty(A^2_0 - A^2_1)}\otimes A^3_1 + \qty{A^1_1 \otimes \qty(A^2_0+A^2_1) - A^1_0 \otimes \qty(A^2_0-A^2_1)} \otimes A^3_0] \otimes \qty(A^4_0 + A^4_1)    
\end{equation}
The local bound is $8$ and the optimal quantum value is $8 \sqrt{2}$ and is obtained when the shared state and observables are
\begin{equation}
            \ket{\psi} = \frac{1}{\sqrt{2}}\Big[\ket{0000}+\ket{1111}\Big] \nonumber
\end{equation}
\begin{equation}\label{swapobsro}
        \begin{aligned}
        A^1_0 &= \sigma_y  \ ;   &A^2_0 &= \frac{\sigma_x+\sigma_y}{\sqrt{2}} \; & A^3_0 &= \sigma_x \; & A^4_0 &= \sigma_x  \\
        A^1_1 &= \sigma_x  \ ;   &A^2_1 &= \frac{\sigma_x-\sigma_y}{\sqrt{2}} \; & A^3_1 &= -\sigma_y \;  &A^4_1 &= \sigma_x  
        \end{aligned}
\end{equation}

Hence, it follows from Eq.~(\ref{swapobsro}) that the four-partite functional will achieve the quantum optimum value of $8\sqrt{2}$ even if the fourth party measures commuting observables. In general, our analysis shows that using Eqs.~(\ref{obsrm11})-(\ref{obsrm14}), the $m-$partite functional given by Eq.~(\ref{eqbancal}) will reach the optimal quantum value of $2^{m-1}\sqrt{2}$ even if all the $m>4$ parties perform commuting observables. This result establishes that the $m-$partite functional proposed in \cite{Bancal2011} does not capture the genuine multiparty nonlocality. This finding is in conformity with the result pointed out in \cite{Collins2002}, which states that while the Mermin inequality can be generalised from the tripartite to the multipartite case according to the method proposed in \cite{Bancal2011} for detecting standard multipartite nonlocality, the Svetlichny inequality cannot be generalised in the same way for genuine multipartite nonlocality. To generalise the Svetlichny inequality for multipartite case, we need to adopt the procedure demonstrated in \cite{Collins2002,Seevinck2002}.

\section{Evaluation of Isometry output for $m$=3 case using swap circuit} \label{isoo3}
For $m=3$, the isometry presented in Fig. \ref{figswap} outputs the following
\begin{equation}
    \Phi(\ket{\psi}_3\ket{000}_{1',2',3'}) = \textbf{C}_X \cdot \textbf{H} \cdot \textbf{C}_Z \cdot \textbf{H} \ket{\psi}_3\ket{000}_{1',2',3'} 
\end{equation}
where $\textbf{C}_X=C_{X_1}(1') \otimes C_{X_2}(2') \otimes C_{X_3}(3')$, $\textbf{C}_Z=C_{Z_1}(1') \otimes C_{Z_2}(2') \otimes C_{Z_3}(3')$ and $\textbf{H}=H_{1'}\otimes H_{2'} \otimes H_{3'}$. Here $C_{X_i}(j')$ and $C_{Z_i}(j')$ are the controlled gates operating on the physical system's appropriate Hilbert space with control on the $j'$-th ancillary system. Applying the Hadamard on each ancilla, we get 
\begin{equation}
\textbf{H} \ket{\psi}_3\ket{000}_{1',2',3'} = \frac{1}{2\sqrt{2}}\ket{\psi}_3(\ket{000}+\ket{001}+\ket{010}+\ket{011}+\ket{100}+\ket{101}+\ket{110}+\ket{111})
\end{equation}
Further application of $\textbf{C}_{Z}$ gives
\begin{equation}\label{cnots}
\frac{1}{2\sqrt{2}}(\ket{\psi}_3\ket{000} + Z_3\ket{\psi}_3\ket{001} + Z_2\ket{\psi}_3\ket{010} + Z_2 Z_3\ket{\psi}_3\ket{011} + Z_1 \ket{\psi}_3\ket{100} + Z_1 Z_3 \ket{\psi}_3\ket{101} + Z_1 Z_2 \ket{\psi}_3\ket{110} + Z_1Z_2Z_3\ket{\psi}_3\ket{111})
\end{equation}
We need to apply Hadamards $\textbf{H}$ followed by  $\textbf{C}_X$ on the state in Eq. (\ref{cnots}). We start the evaluation by taking each term in Eq. (\ref{cnots}) separately. For the first term $\ket{\psi}_3\ket{000}$ in Eq. (\ref{cnots}), 
\begin{eqnarray}
 \textbf{C}_X \cdot \textbf{H} \ket{\psi}_3\ket{000} &=&  \frac{1}{2\sqrt{2}} \textbf{C}_X \ket{\psi}_3(\ket{000}+\ket{001}+\ket{010}+\ket{011}+\ket{100}+\ket{101}+\ket{110}+\ket{111})  \nonumber \\
  &=& \frac{1}{2\sqrt{2}}\Big(\ket{\psi}_3\ket{000} + X_3\ket{\psi}_3\ket{001} + X_2\ket{\psi}_3\ket{010} + X_2 X_3\ket{\psi}_3\ket{011} + X_1 \ket{\psi}_3\ket{100} \nonumber \\
  &&  \hspace{1 cm} + X_1 X_3 \ket{\psi}_3\ket{101} + X_1 X_2 \ket{\psi}_3\ket{110} + X_1X_2X_3\ket{\psi}_3\ket{111}\Big) 
\end{eqnarray}
Similarly, other terms can be evaluated. Lets give a detailed evaluation of action of $\textbf{C}_X \cdot \textbf{H}$ on one more term in Eq.~(\ref{cnots}) for clarity.
\begin{eqnarray}
 \textbf{C}_X \cdot \textbf{H} \ket{\psi}_3\ket{111} &=& \frac{1}{2\sqrt{2}}\Big(Z_1Z_2Z_3 \ket{\psi}_3\ket{000} - Z_1Z_2X_3Z_3 \ket{\psi}_3\ket{001} - Z_1X_2Z_2Z_3 \ket{\psi}_3\ket{010} + Z_1X_2Z_2X_3Z_3 \ket{\psi}_3 \ket{011} \nonumber \\
  &&  - X_1Z_1Z_2Z_3 \ket{\psi}_3\ket{100} + X_1Z_1Z_2X_3Z_3 \ket{\psi}_3\ket{101} + X_1Z_1X_2Z_2Z_3 \ket{\psi}_3\ket{110} - X_1Z_1X_2Z_2X_3Z_3 \ket{\psi}_3\ket{111}\Big) 
\end{eqnarray}
After algebraic simplification of all the terms, we obtain the final output state $ \ket{\psi}_3=\Phi(\ket{\psi}_3\ket{000}_{1',2',3'})$(before the application of $U_i$ on ancillary qubits) as
\begin{eqnarray}
    \ket{\psi}_3 &=& \frac{1}{8} \Big[ (\openone + Z_1)(\openone + Z_2)(\openone + Z_3)\ket{\psi}_3\ket{000} +  (\openone + Z_1)(\openone + Z_2)X_3(\openone - Z_3)\ket{\psi}_3\ket{001}  \nonumber\\
        &&+ (\openone + Z_1)X_2(\openone - Z_2)(\openone + Z_3)\ket{\psi}_3\ket{010} + (\openone + Z_1)X_2(\openone - Z_2)X_3(\openone - Z_3)\ket{\psi}_3\ket{011} \nonumber \\
        &&+ X_1(\openone - Z_1)(\openone + Z_2)(\openone + Z_3)\ket{\psi}_3\ket{100} + X_1(\openone - Z_1)(\openone + Z_2)X_3(\openone - Z_3)\ket{\psi}_3\ket{101} \nonumber \\
        &&+ X_1(\openone - Z_1)X_2(\openone - Z_2)(\openone + Z_3)\ket{\psi}_3\ket{110}  + X_1(\openone - Z_1)X_2(\openone - Z_2)X_3(\openone - Z_3)\ket{\psi}_3\ket{111}\Big] \label{iso3a}
\end{eqnarray}
This can be expressed in a more compact form as
\begin{equation}\label{isometry_output_apx}
    \ket{\Psi} =  \Phi \ \qty[\ket{\psi}_{3} \otimes \ket{0}^{\otimes 3}] = \frac{1}{2^3} \sum_{a_{k'} \in \{0,1\}}  \qty(\Motimes_{k=1}^3   \qty(X_k)^{a_{k'=k}}\qty[\openone + (-1)^{a_{k'=k}}Z_k]) \ket{\psi}_{3}\qty(\Motimes_{k'={1'}}^{3'} \ket{a_{k'}}_{k'})
\end{equation}
Following the same technique, the output can be evaluated to self-testing of observables when the input of the isometry is $\mathscr{A}^l \ket{\psi}_{m}$ where $\mathscr{A}^{l=0} = \openone$, $\mathscr{A}^{l=1} = Z_k$, $\mathscr{A}^{l=2} = X_k$ with $k\in[3]$
\begin{equation}\label{isooutapx}
    \ket{\Psi} =  \Phi \ \qty[\mathscr{A}^l \ket{\psi}_{3} \otimes \ket{0}^{\otimes 3}] = \frac{1}{2^3} \sum_{a_{k'} \in \{0,1\}}  \qty(\Motimes_{k=1}^3   \qty(X_k)^{a_{k'=k}}\qty[\openone + (-1)^{a_{k'=k}}Z_k]) \mathscr{A}^l \ket{\psi}_{3}\qty(\Motimes_{k'={1'}}^{3'} \ket{a_{k'}}_{k'})
\end{equation}


\section{Self-testing using swap circuit for $m=3$} \label{sss3}

Using SOS method we have already derived self-testing relations (See Eq.~(\ref{three_party_L}). For $m=3$, rewriting these relations in terms of $X_i$ and $Z_i$ as defined in Eq.~(\ref{XZ_relation}), we obtain the following relations
\begin{equation} \label{st1}
    (i)  \ X_1 \ket{\psi}_3 = Z_2  Z_3\ket{\psi}_3 \ ; \ (ii) \ Z_1 \ket{\psi}_3 = -Z_2 X_3\ket{\psi}_3 \ ; \ (iii)  \ Z_1  \ket{\psi}_3 = -X_2 Z_3\ket{\psi}_3 \ ; \ (iv) \ X_1 \ket{\psi}_3 = -X_2 X_3\ket{\psi}_3
\end{equation}
Using these relations given by Eq.~(\ref{st1}), we further derive the following relations 
\begin{equation} \label{st2}
\begin{aligned}
 (v) & \ X_2 \ket{\psi}_3 = -Z_1Z_3 \ket{\psi}_3 = -X_1  X_3\ket{\psi}_3 \ ;  & (vi) \ X_3 \ket{\psi}_3 = -Z_1 Z_3\ket{\psi}_3=-X_1 X_3\ket{\psi}_3 \ ; \\
 (vii) & \ Z_2 \ket{\psi}_3 = X_1 Z_3\ket{\psi}_3 = -Z_1X_3 \ket{\psi}_3\ ;  & (viii) \  Z_3  \ket{\psi}_3 = -Z_1X_2\ket{\psi}_3 =  X_1 Z_2\ket{\psi}_3 \ ;
\end{aligned}
\end{equation}

\subsection{Self-testing of state for $m=3$}
Employing eight relations given by Eqs.~(\ref{st1}) and (\ref{st2}), we proceed to evaluate the isometry output in Eq.~(\ref{iso3a}) term by term in the following way.
\begin{equation}\label{iso3a1}
    \begin{aligned}
     \qty(\openone + Z_1)\qty(\openone + Z_2)X_3\qty(\openone - Z_3)\ket{\psi}_{3}\ket{001} &=
     \qty(\openone + Z_1)\qty(\openone + Z_2)\qty(X_3 - X_3Z_3)\ket{\psi}_{3}\ket{001} \\
    &=\qty(\openone + Z_1)\qty(\openone + Z_2)\qty(\openone + Z_3)X_3\ket{\psi}_{3}\ket{001}\ \ \ \text{[from $\{Z_3,X_3\} = 0$]}\\
    &=\qty(\openone + Z_2)\qty(\openone + Z_3)X_3\qty(\openone + Z_1)\ket{\psi}_{3}\ket{001} \\
    &= \qty(\openone + Z_2)\qty(\openone + Z_3)\qty(X_3 + Z_1X_3)\ket{\psi}_{3}\ket{001} \\
    &= \qty(\openone + Z_2)\qty(\openone + Z_3)\qty(-Z_1Z_2 - Z_2)\ket{\psi}_{3}\ket{001} \ \ \ \text{[from (vi), (vii) in \ref{st2}]} \\
    &= -\qty(\openone + Z_2)\qty(\openone + Z_3)\qty(\openone + Z_1)Z_2\ket{\psi}_{3}\ket{001}\\
    &= -\qty(\openone + Z_1)\qty(\openone + Z_2)Z_2\qty(\openone + Z_3)\ket{\psi}_{3}\ket{001}
    \end{aligned}
\end{equation}
The other simplified terms of the isometry output are given as follows
\begin{equation}\label{iso3a2}
    \begin{aligned}
     (\openone + Z_1)X_2(\openone - Z_2)(\openone + Z_3)\ket{\psi}_3\ket{010}
    & \ = \ -\qty(\openone + Z_1)\qty(\openone + Z_2)\qty(\openone + Z_3)\ket{\psi}_3\ket{010}\\
    (\openone + Z_1)X_2(\openone - Z_2)X_3(\openone - Z_3)\ket{\psi}_3\ket{011}
    & \ = \ -\qty(\openone + Z_1)\qty(\openone + Z_2)\qty(\openone + Z_3)\ket{\psi}_3\ket{011} \\
    X_1(\openone - Z_1)(\openone + Z_2)(\openone + Z_3)\ket{\psi}_3\ket{100}
    & \ = \ \qty(\openone + Z_1)\qty(\openone + Z_2)\qty(\openone + Z_3)\ket{\psi}_3\ket{100}\\
    X_1(\openone - Z_1)(\openone + Z_2)X_3(\openone - Z_3)\ket{\psi}_3\ket{101}
    &  \ = \ \qty(\openone + Z_1)\qty(\openone + Z_2)\qty(\openone + Z_3)\ket{\psi}_3\ket{101}\\
    X_1(\openone - Z_1)X_2(\openone - Z_2)(\openone + Z_3)\ket{\psi}_3\ket{110}
    & \ = \ \qty(\openone + Z_1)\qty(\openone + Z_2)\qty(\openone + Z_3) \ket{\psi}_3\ket{110}\\
    X_1(\openone - Z_1)X_2(\openone - Z_2)X_3(\openone - Z_3)\ket{\psi}_3\ket{111}
    & \ = \ -\qty(\openone + Z_1)\qty(\openone + Z_2)\qty(\openone + Z_3) \ket{\psi}_3\ket{111}
    \end{aligned}
\end{equation}
Combining the simplified terms given by Eqs.~(\ref{iso3a1}) and (\ref{iso3a2}) result in the following output state of the isometry
\begin{equation}
        \Phi \ \Big(\ket{\psi}_{3}\otimes \ket{000}_{1'2'3'}\Big) = \frac{1}{2\sqrt{2}}\ket{\chi}_{3}\qty(\ket{000} - \ket{001} - \ket{010} - \ket{011} + \ket{100} + \ket{101} + \ket{110}- \ket{111})_{1'2'3'}
\end{equation}
where $\ket{\chi}_{3} = \frac{1}{2\sqrt{2}}\qty(\openone + Z_1)\qty(\openone + Z_2)\qty(\openone + Z_3)\ket{\psi}_{3}$. To show that the state obtained in the ancillary system is unitarily equivalent to the GHZ state, we construct the unitary $U = U_{1'}\otimes U_{2'} \otimes U_{3'}$ as
\begin{equation}\label{locuni}
    U_{1'} = \frac{1}{\sqrt{2}}\qty( \sigma_y + \sigma_z) \ \ ; \ \ U_{2'} = U_{3'} = \frac{1}{\sqrt{2}}\qty(\openone + i\sigma_x)
\end{equation}
Applying this unitary at the output of the ancillary systems, the output of the isometry becomes $U \Phi \ \Big(\ket{\psi}_{3}\otimes \ket{000}_{1'2'3'}\Big)=e^{-i\frac{\pi}{4}}\ket{\chi}_3\ket{GHZ}$ with $e^{-i\frac{\pi}{4}}$ being a global phase.


\subsection{Self-testing of observables for $m=3$} \label{sso3}

Let us first consider the self-testing of first party's observables. Using Eq.~(\ref{isooutapx}) and taking input state as $X_1 \ket{\psi}_{3}$, we derive
\begin{equation}
    \begin{aligned}
        \ket{\Psi} = \Phi\qty(X_1\ket{\psi}_{3}\ket{000}_{1'2'3'}) &=\frac{1}{8} \Big[ \qty(\openone + Z_1)\qty(\openone + Z_2)\qty(\openone + Z_3)X_1\ket{\psi}_3\ket{000} +  \qty(\openone + Z_1)\qty(\openone + Z_2)X_3\qty(\openone - Z_3)X_1\ket{\psi}_3\ket{001} \\
        &+ \qty(\openone + Z_1)X_2\qty(\openone - Z_2)\qty(\openone + Z_3)X_1\ket{\psi}_3\ket{010}+ \qty(\openone + Z_1)X_2\qty(\openone - Z_2)X_3\qty(\openone - Z_3)X_1\ket{\psi}_3\ket{011}\\
        &+ X_1\qty(\openone - Z_1)\qty(\openone + Z_2)\qty(\openone + Z_3)X_1\ket{\psi}_3\ket{100} + X_1\qty(\openone - Z_1)\qty(\openone + Z_2)X_3\qty(\openone - Z_3)X_1\ket{\psi}_3\ket{101} \\
        &+ X_1\qty(\openone - Z_1)X_2\qty(\openone - Z_2)\qty(\openone + Z_3)X_1\ket{\psi}_3\ket{110}+ X_1\qty(\openone - Z_1)X_2\qty(\openone - Z_2)X_3\qty(\openone - Z_3)X_1\ket{\psi}_3 \ket{111}\Big]
    \end{aligned}
\end{equation}

Employing the self-testing relations given by Eqs.~(\ref{st1}) and (\ref{st2}), we evaluate each of the term in the output and obtain
\begin{equation}
\begin{aligned}
    \left(\openone + Z_1\right)\left(\openone + Z_2\right)\left(\openone + Z_3\right)X_1\ket{\psi}_3\ket{000} &=2\sqrt{2}\ket{\chi}\ket{000}\\
    \qty(\openone + Z_1)\qty(\openone + Z_2)X_3\qty(\openone - Z_3)X_1\ket{\psi}_3\ket{001}
     &= 2\sqrt{2}\ket{\chi}\ket{001}\\
     \qty(\openone + Z_1)X_2\qty(\openone - Z_2)\qty(\openone + Z_3)X_1\ket{\psi}_3\ket{010}
        &= 2\sqrt{2}\ket{\chi}\ket{010}\\
        \qty(\openone + Z_1)X_2\qty(\openone - Z_2)X_3\qty(\openone - Z_3)X_1\ket{\psi}_3\ket{011}
        &= -2\sqrt{2}\ket{\chi}\ket{011}\\
        X_1\qty(\openone - Z_1)\qty(\openone + Z_2)\qty(\openone + Z_3)X_1\ket{\psi}_3\ket{100}
        &= 2\sqrt{2}\ket{\chi}\ket{100}\\
        X_1\qty(\openone - Z_1)\qty(\openone + Z_2)X_3\qty(\openone - Z_3)X_1\ket{\psi}_3\ket{101}
        &= -2\sqrt{2}\ket{\chi}\ket{101}\\
        X_1\qty(\openone - Z_1)X_2\qty(\openone - Z_2)\qty(\openone + Z_3)X_1\ket{\psi}_3\ket{110}\
        &= -2\sqrt{2}\ket{\chi}\ket{110}\\
        X_1\qty(\openone - Z_1)X_2\qty(\openone - Z_2)X_3\qty(\openone - Z_3)X_1\ket{\psi}_3\ket{111}
        &= -2\sqrt{2}\ket{\chi}\ket{111}
\end{aligned}
\end{equation}
The output then simplifies to 
\begin{equation}
    \ket{\Psi} = \Phi \ \Big(X_1\ket{\psi}_{3}\otimes \ket{000}_{1'2'3'}\Big) = \frac{1}{2\sqrt{2}}\ket{\chi}_{3}\qty(\ket{000} + \ket{001} + \ket{010} - \ket{011} + \ket{100} - \ket{101} - \ket{110} - \ket{111})_{1'2'3'}
\end{equation}
Applying the unitary $U$ as defined in Eq.~(\ref{locuni}) on the output, we obtain $U \Phi \ \Big(\ket{\psi'}_{123}\otimes \ket{000}_{1'2'3'}\Big)=e^{-i\frac{5\pi}{4}}\ket{\chi}\qty(\sigma_x\otimes\openone\otimes\openone)\ket{GHZ}_{1'2'3'}$ with $e^{-i\frac{5\pi}{4}}$ being a global phase.

Now, considering the input state as $Z_1 \ket{\psi}_3$, we derive
\begin{equation}
    \begin{aligned}
        \ket{\Psi} = \Phi\qty(Z_1\ket{\psi}_3\ket{000}) \ = \ & \frac{1}{8} \Big[ \qty(\openone + Z_1)\qty(\openone + Z_2)\qty(\openone + Z_3)Z_1\ket{\psi}_3\ket{000} +  \qty(\openone + Z_1)\qty(\openone + Z_2)X_3\qty(\openone - Z_3)Z_1\ket{\psi}_3\ket{001} \\
        &+ \qty(\openone + Z_1)X_2\qty(\openone - Z_2)\qty(\openone + Z_3)Z_1\ket{\psi}_3\ket{010}+ \qty(\openone + Z_1)X_2\qty(\openone - Z_2)X_3\qty(\openone - Z_3)Z_1\ket{\psi}_3\ket{011}\\
        &+ X_1\qty(\openone - Z_1)\qty(\openone + Z_2)\qty(\openone + Z_3)Z_1\ket{\psi}_3\ket{100} + X_1\qty(\openone - Z_1)\qty(\openone + Z_2)X_3\qty(\openone - Z_3)Z_1\ket{\psi}_3\ket{101} \\
        &+ X_1\qty(\openone - Z_1)X_2\qty(\openone - Z_2)\qty(\openone + Z_3)Z_1\ket{\psi}_3\ket{110}+ X_1\qty(\openone - Z_1)X_2\qty(\openone - Z_2)X_3\qty(\openone - Z_3)Z_1\ket{\psi}_3 \ket{111}\Big]
    \end{aligned}
\end{equation}
Employing the self-testing statements given by Eqs.~(\ref{st1}) and (\ref{st2}) , we obtain
\begin{equation}
\begin{aligned}
    \left(\openone + Z_1\right)\left(\openone + Z_2\right)\left(\openone + Z_3\right)Z_1\ket{\psi}_3\ket{000} &= 2\sqrt{2}\ket{\chi}\ket{000}\\
    \qty(\openone + Z_1)\qty(\openone + Z_2)X_3\qty(\openone - Z_3)Z_1\ket{\psi}_3\ket{001}
    &= -2\sqrt{2}\ket{\chi}\ket{001}\\
    \qty(\openone + Z_1)X_2\qty(\openone - Z_2)\qty(\openone + Z_3)Z_1\ket{\psi}_3\ket{010}
    &= -2\sqrt{2}\ket{\chi}\ket{010}\\
    \qty(\openone + Z_1)X_2\qty(\openone - Z_2)X_3\qty(\openone - Z_3)Z_1\ket{\psi}_3\ket{011}
    &= -2\sqrt{2}\ket{\chi}\ket{011}\\
    X_1\qty(\openone - Z_1)\qty(\openone + Z_2)\qty(\openone + Z_3)Z_1\ket{\psi}_3\ket{100}
    &= -2\sqrt{2}\ket{\chi}\ket{100}\\
    X_1\qty(\openone - Z_1)\qty(\openone + Z_2)X_3\qty(\openone - Z_3)Z_1\ket{\psi}_3\ket{101}
    &= -2\sqrt{2}\ket{\chi}\ket{101}\\
    X_1\qty(\openone - Z_1)X_2\qty(\openone - Z_2)\qty(\openone + Z_3)Z_1\ket{\psi}_3\ket{110}
    &= -2\sqrt{2}\ket{\chi}\ket{110}\\
    X_1\qty(\openone - Z_1)X_2\qty(\openone - Z_2)X_3\qty(\openone - Z_3)Z_1\ket{\psi}_3\ket{111}
    &= 2\sqrt{2}\ket{\chi}\ket{111}
\end{aligned}
\end{equation}
The output then simplifies to 
\begin{equation}
    \ket{\Psi} = \Phi \ \Big(Z_1\ket{\psi}_{3}\otimes \ket{000}_{1'2'3'}\Big) = \frac{1}{2\sqrt{2}}\ket{\chi}_{3}\qty(\ket{000} - \ket{001} - \ket{010} - \ket{011} - \ket{100} - \ket{101} - \ket{110}+ \ket{111})_{1'2'3'}
\end{equation}
Applying the unitary $U$ as defined in Eq.~(\ref{locuni}) on the output, we obtain the output of the isometry as $U \Phi \ \Big(\ket{\psi'}_{3}\otimes \ket{000}_{1'2'3'}\Big)=e^{-i\frac{\pi}{4}}\ket{\chi}_{3}\qty(\sigma_y\otimes\openone\otimes\openone)\ket{GHZ}_{1'2'3'}$ with $e^{-i\frac{\pi}{4}}$ being a global phase. Following the similar procedure as presented in the preceding evaluation, it is straightforward to derive the output of the isometry corresponding to input states $X_2\ket{\psi}_3$, $Z_2\ket{\psi}_3$, $X_3\ket{\psi}_3$ and $Z_3\ket{\psi}_3$ 

\begin{equation} \label{eq3mo1}
    \begin{aligned}
\Phi \qty(X_1\ket{\psi}_3\ket{000}) &= \ket{\chi} \otimes \qty(\sigma_x \otimes \openone \otimes \openone) \ket{GHZ} \\
\Phi \qty(Z_2\ket{\psi}_3\ket{000}) &= \ket{\chi} \otimes \qty(\openone \otimes \sigma_y  \otimes \openone) \ket{GHZ} \\
\Phi \qty(X_2\ket{\psi}_3\ket{000}) &= \ket{\chi} \otimes \qty(\openone \otimes \sigma_x \otimes \openone ) \ket{GHZ} \\
\Phi \qty(Z_3\ket{\psi}_3\ket{000}) &= \ket{\chi} \otimes \qty(\openone \otimes \openone \otimes \sigma_y) \ket{GHZ} \\              
\Phi \qty(X_3\ket{\psi}_3\ket{000}) &= \ket{\chi} \otimes \qty(\openone \otimes \openone \otimes \sigma_x) \ket{GHZ} 
    \end{aligned}
\end{equation}


\section{Evaluation of Isometry output for $m$=4 case} \label{isoo4}
The isometry presented in Fig. \ref{figswap} for $m=4$ outputs the following
\begin{equation}
    \Phi\qty(\ket{\psi}_4\ket{0000}) = \textbf{C}_X \cdot \textbf{H} \cdot \textbf{C}_Z \cdot \textbf{H} \ket{\psi}_4\ket{0000}
\end{equation}
where $\textbf{C}_X=C_{X_1}(1') \otimes C_{X_2}(2') \otimes C_{X_3}(3') \otimes C_{X_4}(4') $, $\textbf{C}_Z=C_{Z_1}(1') \otimes C_{Z_2}(2') \otimes C_{Z_3}(3')\otimes C_{Z_4}(4')$ and $\textbf{H}=H_{1'}\otimes H_{2'} \otimes H_{3'} \otimes H_{4'}$. Here $C_{X_i}(j')$ and $C_{Z_i}(j')$ are the controlled gates operating on the physical system's appropriate Hilbert space with control on the $j'$-th ancillary system. Applying the Hadamard on each ancilla, we get 
\begin{equation}\label{m4ha}
    \begin{aligned}
        \textbf{H} \ket{\psi}_4\ket{0000} = \frac{1}{4}\ket{\psi}_4\Big(&\ket{0000} + \ket{0001} + \ket{0010} + \ket{0011} + \ket{0100} + \ket{0101} + \ket{0110} + \ket{0111}\\
        &+\ket{1000} + \ket{1001} + \ket{1010} + \ket{1011}+\ket{1100} + \ket{1101} + \ket{1110} + \ket{1111}\Big)
    \end{aligned}
\end{equation}
After application of $\textbf{C}_Z$, the right hand side of Eq.~(\ref{m4ha}) becomes  
    \begin{align*}
        \frac{1}{4}\Big(&\ket{\psi}_4\ket{0000} + Z_4\ket{\psi}_4\ket{0001} + Z_3\ket{\psi}_4\ket{0010} + Z_3Z_4\ket{\psi}_4\ket{0011} + Z_2\ket{\psi}_4\ket{0100} + Z_2Z_4\ket{\psi}_4\ket{0101} + Z_2Z_3\ket{\psi}_4\ket{0110} + Z_2Z_3Z_4\ket{\psi}_4\ket{0111}\\
        &+Z_1\ket{\psi}_4\ket{1000} + Z_1Z_4\ket{\psi}_4\ket{1001} + Z_1Z_3\ket{\psi}_4\ket{1010} + Z_1Z_3Z_4\ket{\psi}_4\ket{1011}+Z_1Z_2\ket{\psi}_4\ket{1100} + Z_1Z_2Z_4\ket{\psi}_4\ket{1101} + Z_1Z_2Z_3\ket{\psi}_4\ket{1110} \\
        &+ Z_1Z_2Z_3Z_4\ket{\psi}_4\ket{1111}\Big)
    \end{align*}
Similar to the $3-$party case, we will perform term-by-term calculations to obtain
\begin{eqnarray}
    \ket{\Psi} &=& \frac{1}{16} \Big[ (\openone + Z_1)(\openone + Z_2)(\openone + Z_3)(\openone + Z_4)\ket{\psi}_4\ket{0000} + (\openone + Z_1)(\openone + Z_2)(\openone + Z_3)X_4(\openone - Z_4)\ket{\psi}_4\ket{0001} \nonumber \\
        &+& (\openone + Z_1)(\openone + Z_2)X_3(\openone - Z_3)(\openone + Z_4)\ket{\psi}_4\ket{0010}  + (\openone + Z_1)(\openone + Z_2)X_3(\openone - Z_3)X_4(\openone - Z_4)\ket{\psi}_4\ket{0011} \nonumber \\
        &+& (\openone + Z_1)X_2(\openone - Z_2)(\openone + Z_3)(\openone + Z_4)\ket{\psi}_4\ket{0100} + (\openone + Z_1)X_2(\openone - Z_2)(\openone + Z_3)X_4(\openone - Z_4)\ket{\psi}_4\ket{0101} \nonumber \\
        &+& (\openone + Z_1)X_2(\openone - Z_2)X_3(\openone + Z_3)(\openone + Z_4)\ket{\psi}_4\ket{0110} + (\openone +Z_1)X_2(\openone - Z_2)X_3(\openone - Z_3)X_4(\openone - Z_4)\ket{\psi}_4\ket{0111} \nonumber \\
        &+& X_1(\openone - Z_1)(\openone + Z_2)(\openone + Z_3)(\openone + Z_4)\ket{\psi}_4\ket{1000} + X_1(\openone - Z_1)(\openone + Z_2)(\openone + Z_3)X_4(\openone - Z_4)\ket{\psi}_4\ket{1001} \nonumber \\
        &+& X_1(\openone - Z_1)(\openone + Z_2)X_3(\openone - Z_3)(\openone + Z_4)\ket{\psi}_4\ket{1010} + X_1(\openone - Z_1)(\openone + Z_2)X_3(\openone - Z_3)X_4(\openone - Z_4)\ket{\psi}_4\ket{1011} \nonumber \\
        &+& X_1(\openone - Z_1)X_2(\openone - Z_2)(\openone + Z_3)(\openone + Z_4)\ket{\psi}_4\ket{1100} + X_1(\openone - Z_1)X_2(\openone - Z_2)(\openone + Z_3)X_4(\openone - Z_4)\ket{\psi}_4\ket{1101} \nonumber\\
        &+& X_1(\openone - Z_1)X_2(\openone - Z_2)X_3(\openone - Z_3)(\openone + Z_4)\ket{\psi}_4\ket{1110} + X_1(\openone - Z_1)X_2(\openone - Z_2)X_3(\openone - Z_3)X_4(\openone - Z_4)\ket{\psi}_4\ket{1111}\Big] \label{rhoiso4}
\end{eqnarray}
The above Eq.~(\ref{rhoiso4}) can be expressed in the following compact form
\begin{equation}\label{isooutapx}
    \ket{\Psi} =  \Phi \ \qty[\ket{\psi}_{4} \otimes \ket{0}^{\otimes 4}] = \frac{1}{2^4} \sum_{a_{k'} \in \{0,1\}}  \qty(\Motimes_{k=1}^4   \qty(X_k)^{a_{k'=k}}\qty[\openone + (-1)^{a_{k'=k}}Z_k]) \ket{\psi}_{4}\qty(\Motimes_{k'={1'}}^{4'} \ket{a_{k'}}_{k'})
\end{equation}
This output can be generalised to self-testing observables when the input of the isometry is $\mathscr{A}^l \ket{\psi}_{m}$ where $\mathscr{A}^{l=0} = \openone$, $\mathscr{A}^{l=1} = Z_k$, $\mathscr{A}^{l=2} = X_k$ with $k\in[4]$
\begin{equation}\label{isooutoapx}
    \ket{\Psi} =  \Phi \ \qty[\mathscr{A}^l \ket{\psi}_{4} \otimes \ket{0}^{\otimes 4}] = \frac{1}{2^4} \sum_{a_{k'} \in \{0,1\}}  \qty(\Motimes_{k=1}^4   \qty(X_k)^{a_{k'=k}}\qty[\openone + (-1)^{a_{k'=k}}Z_k]) \mathscr{A}^l \ket{\psi}_{4}\qty(\Motimes_{k'={1'}}^{4'} \ket{a_{k'}}_{k'})
\end{equation}


\section{Self-testing using swap circuit for $m=4$} \label{sss4}

Similar to $3-$party case, to evaluate the isometry, the following self-testing relations are required which where obtained by restating Eq.~(\ref{four_party_L}) in terms of $X_i$ and $Z_i$ as defined in Eq.~(\ref{XZ_relation})
\begin{equation} \label{4party_self_testing}
    \begin{aligned}
        (i)  \ X_1 \ket{\psi}_4 = Z_2  Z_3 Z_4\ket{\psi}_4 \ ; \ (ii) \ Z_1 \ket{\psi}_4 = -Z_2 Z_3 X_4\ket{\psi}_4 \ ; \ (iii)  \ Z_1  \ket{\psi}_4 = -Z_2 X_3 Z_4\ket{\psi}_4 \ ; \ (iv) \ X_1 \ket{\psi}_4 = - Z_2 X_3 X_4\ket{\psi}_4 \\ 
        (v)  \ Z_1 \ket{\psi}_4 = -X_2  Z_3 Z_4\ket{\psi}_4 \ ; \ (vi) \ X_1 \ket{\psi}_4 = -X_2 Z_3 X_4\ket{\psi}_4 \ ; \ (vii)  \ X_1  \ket{\psi}_4 = -X_2 X_3 Z_4\ket{\psi}_4 \ ; \ (viii) \ Z_1 \ket{\psi}_4 =  X_2 X_3 X_4\ket{\psi}_4 \\ 
    \end{aligned}
\end{equation}
Using self-testing relations given by Eq.~(\ref{st1}), we further obtain the following relations
\begin{equation} \label{st21}
\begin{aligned}
 &(v)  \ X_2 \ket{\psi}_4 = -Z_1Z_3Z_4 \ket{\psi}_4 = -X_1Z_3Z_4\ket{\psi}_4 = -X_1X_3Z_4\ket{\psi}_4 = Z_1X_3X_4\ket{\psi}_4\ ;   \\
 &(vi) \ Z_2 \ket{\psi}_4 = X_1 Z_3 Z_4\ket{\psi}_4 = -Z_1Z_3X_4 \ket{\psi}_4  = -Z_1Z_3X_4 \ket{\psi}_4 = -X_1X_3X_4\ket{\psi}_4\ ; \\
  & (vii)  \ X_3 \ket{\psi}_4 = -Z_1Z_2 Z_4\ket{\psi}_4 = -X_1 Z_2 X_4\ket{\psi}_4 = -X_1X_2 Z_4\ket{\psi}_4 = Z_1 X_2 X_4\ket{\psi}_4 \ ; \\
  & (viii)  \  Z_3  \ket{\psi}_4 = X_1Z_2Z_4\ket{\psi}_4 =  -Z_1Z_2X_4\ket{\psi}_4 = -Z_1X_2Z_4\ket{\psi}_4 =  -X_1X_2X_4\ket{\psi}_4 \ ; \\
  & (ix)  \ X_4 \ket{\psi}_4 = -Z_1Z_2 Z_3\ket{\psi}_4 = -X_1 Z_2 X_3\ket{\psi}_4 = -X_1X_2 Z_3\ket{\psi}_4 = X_2 X_3 X_4 \ket{\psi}_4 \ ; \\
  & (x)  \  Z_4 \ket{\psi}_4 = X_1Z_2Z_3\ket{\psi}_4 =  -Z_1Z_2X_3\ket{\psi}_4 = -Z_1X_2Z_3\ket{\psi}_4 =  -X_1X_2X_3\ket{\psi}_4 \ ;
\end{aligned}
\end{equation}


\subsection{Self-testing of state for $m=4$}

Employing the relations given by Eqs.~(\ref{4party_self_testing}) and (\ref{st21}), we evaluate the isometry output given in Eq.~(\ref{rhoiso4}) as
\begin{equation}\label{Cfirsteqn}
\begin{aligned}
    \left(\openone + Z_1\right)\left(\openone + Z_2\right)\left(\openone + Z_3\right)\left(\openone + Z_4\right)\ket{\psi}_4\ket{0000} &=4\ket{\chi}_4\ket{0000}\\
    \qty(\openone + Z_1)\qty(\openone + Z_2)\qty(\openone + Z_3)X_4\qty(\openone - Z_4)\ket{\psi}_4\ket{0001} &= -4\ket{\chi}_4\ket{0001}\\
    \qty(\openone + Z_1)\qty(\openone + Z_2)X_3\qty(\openone - Z_3)\qty(\openone + Z_4)\ket{\psi}_4\ket{0010} &= -4\ket{\chi}_4\ket{0010}\\
    \qty(\openone + Z_1)\qty(\openone + Z_2)X_3\qty(\openone - Z_3)X_4\qty(\openone - Z_4)\ket{\psi}_4\ket{0011} 
     &= -4\ket{\chi}_4\ket{0011}\\
     \qty(\openone + Z_1)Z_2\qty(\openone - Z_2)X_3\qty(\openone + Z_3)\qty(\openone + Z_4)\ket{\psi}_4\ket{0100} 
     &= -4\ket{\chi}_4\ket{0100}\\
     \qty(\openone + Z_1)Z_2\qty(\openone - Z_2)\qty(\openone + Z_3)X_4\qty(\openone - Z_4)\ket{\psi}_4\ket{0101}
     &= -4\ket{\chi}_4\ket{0101}\\
     \qty(\openone + Z_1)Z_2\qty(\openone - Z_2)X_3\qty(\openone - Z_3)\qty(\openone + Z_4)\ket{\psi}_4\ket{0110} 
     &= -4\ket{\chi}_4\ket{0110}\\
     \qty(\openone + Z_1)Z_2\qty(\openone - Z_2)X_3\qty(\openone - Z_3)X_4\qty(\openone - Z_4)\ket{\psi}_4\ket{0111} 
     &= 4\ket{\chi}_4\ket{0111}\\
     X_1\qty(\openone - Z_1)\qty(\openone + Z_2)\qty(\openone + Z_3)\qty(\openone + Z_4)\ket{\psi}_4\ket{1000} 
     &= 4\ket{\chi}_4\ket{1000}\\
     X_1\qty(\openone - Z_1)\qty(\openone + Z_2)\qty(\openone + Z_3)X_4\qty(\openone - Z_4)\ket{\psi}_4\ket{1001} 
     &= 4\ket{\chi}_4\ket{1001}\\
     X_1\qty(\openone - Z_1)\qty(\openone + Z_2)X_3\qty(\openone - Z_3)\qty(\openone + Z_4)\ket{\psi}_4\ket{1010}
     &= 4\ket{\chi}_4\ket{1010}\\
     X_1\qty(\openone - Z_1)\qty(\openone + Z_2)X_3\qty(\openone - Z_3)X_4\qty(\openone - Z_4)\ket{\psi}_4\ket{1011} 
     &= -4\ket{\chi}_4\ket{1011}\\
     X_1\qty(\openone - Z_1)Z_2\qty(\openone - Z_2)X_3\qty(\openone + Z_3)\qty(\openone + Z_4)\ket{\psi}_4\ket{1100} 
     &= 4\ket{\chi}_4\ket{1100}\\
     X_1\qty(\openone - Z_1)Z_2\qty(\openone - Z_2)X_3\qty(\openone + Z_3)X_4\qty(\openone - Z_4)\ket{\psi}_4\ket{1101}
     &= -4\ket{\chi}_4\ket{1101}\\
     X_1\qty(\openone - Z_1)Z_2\qty(\openone - Z_2)X_3\qty(\openone - Z_3)\qty(\openone + Z_4)\ket{\psi}_4\ket{1110} 
     &= -4\ket{\chi}_4\ket{1110}\\
     X_1\qty(\openone - Z_1)Z_2\qty(\openone - Z_2)X_3\qty(\openone - Z_3)X_4\qty(\openone - Z_4)\ket{\psi}_4\ket{1111}
     &= -4\ket{\chi}_4\ket{1111}
\end{aligned}
\end{equation}
where $\ket{\chi}_4 = \frac{1}{4}\left(\openone + Z_1\right)\left(\openone + Z_2\right)\left(\openone + Z_3\right)\left(\openone + Z_4\right)\ket{\psi}_4$. Combining the terms in Eq.~(\ref{Cfirsteqn}), we obtain
\begin{equation}
\begin{aligned}
\Phi\qty(\ket{\psi}_{4}\otimes\ket{0000}) &= \frac{1}{4} \ket{\chi}_4\big( \ket{0000} - \ket{0001} - \ket{0010} - \ket{0011} - \ket{0100} - \ket{0101} - \ket{0110} + \ket{0111}  \\& + \ket{1000} + \ket{1001} + \ket{1010} - \ket{1011} + \ket{1100} - \ket{1101} - \ket{1110} - \ket{1111}\big)
\end{aligned}
\end{equation}
Taking the unitary $U=U_{1'}\otimes U_{2'}\otimes U_{3'} \otimes U_{4'}$ with
\begin{equation}\label{locuni4}
    U_{1'} = \frac{1}{\sqrt{2}}\qty( \sigma_y + \sigma_z) \ \ ; \ \ U_{2'} = U_{3'} = \frac{1}{\sqrt{2}}\qty(\openone + i\sigma_x) \ \ ; \ \ U_{4'} = \frac{1}{2} \qty(\openone + i\sigma_x - i\sigma_y + i\sigma_z)  
\end{equation}
Applying this unitary at the output of the ancillary systems, the output of the isometry becomes $U \Phi\qty(\ket{\psi}_{4}\otimes\ket{0000}) = \ket{\chi}_4\ket{GHZ}$, where $\ket{GHZ} = \frac{1}{\sqrt{2}}\qty(\ket{0000}+\ket{1111})$.


\subsection{Self-testing of observables for $m=4$} \label{sso4}

To self-test the observables $Z_1$, From Eq.~(\ref{isooutoapx}), we obtain
\begin{eqnarray}
        \ket{\Psi} &=& \qty[Z_1 \ket{\psi}_4 \otimes \ket{0000}] = \frac{1}{16} \Big[ (\openone + Z_1)(\openone + Z_2)(\openone + Z_3)(\openone + Z_4)Z_1\ket{\psi}_4\ket{0000} + (\openone + Z_1)(\openone + Z_2)(\openone + Z_3)X_4(\openone - Z_4)Z_1\ket{\psi}_4\ket{0001} \nonumber\\
        &+& (\openone + Z_1)(\openone + Z_2)X_3(\openone - Z_3)(\openone + Z_4)Z_1\ket{\psi}_4\ket{0010}  + (\openone + Z_1)(\openone + Z_2)X_3(\openone - Z_3)X_4(\openone - Z_4)Z_1\ket{\psi}_4\ket{0011} \nonumber \\
        &+& (\openone + Z_1)X_2(\openone - Z_2)(\openone + Z_3)(\openone + Z_4)Z_1\ket{\psi}_4\ket{0100} + (\openone + Z_1)X_2(\openone - Z_2)(\openone + Z_3)X_4(\openone - Z_4)Z_1\ket{\psi}_4\ket{0101} \nonumber \\
        &+& (\openone + Z_1)X_2(\openone - Z_2)X_3(\openone + Z_3)(\openone + Z_4)Z_1\ket{\psi}_4\ket{0110} + (\openone +Z_1)X_2(\openone - Z_2)X_3(\openone - Z_3)X_4(\openone - Z_4)Z_1\ket{\psi}_4\ket{0111}  \nonumber\\
        &+& X_1(\openone - Z_1)(\openone + Z_2)(\openone + Z_3)(\openone + Z_4)Z_1\ket{\psi}_4\ket{1000} + X_1(\openone - Z_1)(\openone + Z_2)(\openone + Z_3)X_4(\openone - Z_4)Z_1\ket{\psi}_4\ket{1001}  \\
        &+& X_1(\openone - Z_1)(\openone + Z_2)X_3(\openone - Z_3)(\openone + Z_4)Z_1\ket{\psi}_4\ket{1010} + X_1(\openone - Z_1)(\openone + Z_2)X_3(\openone - Z_3)X_4(\openone - Z_4)Z_1\ket{\psi}_4\ket{1011} \nonumber \\
        &+& X_1(\openone - Z_1)X_2(\openone - Z_2)(\openone + Z_3)(\openone + Z_4)Z_1\ket{\psi}_4\ket{1100} + X_1(\openone - Z_1)X_2(\openone - Z_2)(\openone + Z_3)X_4(\openone - Z_4)Z_1\ket{\psi}_4\ket{1101} \nonumber\\
        &+& X_1(\openone - Z_1)X_2(\openone - Z_2)X_3(\openone - Z_3)(\openone + Z_4)Z_1\ket{\psi}_4\ket{1110} + X_1(\openone - Z_1)X_2(\openone - Z_2)X_3(\openone - Z_3)X_4(\openone - Z_4)Z_1\ket{\psi}_4\ket{1111}\Big] \nonumber
\end{eqnarray}
Each of the term in the above expression can be simplified using Eq.~(\ref{4party_self_testing}) to obtain the following expression for the output
\begin{equation}
    \begin{aligned}
        \Phi\qty(Z_1\ket{\psi}_4\otimes\ket{0000}) &= \frac{1}{4} \ket{\chi}_4\big( \ket{0000} - \ket{0001} - \ket{0010} - \ket{0011} - \ket{0100} - \ket{0101} - \ket{0110} + \ket{0111}  \\& - \ket{1000} - \ket{1001} - \ket{1010} + \ket{1011} - 
        \ket{1100} + \ket{1101} + \ket{1110} + \ket{1111}\big)
    \end{aligned}
\end{equation}
Applying the local unitary defined in Eq.~(\ref{locuni4}), we obtain the following output 
\begin{equation}
    U\Phi\qty(Z_1\ket{\psi}_4\otimes\ket{0000}) = \ket{\chi}_4\qty(\sigma_y \otimes \openone \otimes \openone \otimes \openone) \frac{1}{\sqrt{2}}\qty(\ket{0000}+\ket{1111})
\end{equation}
To self-test the observables $X_1$ we take the input state of the isometry as $X_1\ket{\psi}_4$ which gives the following
\begin{equation}
    \begin{aligned}
        \ket{\Psi} &= \frac{1}{16} \Big[ (\openone + Z_1)(\openone + Z_2)(\openone + Z_3)(\openone + Z_4)X_1\ket{\psi}_4\ket{0000} + (\openone + Z_1)(\openone + Z_2)(\openone + Z_3)X_4(\openone - Z_4)X_1\ket{\psi}_4\ket{0001} \\
        &+ (\openone + Z_1)(\openone + Z_2)X_3(\openone - Z_3)(\openone + Z_4)X_1\ket{\psi}_4\ket{0010}  + (\openone + Z_1)(\openone + Z_2)X_3(\openone - Z_3)X_4(\openone - Z_4)X_1\ket{\psi}_4\ket{0011}  \\
        &+ (\openone + Z_1)X_2(\openone - Z_2)(\openone + Z_3)(\openone + Z_4)X_1\ket{\psi}_4\ket{0100} + (\openone + Z_1)X_2(\openone - Z_2)(\openone + Z_3)X_4(\openone - Z_4)X_1\ket{\psi}_4\ket{0101}  \\
        &+ (\openone + Z_1)X_2(\openone - Z_2)X_3(\openone + Z_3)(\openone + Z_4)X_1\ket{\psi}_4\ket{0110} + (\openone +Z_1)X_2(\openone - Z_2)X_3(\openone - Z_3)X_4(\openone - Z_4)X_1\ket{\psi}_4\ket{0111}  \\
        &+ X_1(\openone - Z_1)(\openone + Z_2)(\openone + Z_3)(\openone + Z_4)X_1\ket{\psi}_4\ket{1000} + X_1(\openone - Z_1)(\openone + Z_2)(\openone + Z_3)X_4(\openone - Z_4)X_1\ket{\psi}_4\ket{1001}  \\
        &+ X_1(\openone - Z_1)(\openone + Z_2)X_3(\openone - Z_3)(\openone + Z_4)X_1\ket{\psi}_4\ket{1010} + X_1(\openone - Z_1)(\openone + Z_2)X_3(\openone - Z_3)X_4(\openone - Z_4)X_1\ket{\psi}_4\ket{1011}  \\
        &+ X_1(\openone - Z_1)X_2(\openone - Z_2)(\openone + Z_3)(\openone + Z_4)X_1\ket{\psi}_4\ket{1100} + X_1(\openone - Z_1)X_2(\openone - Z_2)(\openone + Z_3)X_4(\openone - Z_4)X_1\ket{\psi}_4\ket{1101} \\
        &+ X_1(\openone - Z_1)X_2(\openone - Z_2)X_3(\openone - Z_3)(\openone + Z_4)X_1\ket{\psi}_4\ket{1110} + X_1(\openone - Z_1)X_2(\openone - Z_2)X_3(\openone - Z_3)X_4(\openone - Z_4)X_1\ket{\psi}_4\ket{1111}\Big] 
    \end{aligned}
\end{equation}
Again, simplifying the output using the self-testing relations, we get 
\begin{equation}
    \begin{aligned}
        \Phi\qty(Z_1\ket{\psi}_4\otimes\ket{0000}) &= \frac{1}{4} \ket{\chi}_4\big( \ket{0000} + \ket{0001} + \ket{0010} - \ket{0011} - \ket{0100} - \ket{0101} - \ket{0110} - \ket{0111}  \\& + \ket{1000} - \ket{1001} - \ket{1010} - \ket{1011} - 
        \ket{1100} - \ket{1101} - \ket{1110} + \ket{1111}\big)
    \end{aligned}
\end{equation}
Applying the local unitary defined in Eq.~(\ref{locuni4}), we obtain the following output. 
\begin{equation}
    U\Phi\qty(X_1\ket{\psi}_4\otimes\ket{0000}) = \ket{\chi}_4\qty(\sigma_x \otimes \openone \otimes \openone \otimes \openone) \frac{1}{\sqrt{2}}\qty(\ket{0000}+\ket{1111})
\end{equation}
Similarly, other observables can be self-tested using the swap circuit.
\begin{equation} \label{eq4mo1}
    \begin{aligned}
        \Phi \qty(Z_1\ket{\psi}_4\ket{0000}) &= \ket{\chi}_4 \otimes \qty(\sigma_y \otimes \openone \otimes \openone\otimes \openone) \ket{GHZ} \\
        \Phi \qty(X_1\ket{\psi}_4\ket{0000}) &= \ket{\chi}_4 \otimes \qty(\sigma_x \otimes \openone \otimes \openone\otimes \openone) \ket{GHZ} \\
        \Phi \qty(Z_2\ket{\psi}_4\ket{0000}) &= \ket{\chi}_4 \otimes \qty(\openone \otimes \sigma_x \otimes \openone\otimes \openone ) \ket{GHZ} \\
        \Phi \qty(X_2\ket{\psi}_4\ket{0000}) &= \ket{\chi}_4 \otimes \qty(\openone \otimes \sigma_y \otimes \openone \otimes \openone) \ket{GHZ} \\              
        \Phi \qty(Z_3\ket{\psi}_4\ket{0000}) &= \ket{\chi}_4 \otimes \qty(\openone \otimes \openone \otimes \sigma_x\otimes \openone) \ket{GHZ} \\
        \Phi \qty(X_3\ket{\psi}_4\ket{0000}) &= \ket{\chi}_4 \otimes \qty( \openone \otimes \openone \otimes \sigma_y \otimes \openone) \ket{GHZ} \\
        \Phi \qty(Z_4\ket{\psi}_4\ket{0000}) &= \ket{\chi}_4 \otimes \qty(\openone \otimes \openone \otimes \openone\otimes \sigma_x) \ket{GHZ} \\
        \Phi \qty(X_4\ket{\psi}_4\ket{0000}) &= \ket{\chi}_4 \otimes \qty(\openone \otimes \openone \otimes \openone\otimes \sigma_y) \ket{GHZ}
    \end{aligned}
\end{equation}


\section{Detailed calculations for robust self-testing of state and observables} \label{rsca}

\begin{equation}\label{h1}
 \norm{\qty(\tilde{X}_i - X_i)\ket{\psi}} \leq \xi_i \implies \tilde{X}_i \approx \xi_i\openone + X_i \ ;  \ \ \ \ \ \ \norm{\qty(\tilde{Z}_i - Z_i)\ket{\psi}} \leq \zeta_i \implies \tilde{Z}_i \approx \zeta_i\openone + Z_i \ \ \forall i\in[m] \ ;
\end{equation}
Using Eq.~(\ref{h1}), following relations are derived
\begin{eqnarray}
     \norm{\qty(\tilde{X}_\gamma \tilde{X}_\delta - X_\gamma X_\delta)\ket{\psi}} &\approx& \norm{\qty(\tilde{X}_\gamma(\xi_\delta \openone + X_\delta) - X_\gamma X_\delta)\ket{\psi}} = \norm{\tilde{X}_\gamma \xi_\delta \ket{\psi} + \qty(\tilde{X}_\gamma X_\delta - X_\gamma X_\delta)\ket{\psi}} \nonumber \\
        &&\leq \xi_\delta \norm{\tilde{X}_\gamma \ket{\psi}} + \norm{\qty(\tilde{X}_\gamma X_\delta - X_\gamma X_\delta)\ket{\psi}} = \xi_\delta \norm{\tilde{X}_\gamma \ket{\psi}} + \norm{\qty(\tilde{X}_\gamma  - X_\gamma) X_\delta\ket{\psi}} = \xi_\gamma + \xi_\delta \norm{\tilde{X}_\gamma \ket{\psi}} \nonumber \\
       \norm{\qty(\tilde{X}_\gamma \tilde{X}_\delta\tilde{X}_\eta - X_\gamma X_\delta X_\eta)\ket{\psi}} &\approx& \norm{\qty(\tilde{X}_\gamma\tilde{X}_\delta(\xi_\eta \openone + X_\eta) - X_\gamma X_\delta X_\eta)\ket{\psi}} = \norm{\qty(\tilde{X}_\gamma\tilde{X}_\delta\xi_\eta)\ket{\psi}  + \qty(\tilde{X}_\gamma\tilde{X}_\delta - X_\gamma X_\delta )X_\eta)\ket{\psi}} \nonumber \\
        &&\leq \xi_\eta \norm{\tilde{X}_\gamma \tilde{X}_\delta \ket{\psi}} + \norm{\qty(\tilde{X}_\gamma \tilde{X}_\delta - X_\gamma X_\delta)\ket{\psi}}= \xi_\gamma + \xi_\delta \norm{\tilde{X}_\gamma \ket{\psi}} + \xi_\eta \norm{\tilde{X}_\gamma \tilde{X}_\delta \ket{\psi}}
\end{eqnarray}
where $\gamma \neq \delta \neq \eta \in [m]$. Now, extending similar calculation, we get
\begin{equation}
        \norm{\qty(\boldsymbol{\mathcal{\tilde{X}}} - \boldsymbol{\mathcal{X}})\ket{\psi}} \leq f_1(\boldsymbol{\xi}) \ \ \ \text{where $\boldsymbol{\xi} = (\xi_1,\xi_2,\dots,\xi_m)$ and $f_1:\mathbb{R}^n\rightarrow\mathbb{R}$}
\end{equation}
Next, considering the terms involving $\tilde{Z}$ and $Z$,
\begin{eqnarray}
        \norm{\qty(\tilde{Z}_i\tilde{Z}_j - Z_iZ_j)\ket{\psi}} &\approx&\norm{\qty(\tilde{Z}_i(\zeta_j \openone + Z_j) - Z_iZ_j)\ket{\psi}} = \norm{\tilde{Z}_i\zeta_j\ket{\psi} + \qty(\tilde{Z}_iZ_j - Z_iZ_j)\ket{\psi}} \leq \zeta_i + \zeta_j\norm{\tilde{Z}_i\ket{\psi}}  \nonumber \\
        \norm{\qty(\tilde{Z}_i\tilde{Z}_j\tilde{Z}_k - Z_iZ_jZ_k)\ket{\psi}} &\approx& \norm{\qty(\tilde{Z}_i\tilde{Z}_j(\zeta_k \openone + Z_k) - Z_iZ_jZ_k)\ket{\psi}} = \norm{\tilde{Z}_i\tilde{Z}_j\zeta_k\ket{\psi} + \qty(\tilde{Z}_i\tilde{Z}_jZ_k - Z_iZ_jZ_k)\ket{\psi}}  \nonumber \\
        &&\leq \zeta_k\norm{\tilde{Z}_i\tilde{Z}_j\ket{\psi}} + \norm{\qty(\tilde{Z}_i\tilde{Z}_j - Z_iZ_j)Z_k\ket{\psi}} \leq \zeta_i + \zeta_j\norm{\tilde{Z}_i\ket{\psi}} + \zeta_k\norm{\tilde{Z}_i\tilde{Z}_j\ket{\psi}}
\end{eqnarray}
Extending similar calculation, we get
\begin{equation}
    \norm{\qty(\boldsymbol{\tilde{\mathcal{Z}}} - \boldsymbol{\mathcal{Z}})\ket{\psi}} \leq f_2(\boldsymbol{\zeta}) \ \ \ \text{where $\boldsymbol{\zeta} = (\zeta_1,\zeta_2,\dots,\zeta_m)$ and $f_2: \mathbb{R}^n\rightarrow\mathbb{R}$}
\end{equation}
Taking terms of the following types
\begin{eqnarray}
        \norm{\qty(\tilde{X}_\gamma \tilde{Z}_i - X_\gamma Z_i)\ket{\psi}} &\approx &\norm{\qty(\tilde{X}_\gamma(\zeta_i \openone + Z_i) - X_\gamma Z_j)\ket{\psi}} = \norm{\tilde{X}_\gamma \zeta_j \ket{\psi} + \qty(\tilde{X}_\gamma Z_j - X_\gamma Z_j)\ket{\psi}} \nonumber \\
        &&\leq \zeta_j\norm{\tilde{Z}_i\ket{\psi}} + \norm{\qty(\tilde{X}_\gamma - X_\gamma)Z_j\ket{\psi}} \leq \xi_\gamma + \zeta_j\norm{\tilde{Z}_i\ket{\psi}} \nonumber \\
         \norm{\qty(\tilde{X}_\gamma \tilde{X}_\delta \tilde{Z}_i - X_\gamma X_\delta Z_i)\ket{\psi}} &\approx& \norm{\qty(\tilde{X}_\gamma \tilde{X}_\delta(\zeta_i \openone + Z_i) - X_\gamma X_\delta Z_i)\ket{\psi}} = \norm{\tilde{X}_\gamma \tilde{X}_\delta \zeta_i \ket{\psi} + \qty(\tilde{X}_\gamma \tilde{X}_\delta Z_i - X_\gamma X_\delta Z_i)\ket{\psi}} \nonumber \\
        &&\leq \zeta_i\norm{\tilde{X}_\gamma \tilde{X}_\delta \ket{\psi}} + \norm{\qty(\tilde{X}_\gamma \tilde{X}_\delta - X_\gamma X_\delta)Z_i\ket{\psi}} \leq \xi_\gamma + \xi_\delta \norm{\tilde{X}_\gamma \ket{\psi}} + \zeta_i\norm{\tilde{X}_\gamma \tilde{X}_\delta \ket{\psi}} 
\end{eqnarray}
Thus, we obtain
\begin{equation}
    \norm{\qty(\boldsymbol{\tilde{\mathcal{X}}}\tilde{Z}_i - \boldsymbol{\mathcal{X}} Z_i)\ket{\psi}} \leq f_{3,i}(\boldsymbol{\zeta},\boldsymbol{\xi}) \ \ \text{where $f_{3,i}: \mathbb{R}^n\times\mathbb{R}^n\rightarrow\mathbb{R}$} \ \forall i \in [m]
\end{equation}
Now evaluating terms of the following type
\begin{eqnarray}
\norm{\qty(\tilde{X}_\gamma\tilde{Z}_i\tilde{Z}_j - X_\gamma Z_iZ_j)\ket{\psi}} &\approx& \norm{\qty(\tilde{X}_\gamma\tilde{Z}_i(\zeta_j \openone + Z_j) - X_\gamma Z_iZ_j)\ket{\psi}} = \norm{\qty(\tilde{X}_\gamma\tilde{Z}_i\zeta_j  + \tilde{X}_\gamma\tilde{Z}_iZ_j) - X_\gamma 
        Z_iZ_j)\ket{\psi}} \nonumber \\
        &&\leq \zeta_j \norm{\tilde{X}_\gamma\tilde{Z}_i\ket{\psi}}  + \norm{\qty(\tilde{X}_\gamma\tilde{Z}_iZ_j - X_\gamma Z_i Z_j)\ket{\psi}} \leq \xi_\gamma + \zeta_i\norm{\tilde{Z}_i\ket{\psi}} + \zeta_j \norm{\tilde{X}_\gamma\tilde{Z}_i\ket{\psi}} \nonumber \\
         \norm{\qty(\tilde{X}_\gamma \tilde{X}_\delta \tilde{Z}_i\tilde{Z}_j - X_\gamma X_\delta Z_iZ_j)\ket{\psi}} &\approx& \norm{\qty(\tilde{X}_\gamma \tilde{X}_\delta \tilde{Z}_i(\zeta_j \openone + Z_j) - X_\gamma X_\delta Z_iZ_j)\ket{\psi}} = \norm{\tilde{X}_\gamma\tilde{X}_\delta \tilde{Z}_i\zeta_j\ket{\psi}  + \qty(\tilde{X}_\gamma\tilde{X}_\delta \tilde{Z}_iZ_j - X_\gamma X_\delta  
        Z_iZ_j)\ket{\psi}} \nonumber \\
        &&\leq \xi_\gamma + \xi_\delta \norm{\tilde{X}_\gamma \ket{\psi}} + \zeta_i\norm{\tilde{X}_\gamma \tilde{X}_\delta \ket{\psi}} + \zeta_j \norm{\tilde{X}_\gamma\tilde{X}_\delta \tilde{Z}_i\ket{\psi}}
\end{eqnarray}
Then we obtain
\begin{equation}
        \norm{\qty(\boldsymbol{\tilde{\mathcal{X}}}\tilde{Z}_i\tilde{Z}_j - \boldsymbol{\mathcal{X}}Z_iZ_j)\ket{\psi}} \leq f_{i,j}(\boldsymbol{\zeta},\boldsymbol{\xi}) \ \ \text{where $f_{i,j}: \mathbb{R}^n\times\mathbb{R}^n\rightarrow\mathbb{R}$} \ \forall i\neq j \in [m]
\end{equation}
Similar to the aforementioned expressions, the generalised expression shall be 
\begin{equation}
    \norm{\qty(\boldsymbol{\tilde{\mathcal{X}}}\boldsymbol{\tilde{\mathcal{Z}}} - \boldsymbol{\mathcal{X}}\boldsymbol{\mathcal{Z}})\ket{\psi}} \leq f_4(\boldsymbol{\zeta},\boldsymbol{\xi}) \ \ \ \text{where $f_4: \mathbb{R}^n\times\mathbb{R}^n\rightarrow\mathbb{R}$}
\end{equation}


\subsection{Robust self-testing of state}\label{rssap}

Now, to evaluate the trace distance given by Eq.~(\ref{gentradis}) in the main text, we proceed as follows.
\begin{eqnarray}\label{triin}
            &&\norm{\tilde{\Phi}\qty({\ket{\psi}\otimes\ket{0}^{\otimes m}}) - \Phi\qty({\ket{\psi}\otimes\ket{0}^{\otimes m}})} \nonumber \\
            &=& \bigg\Vert \frac{1}{2^m} \sum_{a_{k'} \in \{0,1\}}  \qty(\Motimes_{k=1}^m   \qty(\tilde{X}_k)^{a_{k'=k}}\qty(\openone + (-1)^{a_{k'=k}}\tilde{Z}_k))\ket{\psi}\Motimes_{k' = 1}^m\ket{a_{k'}} -\frac{1}{2^m} \sum_{a_{k'} \in \{0,1\}}  \qty(\Motimes_{k=1}^m   \qty(X_i)^{a_{k'=k}}\qty(\openone + (-1)^{a_{k'=k}}Z_i))\ket{\psi}\Motimes_{k'=1}^m\ket{a_{k'}} \bigg\Vert \nonumber \\     
            & \leq& \frac{1}{2^m} \sum_{a_{k'} \in \{0,1\}} \Bigg\Vert \Bigg\{ \Motimes_{k=1}^m   \Bigg[\qty(\tilde{X}_k)^{a_{k'=k}}\qty(\openone + (-1)^{a_{k'=k}}\tilde{Z}_k)\Bigg]  - \Motimes_{k=1}^m   \Bigg[\qty(X_k)^{a_{k'=k}}\qty(\openone + (-1)^{a_{k'=k}}Z_k)\Bigg]\Bigg\} \ket{\psi}\Motimes_{k'=1}^m\ket{a_{k'}} \Bigg\Vert\ \ \text{(using $\norm{C+D}\leq \norm{C}+\norm{D}$)} \nonumber \\
             &=& \frac{1}{2^m}  \sum_{a_{k'} \in \{0,1\}}\Bigg\Vert \  \Bigg[\boldsymbol{\tilde{\mathcal{X}}} \ \Big\{\openone + \sum_{i=1}^m (-1)^{a_i}\tilde{Z}_i +\sum_{i<j} (-1)^{a_i+a_j}\tilde{Z}_i\tilde{Z}_j + \sum_{i<j<k} (-1)^{a_i+a_j+a_k}\tilde{Z}_i\tilde{Z}_j\tilde{Z}_k + \dots  \Big\} \nonumber \\ 
        && - \boldsymbol{\mathcal{X}} \  \Big\{\openone + \sum_{i=1}^m(-1)^{a_i} Z_i + \sum_{i<j} (-1)^{a_i+a_j}Z_iZ_j + \sum_{i<j<k} (-1)^{a_i+a_j+a_k}Z_iZ_jZ_k + \dots\Big\}\Bigg]\ket{\psi}\ket{a_1,a_2,\dots,a_m} \ \Bigg\Vert
\end{eqnarray}
Collecting terms with same indices, we re-express the right hand side of above Eq.~(\ref{triin}) as follows
\begin{eqnarray} \label{complieqcontd}
        &=& \frac{1}{2^m}  \sum_{a_{k'} \in \{0,1\}}\Bigg\Vert \qty(\boldsymbol{\tilde{\mathcal{X}}} - \boldsymbol{\mathcal{X}}) \ket{\psi}\ket{a_1,a_2,\dots,a_m} + \sum_{i=1}^{m} (-1)^{a_i}\qty(\boldsymbol{\tilde{\mathcal{X}}} \ \tilde{Z}_i - \boldsymbol{\mathcal{X}} \ Z_i)\ket{\psi}\ket{a_1,a_2,\dots,a_m} \nonumber \\
        &&+ \sum_{i<j} (-1)^{a_i+a_j} \qty(\boldsymbol{\tilde{\mathcal{X}}} \ \tilde{Z}_i\tilde{Z}_j - \boldsymbol{\mathcal{X}} \ Z_iZ_j)\ket{\psi}\ket{a_1,a_2,\dots,a_m} + \sum_{i<j<k} (-1)^{a_i+a_j+a_k} \qty(\boldsymbol{\tilde{\mathcal{X}}} \ \tilde{Z}_i\tilde{Z}_j\tilde{Z}_k - \boldsymbol{\mathcal{X}} \ Z_iZ_jZ_k)\ket{\psi}\ket{a_1,a_2,\dots,a_m} + \dots \Bigg\Vert \nonumber \\
        &\leq& \frac{1}{2^m}  \sum_{a_{k'} \in \{0,1\}} \Biggl[ \big\Vert\qty(\boldsymbol{\tilde{\mathcal{X}}} - \boldsymbol{\mathcal{X}}) \ket{\psi}\ket{a_1,a_2,\dots,a_m} \big\Vert + \sum_{i=1}^{m} \norm{(-1)^{a_i}\qty(\boldsymbol{\tilde{\mathcal{X}}} \ \tilde{Z}_i - \boldsymbol{\mathcal{X}} \ Z_i)\ket{\psi}\ket{a_1,a_2,\dots,a_m}} \nonumber \\
        & & + \sum_{i<j} \norm{(-1)^{a_i+a_j} \qty(\boldsymbol{\tilde{\mathcal{X}}} \ \tilde{Z}_i\tilde{Z}_j - \boldsymbol{\mathcal{X}} \ Z_iZ_j)\ket{\psi}\ket{a_1,a_2,\dots,a_m}} + \sum_{i<j<k} \norm{(-1)^{a_i+a_j+a_k} \qty(\boldsymbol{\tilde{\mathcal{X}}} \ \tilde{Z}_i\tilde{Z}_j\tilde{Z}_k - \boldsymbol{\mathcal{X}} \ Z_iZ_jZ_k)\ket{\psi}\ket{a_1,a_2,\dots,a_m}} + \dots \Biggl] \nonumber \\
        &\leq& \frac{1}{2^m}  \sum_{a_{k'} \in \{0,1\}} \Biggl[ \big\Vert\qty(\boldsymbol{\tilde{\mathcal{X}}} - \boldsymbol{\mathcal{X}}) \ket{\psi}\ket{a_1,a_2,\dots,a_m} \big\Vert + \sum_{i=1}^{m} \norm{\qty(\boldsymbol{\tilde{\mathcal{X}}} \ \tilde{Z}_i - \boldsymbol{\mathcal{X}} \ Z_i)\ket{\psi}\ket{a_1,a_2,\dots,a_m}} \nonumber \\
        & & + \sum_{i<j} \norm{ \qty(\boldsymbol{\tilde{\mathcal{X}}} \ \tilde{Z}_i\tilde{Z}_j - \boldsymbol{\mathcal{X}} \ Z_iZ_j)\ket{\psi}\ket{a_1,a_2,\dots,a_m}} + \sum_{i<j<k} \norm{ \qty(\boldsymbol{\tilde{\mathcal{X}}} \ \tilde{Z}_i\tilde{Z}_j\tilde{Z}_k - \boldsymbol{\mathcal{X}} \ Z_iZ_jZ_k)\ket{\psi}\ket{a_1,a_2,\dots,a_m}} + \dots \Biggl]
\end{eqnarray}
To simplify the proof we show that the upper bound of one of the terms in Eq.~(\ref{complieqcontd}) is some function of $\xi_i, \zeta_i$. Following this, it would be easy to see that each of the terms in the aforementioned equation is some function of $\xi_i, \zeta_i$. Note that, we will be suppressing the $\ket{a_1,a_2,\dots,a_m}$ notation.


\subsubsection{Considering the term $\ket{0}^{\otimes m}$, where $a_{k'} = 0$}
Here $\boldsymbol{\tilde{\mathcal{X}}} = \boldsymbol{\mathcal{X}} = \openone$ and there is only one term of this type
\begin{equation}\label{h13}
    \begin{aligned}
        &\frac{1}{2^m}\qty(\norm{(\boldsymbol{\tilde{\mathcal{X}}}-\boldsymbol{\mathcal{X}})\ket{\psi}} + \sum_{i=1}^m \norm{\qty(\boldsymbol{\tilde{\mathcal{X}}} \ \tilde{Z}_i - \boldsymbol{\mathcal{X}} \ Z_i)\ket{\psi}} +  \sum_{i<j} \norm{\qty(\boldsymbol{\tilde{\mathcal{X}}} \ \tilde{Z}_i\tilde{Z}_j -\boldsymbol{\mathcal{X}} \ Z_iZ_j)\ket{\psi}} + \sum_{i<j<k} \norm{\qty(\boldsymbol{\tilde{\mathcal{X}}} \ \tilde{Z}_i\tilde{Z}_j\tilde{Z}_k - \boldsymbol{\mathcal{X}} \ Z_iZ_jZ_k)\ket{\psi}} + \dots)\\
        &=  \frac{1}{2^m}\qty(  \sum_{i=1}^m \norm{\qty( \tilde{Z}_i -  Z_i)\ket{\psi}} + \sum_{i<j} \norm{\qty( \tilde{Z}_i\tilde{Z}_j -Z_iZ_j)\ket{\psi}} + \sum_{i<j<k} \norm{\qty( \tilde{Z}_i\tilde{Z}_j\tilde{Z}_k - \ Z_iZ_jZ_k)\ket{\psi}} + \dots)\\
        &\leq  \frac{1}{2^m}\qty(\sum_{i=1}^m \zeta_i + \sum_{i<j} \qty(\zeta_i + \zeta_j\norm{\tilde{Z}_i\ket{\psi}} ) + \sum_{i<j<k} \qty(\zeta_i + \zeta_j\norm{\tilde{Z}_i\ket{\psi}} + \zeta_k\norm{\tilde{Z}_i\tilde{Z}_j\ket{\psi}}) + \dots) \leq \frac{1}{2^m} \ F_0(\boldsymbol{\zeta})
    \end{aligned}
\end{equation}


\subsubsection{Considering the term involving one of the $a_{k'}$ to be one and rest $a_{k'}$ to be zero}

Let one of the non zero terms be $a_\gamma = 1$. In this case, $\boldsymbol{\tilde{\mathcal{X}}} = \tilde{X}_\gamma$ and $\boldsymbol{\mathcal{X}} = X_\gamma$. For each choice of $\gamma \in [m]$, each term is upper bounded as follows
\begin{equation}\label{h14}
    \begin{aligned}
        &\frac{1}{2^m}\qty(\norm{(\boldsymbol{\tilde{\mathcal{X}}}-\boldsymbol{\mathcal{X}})\ket{\psi}} + \sum_{i<j} \norm{\qty(\boldsymbol{\tilde{\mathcal{X}}} \ \tilde{Z}_i\tilde{Z}_j -\boldsymbol{\mathcal{X}} \ Z_iZ_j)\ket{\psi}} + \sum_{i<j<k} \norm{\qty(\boldsymbol{\tilde{\mathcal{X}}} \ \tilde{Z}_i\tilde{Z}_j\tilde{Z}_k - \boldsymbol{\mathcal{X}} \ Z_iZ_jZ_k)\ket{\psi}} + \dots)\\
        &=  \frac{1}{2^m}\qty( \norm{(\tilde{X}_\gamma-X_\gamma)\ket{\psi}} + \sum_{i<j} \norm{\qty( \tilde{X}_\gamma\tilde{Z}_i\tilde{Z}_j -X_\gamma Z_iZ_j)\ket{\psi}} + \sum_{i<j<k} \norm{\qty( \tilde{X}_\gamma\tilde{Z}_i\tilde{Z}_j\tilde{Z}_k - X_\gamma Z_iZ_jZ_k)\ket{\psi}} + \dots)\\
        & \leq \frac{1}{2^m}\qty(\xi_\gamma + \sum_{i<j} \qty(\xi_\gamma + \zeta_i \norm{\tilde{X}_\gamma\ket{\psi}} + \zeta_j \norm{\tilde{X}_\gamma\tilde{Z}_i\ket{\psi}} ) + \sum_{i<j<k} \qty(\zeta_i + \zeta_j\norm{\tilde{Z}_i\ket{\psi}} + \zeta_k\norm{\tilde{Z}_i\tilde{Z}_j\ket{\psi}}) + \dots) \leq \frac{1}{2^m}F_{\gamma,1}(\boldsymbol{\zeta})
    \end{aligned}
\end{equation}
There will be $m$ such terms, each upper bounded by some function of $\boldsymbol{\zeta}$. Now extending the analysis presented in Eqs.~(\ref{h13}) and (\ref{h14}) for other combinations of $a_{k'}$, we conclude the following
\begin{equation}
\norm{\tilde{\Phi}\qty({\ket{\psi}\otimes\ket{0}^{\otimes m}}) - \Phi\qty({\ket{\psi}\otimes\ket{0}^{\otimes m}})} \leq F(\boldsymbol{\xi},\boldsymbol{\zeta}) \ \ \ \ \ \text{[$\boldsymbol{\xi},\boldsymbol{\zeta} \rightarrow 0 \implies F(\boldsymbol{\xi},\boldsymbol{\zeta}) \rightarrow 0$]}
\end{equation}


\subsection{Robust self-testing of observables}\label{rsoap}

We begin with the analysis of the robustness of the observables $X_{\alpha}$ and $Z_{\alpha}$ associated with $k=\alpha$-th party. Subsequently, we will argue that by adhering to the same procedure, all observables pertaining to all the parties can be collectively and robustly self-tested. For the robust self-testing of $X_{\alpha}$, we get
    \begin{eqnarray}
        \norm{\tilde{\Phi}\qty({\tilde{X}_\alpha\ket{\psi}\ket{0}^{\otimes m}}) - \Phi\qty(X_\alpha{\ket{\psi}\ket{0}^{\otimes m }})} &\leq& \frac{1}{2^m} \sum_{a_{k'} \in \{0,1\}} \Bigg\Vert \Motimes_{k=1}^m   \Bigg[\qty(\tilde{X}_k)^{a_{k'=k}}\qty(\openone + (-1)^{a_{k'=k}}\tilde{Z}_k)\tilde{X}_\alpha  -   \qty(X_k)^{a_{k'=k}}\qty(\openone + (-1)^{a_{k'=k}}Z_k)X_\alpha\Bigg]\ket{\psi} \Motimes_{k'=1}^m \ket{a_{k'}} \bigg\Vert \nonumber \\
        &\leq& \frac{1}{2^m}  \sum_{a_{k'}\in \{0,1\}} \Biggl[ \big\Vert\qty(\boldsymbol{\mathcal{\tilde{X}}}\tilde{X}_\alpha - \boldsymbol{\mathcal{X}}X_\alpha) \ket{\psi}\Motimes_{k'=1}^m \ket{a_{k'}} \big\Vert + \sum_{i=1}^{m} \norm{\qty(\boldsymbol{\mathcal{\tilde{X}}} \ \tilde{Z}_i\tilde{X}_\alpha - \boldsymbol{\mathcal{X}} \ Z_iX_\alpha)\ket{\psi}\Motimes_{k'=1}^m \ket{a_{k'}}} \nonumber\\
        &+& \sum_{i<j} \norm{ \qty(\boldsymbol{\mathcal{\tilde{X}}} \ \tilde{Z}_i\tilde{Z}_j\tilde{X}_\alpha - \boldsymbol{\mathcal{X}} \ Z_iZ_jX_\alpha)\ket{\psi}\Motimes_{k'=1}^m \ket{a_{k'}}} + \cdots +\sum \norm{ \qty(\boldsymbol{\mathcal{\tilde{X}}} \ \boldsymbol{\mathcal{\tilde{Z}}}\tilde{X}_\alpha - \boldsymbol{\mathcal{X}} \ \boldsymbol{\mathcal{Z}}X_\alpha)\ket{\psi}\Motimes_{k'=1}^m \ket{a_{k'}}}  \Biggl] \nonumber \\
         &\leq& \frac{1}{2^m}  \sum_{a_{k'} \in \{0,1\}} \Biggl[ \xi_\alpha \norm{\boldsymbol{\mathcal{X}}\ket{\psi}\Motimes_{k'=1}^m \ket{a_{k'}}} + \big\Vert\qty(\boldsymbol{\mathcal{\tilde{X}}} - \boldsymbol{\mathcal{X}}) \ket{\psi}\Motimes_{k'=1}^m \ket{a_{k'}} \big\Vert \nonumber\\
         &+& \sum_{i=1}^{m} \qty(\xi_\alpha \norm{\boldsymbol{\mathcal{\tilde{X}}} \ \tilde{Z}_i\ket{\psi}\Motimes_{k'=1}^m \ket{a_{k'}}} + \norm{\qty(\boldsymbol{\mathcal{\tilde{X}}} \ \tilde{Z}_i - \boldsymbol{\mathcal{X}} \ Z_i)\ket{\psi}\Motimes_{k'=1}^m \ket{a_{k'}}}) \nonumber\\
         &+& \sum_{i<j} \qty(\xi_\alpha \norm{\boldsymbol{\mathcal{\tilde{X}}} \ \tilde{Z}_i\tilde{Z}_j\ket{\psi}\Motimes_{k'=1}^m \ket{a_{k'}}} +\norm{ \qty(\boldsymbol{\mathcal{\tilde{X}}} \ \tilde{Z}_i\tilde{Z}_j - \boldsymbol{\mathcal{X}} \ Z_iZ_j)\ket{\psi}\Motimes_{k'=1}^m \ket{a_{k'}}} ) + \cdots \nonumber \\
         &+& \sum \qty( \xi_\alpha \norm{\boldsymbol{\mathcal{\tilde{X}}} \ \boldsymbol{\mathcal{\tilde{Z}}}\Motimes_{k'=1}^m \ket{a_{k'}}} + 
         \norm{ \qty(\boldsymbol{\mathcal{\tilde{X}}} \ \boldsymbol{\mathcal{\tilde{Z}}} - \boldsymbol{\mathcal{X}} \ \boldsymbol{\mathcal{Z}})\ket{\psi}\Motimes_{k'=1}^m \ket{a_{k'}}})  \Biggl] \nonumber \\
         &\leq& F_{1,\alpha}(\boldsymbol{\xi},\boldsymbol{\zeta}) \ \ \ \ \forall \alpha \in [m]
    \end{eqnarray}
Following the similar calculation as done in the previous section, we obtain the norm for robust self-testing of $Z_\alpha$. 
\begin{eqnarray}
\norm{\tilde{\Phi}\qty({\tilde{Z}_\alpha\ket{\psi}\ket{0}^{\otimes m}}) - \Phi\qty(Z_\alpha{\ket{\psi}\ket{0}^{\otimes m }})} &\leq& \frac{1}{2^m} \sum_{a_{k'}\in \{0,1\}} \Bigg\Vert \Motimes_{k=1}^m   \Bigg[\qty(\tilde{X}_k)^{a_{k'=k}}\qty(\openone + (-1)^{a_{k'=k}}\tilde{Z}_k)\tilde{Z}_\alpha  -   \qty(X_k)^{a_{k'=k}}\qty(\openone + (-1)^{a_{k'=k}}Z_k)Z_\alpha\Bigg]\ket{\psi} \Motimes_{k'=1}^m \ket{a_{k'}} \bigg\Vert \nonumber \\
&\leq& \frac{1}{2^m}  \sum_{a_{k'} \in \{0,1\}} \Biggl[ \big\Vert\qty(\boldsymbol{\mathcal{\tilde{X}}}\tilde{Z}_\alpha - \boldsymbol{\mathcal{X}}Z_\alpha) \ket{\psi}\Motimes_{k'=1}^m \ket{a_{k'}} \big\Vert + \sum_{i=1}^{m} \norm{\qty(\boldsymbol{\mathcal{\tilde{X}}} \ \tilde{Z}_i\tilde{Z}_\alpha - \boldsymbol{\mathcal{X}} \ Z_iZ_\alpha)\ket{\psi}\Motimes_{k'=1}^m \ket{a_{k'}}} \nonumber\\
&+& \sum_{i<j} \norm{ \qty(\boldsymbol{\mathcal{\tilde{X}}} \ \tilde{Z}_i\tilde{Z}_j\tilde{Z}_\alpha - \boldsymbol{\mathcal{X}} \ Z_iZ_jZ_\alpha)\ket{\psi}\Motimes_{k'=1}^m \ket{a_{k'}}} + \cdots +\sum \norm{ \qty(\boldsymbol{\mathcal{\tilde{X}}} \ \boldsymbol{\mathcal{\tilde{Z}}}\tilde{Z}_\alpha - \boldsymbol{\mathcal{X}} \ \boldsymbol{\mathcal{Z}}Z_\alpha)\ket{\psi}\Motimes_{k'=1}^m \ket{a_{k'}}}  \Biggl] \nonumber \\
&\leq& F_{2,\alpha}(\boldsymbol{\xi},\boldsymbol{\zeta})  \ \ \forall \alpha \in [m]
\end{eqnarray}


\section{Special case: When only first party implements imperfect observables}

If only first party implements imperfect measurements then
\begin{equation}
\norm{\qty(\tilde{X}_1 - X_1)\ket{\psi}} \leq \xi_1 \ ; \ \ \norm{\qty(\tilde{Z}_1 - Z_1)\ket{\psi}} \leq \zeta_1 \ ;    
\end{equation}
The construction of $L_\mu$ would be affected and constructed as $\tilde{L}_\mu$. Now, there are $2^{m-1}$ $\tilde{L}_{\mu}$ out of which, $2^{m-2}$ are constructed using $\tilde{X}_1$ and the other $2^{m-2}$ are constructed using $\tilde{Z}_1$. This leads to 
\begin{equation}
    \Tr[\tilde{\Gamma}_m \ \rho]=\frac{1}{\sqrt{2}} \sum\limits_{\mu=1}^{2^{m-1}} \bra{\psi} \tilde{L}^\dagger_\mu \tilde{L}_\mu \ket{\psi} \leq 2^{m-\frac{3}{2}} \qty(\zeta_1^2 + \xi_1^2)
\end{equation}
If the maximal violation of Svetlichny functional is $2^{m-1}\sqrt{2} - \epsilon$ then $\epsilon = 2^{m-\frac{3}{2}} \qty(\zeta_1^2 + \xi_1^2)$.
For robust self-testing of the state, we get
    \begin{eqnarray}
        \norm{\tilde{\Phi}\qty({\ket{\psi}\ket{0}^{\otimes m}}) - \Phi\qty({\ket{\psi}\ket{0}^{\otimes m }})} &\leq& \frac{1}{2^m} \sum_{a_{k'} \in \{0,1\}} \Bigg\Vert \Motimes_{k=1}^m   \Bigg[\qty(\tilde{X}_k)^{a_{k'=k}}\qty(\openone + (-1)^{a_{k'=k}}\tilde{Z}_k)  -   \qty(X_k)^{a_{k'=k}}\qty(\openone + (-1)^{a_{k'=k}}Z_k)\Bigg]\ket{\psi} \Motimes_{k'=1}^m \ket{a_{k'}} \bigg\Vert \nonumber \\ 
        &\leq& \frac{1}{2^m} \sum_{a_{k'} \in \{0,1\}} \Bigg\Vert   \Bigg[\qty(\tilde{X}_1)^{a_1}\qty(\openone + (-1)^{a_1}\tilde{Z}_1)  -   \qty(X_1)^{a_1}\qty(\openone + (-1)^{a_1}Z_1)\Bigg]\Motimes_{k=2}^m \qty(X_k)^{a_{k'=k}}\qty(\openone + (-1)^{a_{k'=k}}Z_k)\ket{\psi} \Motimes_{k'=1}^m \ket{a_{k'}} \bigg\Vert \nonumber\\
        &\leq& \frac{1}{2^m} \sum_{a_2,\dots,a_m \atop a_{k'} \in \{0,1\}} \Bigg\Vert   \Bigg[\qty(\openone + \tilde{Z}_1) - \qty(\openone + Z_1)\Bigg]\Motimes_{k=2}^m \qty(X_k)^{a_{k'=k}}\qty(\openone + (-1)^{a_{k'=k}}Z_k)\ket{\psi} \otimes\ket{0}\Motimes_{k'=2}^m \ket{a_{k'}} \bigg\Vert \nonumber \\
         &+& \frac{1}{2^m} \sum_{a_2,\dots,a_m \atop a_{k'} \in \{0,1\}} \Bigg\Vert   \Bigg[\tilde{X}_1\qty(\openone + \tilde{Z}_1) - X_1\qty(\openone + Z_1)\Bigg]\Motimes_{k=2}^m \qty(X_k)^{a_{k'=k}}\qty(\openone + (-1)^{a_{k'=k}}Z_k)\ket{\psi} \otimes\ket{1}\Motimes_{k'=2}^m \ket{a_{k'}} \bigg\Vert \nonumber \\
         &\leq& \frac{1}{2} \qty[\norm{\qty(\tilde{Z}_1 - Z_1)\ket{\psi}} + \norm{\qty(\tilde{X}_1 - X_1)\ket{\psi}}+ \norm{\qty(\tilde{X}_1\tilde{Z}_1 - X_1Z_1)\ket{\psi}}] \nonumber \\
         &\leq& \frac{1}{2} \qty[\zeta_1\qty(1+\norm{\tilde{X}_1\ket{\psi}}) + 2 \xi_1  ] \ \  \leq \frac{1}{4} \qty[\zeta_1\qty(2+\xi_1) + 2 \xi_1  ]
    \end{eqnarray}
For robust self-testing of $Z_1$, we have
    \begin{eqnarray}
        \norm{\tilde{\Phi}\qty({\tilde{Z}_1\ket{\psi}\ket{0}^{\otimes m}}) - \Phi\qty({Z_1\ket{\psi}\ket{0}^{\otimes m }})} &\leq& \frac{1}{2^m} \sum_{a_{k'} \in \{0,1\}} \Bigg\Vert \Motimes_{k=1}^m   \Bigg[\qty(\tilde{X}_k)^{a_{k'=k}}\qty(\openone + (-1)^{a_{k'=k}}\tilde{Z}_k)\tilde{Z}_1  -   \qty(X_k)^{a_{k'=k}}\qty(\openone + (-1)^{a_{k'=k}}Z_k)Z_1\Bigg]\ket{\psi} \Motimes_{k'=1}^m \ket{a_{k'}} \bigg\Vert \\
        &\leq& \frac{1}{2^m} \sum_{a_{k'} \in \{0,1\} \atop k' \neq 1} \Bigg\Vert   \Bigg[\qty(\tilde{Z}_1 + \tilde{Z}_1\tilde{Z}_1) - \qty(Z_1 + Z_1Z_1)\Bigg]\Motimes_{k=2}^m \qty(X_k)^{a_{k'=k}}\qty(\openone + (-1)^{a_{k'=k}}Z_k)\ket{\psi} \otimes\ket{0}\Motimes_{k'=2}^m \ket{a_{k'}} \bigg\Vert \nonumber \\
         &+& \frac{1}{2^m} \sum_{a_{k'} \in \{0,1\} \atop k' \neq 1} \Bigg\Vert   \Bigg[\tilde{X}_1\qty(\tilde{Z}_1 - \tilde{Z}_1\tilde{Z}_1) - X_1\qty(\openone - Z_1)Z_1\Bigg]\Motimes_{k=2}^m \qty(X_k)^{a_{k'=k}}\qty(\openone + (-1)^{a_{k'=k}}Z_k)\ket{\psi} \otimes\ket{1}\Motimes_{k'=2}^m \ket{a_{k'}} \bigg\Vert \nonumber \\
         &\leq& \frac{1}{2} \Bigg[\norm{\qty(\tilde{Z}_1 - Z_1)\ket{\psi}} + \norm{\qty(\tilde{Z}_1\tilde{Z}_1 - Z_1Z_1)\ket{\psi}} \nonumber\\
         &+& \norm{\qty(\tilde{X}_1\tilde{Z}_1 - X_1Z_1)\ket{\psi}} + \norm{\qty(\tilde{X}_1\tilde{Z}_1\tilde{Z}_1 - X_1Z_1Z_1)\ket{\psi}}\Bigg] \nonumber\\
         &\leq& \frac{1}{2} \qty[2 \zeta_1 + \zeta_1 \norm{\tilde{Z}_1\ket{\psi}} + \zeta_1 \norm{\tilde{X}_1\ket{\psi}} + 2 \xi_1 + \xi_1 \norm{\tilde{X}_1\ket{\psi}} + \xi_1\norm{\tilde{X}_1\tilde{Z}_1\ket{\psi}}] \nonumber\\
         &\leq& \frac{1}{2} \qty[\zeta_1\qty(2 + \norm{\tilde{Z}_1\ket{\psi}} + \norm{\tilde{X}_1\ket{\psi}}) + \xi_1 \qty(2 + \norm{\tilde{X}_1\ket{\psi}} + \norm{\tilde{X}_1\tilde{Z}_1\ket{\psi}})]\nonumber \\
         &\leq& \frac{1}{2} \qty[6 \zeta_1 + 2 \zeta_1^2 + \zeta_1^2\xi_1 + 3\zeta_1 \xi_1 + 2\xi_ 1]
    \end{eqnarray}
Similar analysis for $X_1$ gives us
\begin{equation}
        \norm{\tilde{\Phi}\qty({\tilde{X}_1\ket{\psi}\ket{0}^{\otimes m}}) - \Phi\qty({X _1\ket{\psi}\ket{0}^{\otimes m }})} \leq \frac{1}{2} \qty(6\xi_1 + 2\xi_1^2 + \xi_1^2\zeta_1 + 3\zeta_1\xi_1 + 2\zeta_1)
\end{equation}

\end{document}